\documentclass[twocolumn,showpacs,preprintnumbers,amsmath,amssymb]{revtex4}
\usepackage{epsfig}
\usepackage{dcolumn}
\usepackage{bm}
\usepackage{amsthm}

\newcommand{\remove}[1]{}
\newcommand{\dd}{\mathrm{d}}

\def\be{\begin{equation}}
\def\ee{\end{equation}}

\newcommand{\beq}{\begin{equation}}
\newcommand{\eeq}{\end{equation}}
\newcommand{\beqa}{\begin{eqnarray}}
\newcommand{\eeqa}{\end{eqnarray}}

\renewcommand{\pl}{\partial}

\newcommand{\vu}{{\bf u}}
\newcommand{\vv}{{\bf v}}

\newcommand{\vx}{{\bf x}}
\newcommand{\vk}{{\bf k}}

\renewcommand{\vr}{{\bf r}}

\newcommand{\cG}{{\cal G}}

\newcommand{\bea}{\begin{array}}
\newcommand{\ea}{\end{array}}

\begin{document}


\title{K-mouflage effects on clusters of galaxies}

\author{Philippe Brax}
\affiliation{Institut de Physique Th\'eorique,\\
Universit\'e Paris-Saclay CEA, CNRS, F-91191 Gif-sur-Yvette, C\'edex, France\\}

\author{Luca Alberto Rizzo}
\affiliation{Institut de Physique Th\'eorique,\\
Universit\'e Paris-Saclay CEA, CNRS, F-91191 Gif-sur-Yvette, C\'edex, France\\}

\author{Patrick Valageas}
\affiliation{Institut de Physique Th\'eorique,\\
Universit\'e Paris-Saclay CEA, CNRS, F-91191 Gif-sur-Yvette, C\'edex, France\\}
\vspace{.2 cm}

\date{\today}
\vspace{.2 cm}

\begin{abstract}
We investigate the effects of a K-mouflage modification of gravity on the dynamics of clusters
of galaxies. We extend the description of K-mouflage to situations where the scalar field responsible
for the modification of gravity is coupled to a perfect fluid with pressure. We describe the coupled
system at both the background cosmology and cosmological perturbations levels, focusing on cases
where the pressure emanates from small-scale nonlinear physics.
We derive these properties in both the Einstein and Jordan frames, as these two frames already
differ by a few percents at the background level for K-mouflage scenarios, and next compute
cluster properties in the Jordan frame that is better suited to these observations.
Galaxy clusters are not screened by the K-mouflage mechanism and therefore feel the modification of
gravity in a maximal way. This implies that the halo mass function deviates from $\Lambda$-CDM by
a factor of order one for masses $M\gtrsim 10^{14} \ h^{-1} M_\odot$. We then consider the
hydrostatic equilibrium of gases embedded in galaxy clusters and  the consequences of K-mouflage
on the X-ray cluster luminosity, the gas temperature, and the Sunyaev-Zel'dovich  effect. We find that
the cluster temperature function, and more generally number counts, are largely affected by
K-mouflage, mainly due to the increased cluster abundance in these models. Other scaling
relations such as the mass-temperature and the temperature-luminosity relations are only modified
at the percent level due to the constraints on K-mouflage from local Solar System tests.

\end{abstract}

 \vskip2pc

\maketitle

\section{Introduction}
\label{sec:Introduction}

K-mouflage \cite{Babichev:2009ee,Brax:2012jr} is one of the four types of screening
mechanisms, together with the chameleon \cite{Khoury:2003aq,Khoury:2003rn}, the
Damour-Polyakov \cite{Damour:1994zq}
and the
Vainshtein \cite{Vainshtein:1972sx} ones, which are compatible with second-order
equations of motion for single scalar field models \cite{Brax:2014a}. The properties
of K-mouflage have already been thoroughly studied both at the background
cosmology \cite {Brax:2014a} and perturbation levels
\cite{Brax:2014b,Barreira:2014a}, see also \cite{Sawicki:2013} for a more general
analysis within an ``imperfect-fluid'' formalism.
The small-scale dynamics have been studied in \cite{Brax:2014c}, and models that
satisfy both cosmological and Solar System constraints have been devised
in \cite{Barreira2015}.

In this paper, we extend these studies by including fluids with pressure as befitting
the description of gases in galaxy clusters. We do so for both the background and
perturbations. We also present the dynamics of the system in both the Einstein
frame (used in previous works) and the Jordan frame, and discuss the relations
between both frames.
Because the properties of gases, such as the X-ray luminosity or the
Sunyaev-Zel'dovich effect \cite{Sunyaev1972}, or the wavelength of atomic emission
or absorption lines, are tied to the frame in which atomic physics is described
without any modification due to the scalar field, we  work in the Jordan frame
to describe clusters of galaxies.
We use the fact that galaxy clusters are not screened by the K-mouflage mechanism
and that their number would be increased as compared with $\Lambda$-CDM in this
scenario. As a result, clusters appear as a useful testing ground for K-mouflage
and its effects on the growth of structure.

We take into account the modification of the hydrostatic equilibrium in K-mouflage
models, together with the change of the matter density profiles, which we find
to become slightly more compact. This allows us to investigate the X-ray luminosity
as a function of the gas temperature. The deviation from $\Lambda$-CDM is at the
percent level and is set by the tests of gravity in the Solar System,
which strongly constrain the coupling constant that defines these models.
In a similar fashion, the temperature-mass relation is affected at the same level.
As particular examples, we  focus on a simple ``cubic'' K-mouflage model
(that agrees with cosmological constraints) and on an ``arctan'' model which satisfies
cosmological constraints as well as all Solar System tests, including the advance of the
perihelion of the Moon \cite{Barreira2015}. The latter gives slightly amplified results
as compared to the former, but both remain at the percent level.
The only observable which deviates significantly from $\Lambda$-CDM is the cluster
temperature function (or more generally, number counts) due to the increased
abundance of clusters for masses $M\gtrsim 10^{14} \ h^{-1} M_\odot$.

The paper is arranged as follows. In section \ref{sec:K-mouflage}, we define the K-mouflage
models and detail some of their properties for a  fluid with pressure coupled to K-mouflage
at the background and perturbation levels, working in the Einstein frame.
In section \ref{sec:Jordan-frame}, we reformulate the dynamics in terms of the
Jordan-frame quantities.
In section \ref{sec:Numerical-large-scale}, we present our numerical results for the
background and the growth of structure for two K-mouflage models: the cubic and arctan models.
In section \ref{sec:clusters}, we turn to galaxy clusters and their properties focusing on the
physics of the gas embedded in the clusters.
In section \ref{sec:Comparison}, we discuss in details the similarity and differences between
the K-mouflage scenarios and other modified-gravity theories.
We conclude in section \ref{sec:Conclusions}.

A derivation of the equations of motion in the Einstein frame is given
in appendix~\ref{app:Einstein-frame}, while details on the Einstein-Jordan connection
can be found in the appendix~\ref{sec:E-J-backgrounds}.
We discuss unitarity constraints in appendix~\ref{sec:Unitarity}.

\section{Definition of K-mouflage models}
\label{sec:K-mouflage}

\subsection{Jordan-frame and Einstein-frame metrics}
\label{sec:conformal}

We consider scalar-field models where the action has the form
\cite{Babichev:2009ee,Brax:2012jr}
\beqa
S & = & \int \dd^4 x \; \sqrt{-\tilde{g}} \left[ \frac{\tilde{M}_{\rm Pl}^2}{2} \tilde{R}
+ \tilde{\cal L}_{\varphi}(\varphi) \right] \nonumber \\
&& + \int \dd^4 x \; \sqrt{-g} \, {\cal L}_{\rm m}(\psi^{(i)}_{\rm m},g_{\mu\nu}) ,
\label{S-def}
\eeqa
which involves two metrics, the Jordan-frame metric $g_{\mu\nu}$, with determinant $g$,
and the Einstein-frame metric $\tilde{g}_{\mu\nu}$, with determinant $\tilde{g}$.
The matter Lagrangian density, ${\cal L}_{\rm m}$, where $\psi^{(i)}_{\rm m}$
are various matter fields, is given in the Jordan frame, where it takes the usual
form without explicit coupling to the scalar field (although one could add explicit
couplings to build more complex models).
The gravitational sector is described by the usual Einstein-Hilbert action, but in terms
of the Einstein-frame metric $\tilde{g}_{\mu\nu}$ and the associated reduced Planck
mass $\tilde{M}_{\rm Pl}=1/\sqrt{8\pi\tilde{\cG}}$.
The Lagrangian density $\tilde{\cal L}_{\varphi}(\varphi)$ of the scalar field
is also given in the Einstein frame.

Throughout this paper, we denote Einstein-frame quantities with a tilde, to distinguish
them from their Jordan-frame counterparts (when they are not identical).
We choose this notation, which is the opposite to the one used in our previous papers
\cite{Brax:2014a,Brax:2014b,Brax:2014c,Barreira2015}
where we mostly worked in the Einstein frame, as here  we mostly work in the
Jordan frame.

If the two metrics were identical, this model would be a simple quintessence scenario
\cite{Caldwel1998,Steinhardt1999},
with an additional scalar field to the usual matter and radiation components but with
standard electrodynamics and gravity (General Relativity).
In this paper, we consider modified-gravity models where the two metrics are related
by the conformal transformation \cite{Esposito2001}
\beq
g_{\mu\nu} = A^2(\varphi) \, \tilde{g}_{\mu\nu} .
\label{g-Jordan-def}
\eeq
This gives rise to an explicit coupling between matter and the scalar field.
In the Einstein frame we recover General Relativity (e.g., the Friedmann equations)
but the equations of motion of matter are non-standard (e.g., the continuity equation
shows a source term and matter density is not conserved).
In the Jordan frame the equations of motion of matter take the usual form (e.g., matter
density is conserved) but gravity is modified (e.g., the Friedmann equations are
modified).
In this paper we compute the properties of astrophysical
objects such as clusters of galaxies, including their temperature and X-ray luminosity,
and it is more convenient to work in the Jordan frame.
Then, radiative processes, such as Bremsstrahlung,
are given by the standard results and do not vary with time or space. Moreover,
matter density is conserved. This simplifies the analysis, as the only difference from
a $\Lambda$-CDM scenario will be a change of gravity laws, which can be explicitly
derived from the action (\ref{S-def}).

The conformal transformation (\ref{g-Jordan-def}) actually means that the line
elements are transformed as $\dd s^2 = A^2 \dd \tilde{s}^2$. Using conformal
time $\tau$ and comoving coordinates $\vx$, this local change of distance can be
absorbed in the scale factor for the background universe, as
\beq
\dd s^2 = a^2 (-\dd\tau^2+\dd\vx^2) , \;\; \dd \tilde{s}^2 =
\tilde{a}^2 (-\dd\tau^2+\dd\vx^2) ,
\label{ds2-ds2}
\eeq
with
\beq
a = \bar{A} \tilde{a} , \;\; \tau= \tilde{\tau} , \;\; \vx = \tilde{\vx} .
\label{tilde-a-def}
\eeq
[Throughout this paper, we denote with an overbar mean background quantities, such as
$\bar{A}=A(\bar\varphi)$.]
However, physical time $t$ and distances $\vr$, with $\dd s^2=-\dd t^2 + \dd\vr^2$, are
changed as
\beq
\dd t = \bar{A} \, \dd \tilde{t} , \;\; \vr = a \vx = \bar{A} \, \tilde{\vr} .
\label{tilde-t-def}
\eeq
In particular, the cosmic times $t$ and $\tilde{t}$ are not the same in both frames

\subsection{K-mouflage kinetic function}
\label{sec:kinetic}

In this paper, we consider K-mouflage models
\cite{Babichev:2009ee,Brax:2012jr,Brax:2014a},
which correspond to cases where the scalar-field Lagrangian has a non-standard kinetic term,
\beq
\tilde{\cal L}_{\varphi}(\varphi) = {\cal M}^4 \, K(\tilde{\chi}) \;\;\; \mbox{with} \;\;\;
\tilde{\chi} = - \frac{1}{2{\cal M}^4} \tilde{\nabla}^{\mu}\varphi\tilde{\nabla}_{\mu}\varphi .
\label{K-def}
\eeq
Throughout this paper, $\tilde{\nabla}_{\mu}$ ($\nabla_{\mu}$) is the covariant derivative
associated with the metric $\tilde{g}_{\mu\nu}$ ($g_{\mu\nu}$)
(hence $\chi=A^{-2}\tilde{\chi}$, but we work with $\tilde{\chi}$ in the following).
Here, ${\cal M}^4$ is an energy scale that is of the order of the current dark-energy density,
(i.e., set by the cosmological constant), to recover the late-time accelerated expansion
of the Universe.
Thus, the canonical cosmological behavior, with a cosmological constant
$\tilde{\rho}_{\Lambda} = {\cal M}^4$, is recovered at late time in the weak-$\tilde{\chi}$ limit
if we have
\beq
\tilde{\chi} \rightarrow 0 : \;\;\; K(\tilde{\chi}) \simeq -1 + \tilde{\chi} + ... ,
\label{K-chi=0}
\eeq
where the dots stand for higher-order terms, the zeroth-order factor $-1$ corresponding to
the late-time cosmological constant ${\cal M}^4$. The normalization of the first two terms in
Eq.(\ref{K-chi=0}) defines the normalizations of the constant ${\cal M}^4$ and of the field
 $\varphi$, hence it does not entail any loss of generality (within this class of models).
We only consider models that satisfy this low-$\tilde{\chi}$ expansion in this article, and
where $\bar{\tilde\chi}\rightarrow \infty$ for $\tilde{t}\rightarrow 0$ and
$\bar{\tilde\chi}\rightarrow 0$ for $\tilde{t}\rightarrow\infty$.

Well-behaved K-mouflage scenarios have $K'>0$, where we denote
$K'=\dd K/\dd\tilde{\chi}$, and $W_{\pm}(y)=y K'(\pm y^2/2)$ must
be monotonically increasing functions up to $+\infty$ over $y \geq 0$.
This ensures that the cosmological dynamics are well defined up to arbitrarily high redshift,
where the matter density becomes increasingly large, and that small-scale static solutions
exist for any matter density profile \cite{Brax:2014c}. Moreover, there are no ghosts around the
cosmological background nor small-scale instabilities \cite{Brax:2014a}.

We must point out that the kinetic functions $K(\tilde\chi)$ that we use for numerical
computations and illustrative purposes in this paper are defined by fully nonlinear
expressions, namely Eqs.(\ref{K-arctan-def}) and (\ref{K-cub-def}) below, and as such
go beyond the low-$\tilde\chi$ expansion (\ref{K-chi=0}).
As explained above, the latter expansion is very general and holds for well-behaved models,
where $K'>0$ for all $\tilde\chi$ and $W_{\pm}(y)=y K'(\pm y^2/2)$ are monotonically increasing
functions of $y$. The expansion (\ref{K-chi=0}) would only be violated if $K'$ diverges at
low $\tilde\chi$, e.g. $K(\tilde\chi) = -1 + \tilde\chi^{3/4} + ...$, but we do not consider
such singular cases here.

Then, it happens that at at low redshifts, in the dark-energy era, $\tilde\chi$ [with its
normalization defined by the first two coefficients in the expansion (\ref{K-chi=0})] is small
on cosmological scales, which implies $K' \simeq 1$.
This holds both for the homogeneous background and for the cosmological large-scale
structures.
This property is related to the fact that at low redshifts, in the dark-energy era, we
require the cosmological evolution to remain close to the $\Lambda$-CDM behavior.
From the expressions (\ref{rho-phi-def}), we can see that this implies
$\bar{\tilde\chi} \bar{K}' \ll \bar{K}$ (to recover a dark-energy equation of state
$\bar{p}_{\rm de} \simeq - \bar{\rho}_{\rm de}$) whence $\bar{\tilde\chi} \ll 1$.
In fact, at low $z$ we have the scaling $\bar{\tilde\chi} \sim \beta^2$, where $\beta$ is the
coupling strength defined in Eq.(\ref{beta-def}) below, so that $\bar{\tilde\chi} \sim 0.01$
as we take $\beta=0.1$.
We shall check this behavior in Fig.~\ref{fig_chi_varphi_z} below.

We shall also check in Sec.~\ref{sec:unscreened} and Fig.~\ref{fig_chi_r_z0}
below that this also applies to clusters of galaxies at low redshifts, which are not screened
by the nonlinearities of the scalar-field Lagrangian, in spite of their large mass.
This would not be the case for a coupling $\beta \gg 0.1$, but this would violate some
Solar System and cosmological constraints and we do not consider such models here.

Nevertheless, the nonlinearities of the kinetic function $K(\tilde\chi)$ come into play at
high redshift and are taken into account in our computations, using the explicit nonlinear
examples (\ref{K-arctan-def}) and (\ref{K-cub-def}). This ensures in particular that the
dark-energy density becomes subdominant at high $z$ and that we recover the
Einstein-de Sitter cosmology in the early matter era \cite{Brax:2014a}.
Moreover, the background solution can be shown to be stable and is a tracker solution
\cite{Brax:2014a}.
The nonlinearities on the far negative semiaxis, $-\tilde\chi \gg 1$, also play a critical
role to ensure that Solar System tests of gravity are satisfied by the K-mouflage model,
but we do not consider this regime in this paper.

Although K-mouflage theories involve high-order derivative interactions, they  do not suffer from
quantum-mechanical problems such unitarity violation in their interaction with matter
\cite{Brax:2014d,Brax:2015a}, as explained in the appendix~\ref{sec:Unitarity}.

\subsection{K-mouflage coupling function}
\label{sec:coupling}

The coupling function $A(\varphi)$ has the low-$\varphi$ expansion
\beq
A(\varphi) = 1 + \frac{\beta\varphi}{\tilde{M}_{\rm Pl}} + ... ,
\label{A-phi=0}
\eeq
where the dots stand for higher-order terms. The normalization of the first term does not
entail any loss of generality and only corresponds to a normalization of coordinates.
At early times, $\tilde{t}\rightarrow 0$, we have $\bar\varphi\rightarrow 0$ and
$g_{\mu\nu} \rightarrow \tilde{g}_{\mu\nu}$.
More generally, we define the coupling $\beta$ as
\beq
\beta(\varphi) = \tilde{M}_{\rm Pl} \frac{\dd\ln A}{\dd\varphi} .
\label{beta-def}
\eeq
It is constant for exponential coupling functions,
$A(\varphi) = \exp[\beta\varphi/\tilde{M}_{\rm Pl}]$.
Without loss of generality, we take $\beta>0$ (which simply defines the sign of the scalar
field $\varphi$).

Cosmological and Solar System constraints imply $\beta \lesssim 0.1$,
see Ref. \cite{Barreira2015}.
Moreover, we have the scaling $| \beta\bar\varphi/\tilde{M}_{\rm Pl} | \sim \beta^2 \ll1$,
see Ref.\cite{Brax:2014a}, as we shall check in Fig.~\ref{fig_chi_varphi_z} below
(see also Ref.~\cite{Barreira:2014a}).
Therefore, in realistic models, we have $|\bar{A}-1| \lesssim 0.1$ and the higher-order terms
in the expansion (\ref{A-phi=0}) only have a small quantitative impact.
We shall also check in Fig.~\ref{fig_chi_r_z0} below that the fluctuations of the scalar field are
small, $| \varphi-\bar\varphi | \ll | \bar\varphi |$, so that the coupling function $A(\varphi)$ remains
dominated by the low-order terms of the expansion (\ref{A-phi=0}) in clusters of galaxies
(and at smaller scales).
This can be readily understood from the fact that realistic models should have a fifth force that
is not greater than the standard Newtonian force. This typically implies $|\delta A/A| \lesssim |\Psi_{\rm N}|$,
where $\Psi_{\rm N}$ is the Newtonian potential, whence
$|\beta\delta\varphi/\tilde{M}_{\rm Pl}| \lesssim 10^{-5}$.

\subsection{Equations of motion in the Einstein frame}
\label{sec:Einstein-frame}

Observable effects, such as lensing or two point correlations that can be measured,
are independent of the choice of frame, so that we can work in either the Einstein
or the Jordan frame. As explained in the introduction, for our purposes the Jordan
frame is more convenient and more transparent. Indeed, in this frame both the matter
and radiation components obey their usual equations of motion, e.g. the matter
energy-momentum tensor satisfies $\nabla_{\mu} T^{\mu}_{\nu}=0$ so that the
matter density obeys the usual conservation equation. Moreover, particle masses and
atomic emission or absorption lines do not evolve with the cosmic time (whereas they
do in the Einstein frame). Then, the only effect of the scalar field is to change the
gravitational sector, that is, the Friedmann equations that determine the background
cosmological expansion rate and the relation between the metric gravitational
potentials and the matter density fluctuations (i.e., it leads to modified Poisson equations
that can be interpreted as a fifth force).

Therefore, in this article we work in the Jordan frame and compute observable effects,
in particular the properties of clusters of galaxies, in this frame.
However, to simplify the derivation of the equations of motion, it is convenient to
first derive the Friedmann equations and the equations that govern the growth of
cosmological large-scale structures in the Einstein frame, where gravity takes the
standard form.
In a second step, we will use these results to obtain the equations of motion
in the Jordan frame through a change of variables, in Sec.~\ref{sec:Jordan-frame}.
Afterwards, all our computations will remain in the Jordan frame.

Thus, we describe in the appendix~\ref{app:Einstein-frame} the derivation of the
equations of motion of the scalar field and of the matter component in the
Einstein frame, for a cosmological fluid with a nonzero pressure.
In this section we only give the main results, which will be needed to obtain
the equations of motion in the Jordan frame in Sec.~\ref{sec:Jordan-frame}.

We consider three components of the energy density of the Universe, a matter fluid with
nonzero pressure, radiation, and the scalar field. The Einstein-frame and Jordan-frame
matter energy-momentum tensors are given by
\beq
\tilde{T}_{\mu\nu} = \frac{-2}{\sqrt{-\tilde{g}}}
\frac{\delta S_{\rm m}}{\delta \tilde{g}^{\mu\nu}} , \;\;\;
T_{\mu\nu} = \frac{-2}{\sqrt{-g}} \frac{\delta S_{\rm m}}{\delta g^{\mu\nu}} ,
\label{Tm-def}
\eeq
where we dropped the subscript ``m''. The conformal transformation (\ref{g-Jordan-def})
gives
\beq
T_{\mu\nu} = A^{-2} \tilde{T}_{\mu\nu} , \;\;\; T^{\mu}_{\nu} = A^{-4} \tilde{T}^{\mu}_{\nu} , \;\;\;
T^{\mu\nu} = A^{-6} \tilde{T}^{\mu\nu} ,
\label{tildeT-T}
\eeq
where we use $g^{\mu\nu}$ ($\tilde{g}^{\mu\nu}$) to raise indices in $T$ ($\tilde{T}$),
and the relation $g^{\mu\nu} = A^{-2} \tilde{g}^{\mu\nu}$.
In particular, the Einstein-frame and Jordan-frame densities and pressures are related by
\beq
\tilde{\rho} = A^4 \rho , \;\;\; \tilde{p} = A^4 p .
\label{rho-p-E-J}
\eeq

We work in the non-relativistic limit, $v^2 \ll 1$ where $v$ is the mean fluid peculiar velocity,
and in the weak field regime, $\tilde\Psi_{\rm N} \ll 1$, where $\tilde{\Psi}_{\rm N}$ is
the Newtonian gravitational potential. Moreover, assuming the usual Cold Dark Matter
(CDM) scenario for the dark matter, the matter pressure $\tilde{p}$ is negligible on
cosmological scales and it arises from the small-scale nonlinear processes, such as
the collapse of gas clouds which generate shocks, or the virialization of of dark matter halos
(which generate an effective pressure through the velocity dispersion of the matter
particles).
Then, $\tilde{p} \sim \tilde{\rho} c_s^2$, where $c_s$ is the speed of sound or the
velocity dispersion, and $c_s^2 \sim \tilde\Psi_{\rm N}$ because it is generated
by the gravitational collapse
(as in hydrostatic equilibrium where pressure gradients balance the gravitational force).
In addition, we consider the small-scale (sub-horizon) limit, $k/\tilde{a}\tilde{H} \gg 1$,
where spatial gradients dominate over time derivatives and quasi-static approximations
apply.
Thus, we focus on the regime defined by
\beq
v^2 \ll 1 , \;\; \tilde{\Psi}_{\rm N} \ll 1 , \;\;
\frac{\tilde{p}}{\tilde{\rho}} \sim c_s^2 \sim \tilde{\Psi}_{\rm N} \ll 1 , \;\;\;
\frac{k}{\tilde{a}\tilde{H}} \gg 1 .
\label{approx-non-relativistic}
\eeq

As the background level, the matter background pressure is zero, $\bar{\tilde{p}}=0$,
and the Einstein-frame Friedmann equations read as
\beqa
3 \tilde{M}_{\rm Pl}^2 \tilde{H}^2 & = & \bar{\tilde\rho} + \bar{\tilde\rho}_{\rm (r)}
+ \bar{\tilde\rho}_{\varphi} ,
\label{Friedmann-1-nopress} \\
-2 \tilde{M}_{\rm Pl}^2 \frac{\dd\tilde{H}}{\dd\tilde{t}} & = & \bar{\tilde\rho}
+ \bar{\tilde\rho}_{\rm (r)} + \bar{\tilde{p}}_{\rm (r)} + \bar{\tilde\rho}_{\varphi}
+ \bar{\tilde{p}}_{\varphi} ,
\label{Friedmann-2-nopress}
\eeqa
where $\bar{\tilde{\rho}}$, $\bar{\tilde\rho}_{\rm (r)}$ and
$\bar{\tilde{p}}_{\rm (r)}=\bar{\tilde\rho}_{\rm (r)}/3$ are the background matter and
radiation densities and pressure, which evolve as
\beq
\frac{\dd\bar{\tilde\rho}}{\dd\tilde{t}} = - 3 \tilde{H} \bar{\tilde\rho}
+ \bar{\tilde\rho} \frac{\dd\ln{\bar A}}{\dd\tilde{t}} , \;\;\;
\frac{\dd\bar{\tilde\rho}_{\rm (r)}}{\dd\tilde{t}} =  - 4 \tilde{H} \bar{\tilde\rho}_{\rm (r)} ,
\label{conserv-back-nopress}
\eeq
while $\bar{\tilde\rho}_{\varphi}$ and $\bar{\tilde{p}}_{\varphi}$ are the background
scalar-field density and pressure, given by
\beq
\bar{\tilde{\rho}}_{\varphi} = {\cal M}^4 ( 2 \bar{\tilde{\chi}} \bar{K}' - \bar{K} )  , \;\;\;
\bar{\tilde{p}}_{\varphi} = {\cal M}^4 \bar{K} ,
\label{rho-phi-def}
\eeq
with
\beq
\bar{\tilde{\chi}} = \frac{1}{2{\cal M}^4}
\left( \frac{\dd {\bar\varphi}}{\dd\tilde{t}} \right)^2 .
\label{chi-background}
\eeq
(Throughout this paper, we consider a flat universe with zero background curvature.)
The background scalar field obeys the Klein-Gordon equation
\beq
\frac{\dd}{\dd\tilde{t}}  \left[ \tilde{a}^3 \frac{\dd\bar\varphi}{\dd\tilde{t}} \bar{K}'
\right] = - \tilde{a}^3 \bar{\tilde\rho} \frac{\dd \ln{\bar A}}{\dd\bar\varphi} .
\label{KG-back-nopress}
\eeq

For the matter perturbations, the continuity and Euler equations write as
\beq
\frac{\pl\tilde{\rho}}{\pl\tau} + \nabla \cdot (\tilde{\rho} \vv) + 3 \tilde{\cal H} \tilde{\rho}
= \tilde{\rho} \frac{\dd\ln\bar{A}}{\dd\tau} ,
\label{continuity-small-scale-E}
\eeq
\beq
\frac{\pl\vv}{\pl\tau} + (\vv\cdot\nabla)\vv + \left( \tilde{\cal H}
+ \frac{\dd\ln\bar{A}}{\dd\tau} \right) \vv = - \nabla ( \tilde{\Psi}_{\rm N} + \ln A )
- \frac{\nabla\tilde{p}}{\tilde\rho} ,
\label{Euler-small-scale-E}
\eeq
where the Einstein-frame Newtonian potential is given by the Poisson equation
\beq
\frac{1}{\tilde{a}^2} \nabla^2 \tilde{\Psi}_{\rm N} = \frac{1}{2 \tilde{M}_{\rm Pl}^2}
\delta{\tilde\rho} .
\label{Poisson-small-scale-E}
\eeq
In the small-scale (quasi-static) limit, the Klein-Gordon equation for the scalar field
becomes
\beq
\frac{1}{\tilde{a}^2} \nabla \cdot (\nabla\varphi \; \bar{K}' ) =
\frac{\dd\ln\bar{A}}{\dd\bar\varphi} \delta{\tilde\rho} .
\label{KG-small-scale-E}
\eeq
Here we also used the fact that the fluctuations of $\varphi$ can be neglected in the factor
$K'$, so that the Klein-Gordon equation can actually be linearized in the scalar field, while
keeping the matter density fluctuations nonlinear. See Ref.~\cite{Brax:2014b} for a
detailed discussion and an explicit computation of the matter power spectrum, up to
one-loop order, that includes up to the cubic term in $\varphi$ in
Eq.(\ref{KG-small-scale-E}), which is checked to make no quantitative difference
for cosmological large-scale structures of cluster sizes and beyond.
This corresponds to the fact that clusters are not screened by the nonlinear K-mouflage
mechanism, which comes into play at much smaller scales and higher densities, as in the
Solar System.

\section{Equations of motion in the Jordan frame}
\label{sec:Jordan-frame}

We now derive the equations of motion of the scalar field and of the matter component in the
Jordan frame. To do so, we use the results obtained in Sec.~\ref{sec:Einstein-frame}
in the Einstein frame, and express these equations in terms of Jordan-frame variables.

\subsection{Background dynamics}
\label{sec:Background-Jordan}

The Jordan-frame metric $g_{\mu\nu}$ is related to the Einstein-frame metric
$\tilde{g}_{\mu\nu}$ by the conformal transformation (\ref{g-Jordan-def}).
As seen in Eqs.(\ref{ds2-ds2})-(\ref{tilde-t-def}), this leads to a rescaling of the
scale factor and of physical time and distance, while the conformal coordinates
are unchanged.
The Hubble expansion rates, $H=\dd\ln a/\dd t$ and $\tilde{H}=\dd\ln\tilde{a}/\dd\tilde{t}$,
are also different and related by
\beq
H = \frac{\tilde{H} (1+\tilde{\epsilon}_2)}{\bar{A}} = \frac{\tilde{H}}
{\bar{A} (1-\epsilon_2)} ,
\label{tH-H-def}
\eeq
where $\tilde{\epsilon}_2(t)$ was defined in Eq.(\ref{epsilon-def}) and verifies
$\dd\ln\bar{A}/\dd\tilde{t}= \tilde{\epsilon}_2 \tilde{H}$, and we introduced its
Jordan-frame counterpart,
\beq
\epsilon_2(t) = \frac{\dd\ln\bar{A}}{\dd\ln a} , \;\;
\epsilon_2 = \frac{\tilde{\epsilon}_2}{1+\tilde{\epsilon}_2} , \;\;
\tilde{\epsilon}_2 = \frac{\epsilon_2}{1-\epsilon_2} .
\label{tilde-epsilon2-def}
\eeq

The Einstein-frame and Jordan-frame densities and pressures are related as in
Eq.(\ref{rho-p-E-J}), so that the Friedman equation (\ref{Friedmann-1-nopress}) gives
\beq
3 M_{\rm Pl}^2 H^2 = (1-\epsilon_2)^{-2} \left( \bar{\rho}
+ \bar{\rho}_{\rm (r)} + \bar{\rho}_{\varphi} \right) ,
\label{Friedmann-1-Jordan}
\eeq
where we introduced the Jordan-frame reduced Planck mass,
\beq
M^2_{\rm Pl}(t) = \tilde{M}^2_{\rm Pl} / \bar{A}(t)^2 .
\label{MPl-Jordan}
\eeq
Thus, in the Jordan frame, Newton's constant, $\cG=1/8\pi M_{\rm Pl}^2$, varies with time,
as $\cG(t) = \tilde{\cG} \bar{A}^2 \propto \bar{A}^2$.
Equation (\ref{Friedmann-1-Jordan}) shows how the Friedmann equation is modified
in the Jordan frame, as compared with the usual General Relativity result, because
the gravitational Einstein-Hilbert action is defined in terms of the auxiliary metric
$\tilde{g}_{\mu\nu}$. Substituting for $g_{\mu\nu}$ this effectively corresponds to a change
of the Einstein-Hilbert action. At the background level, this simply introduces the
time-dependent functions $\bar{A}(t)$ and $\epsilon_2(t)$ in Eq.(\ref{Friedmann-1-Jordan}).

We can also write Eq.(\ref{Friedmann-1-Jordan}) in the standard form (albeit with
a time-dependent reduced Planck mass), as
\beq
3 M_{\rm Pl}^2 H^2 = \bar{\rho} + \bar{\rho}_{\rm (r)} + \bar{\rho}_{\rm de} ,
\label{Friedmann-2-Jordan}
\eeq
by defining the dark-energy component as the energy density that is ``missing'' in the
Friedmann equation to match the Hubble rate, after we sum over the other matter
and radiation components. This yields
\beq
\bar{\rho}_{\rm de} \equiv \bar{\rho}_{\varphi} +
\frac{2\epsilon_2 - \epsilon_2^2}{(1-\epsilon_2)^2}
\left( \bar{\rho} + \bar{\rho}_{\rm (r)} + \bar{\rho}_{\varphi} \right) .
\label{rho-de-Jordan}
\eeq
This interpretation corresponds to the case where measurements of the Hubble rate and
of the matter and radiation densities are performed in the Jordan frame and the remaining
part, which explains the accelerated expansion, is ascribed to the dark-energy component,
[as in the usual $\Lambda$-CDM case, where the background dark energy is also
measured from the missing energy density that is required to account for $H(z)$].
This is a natural configuration, as $a(t)$ and $H(t)$ are obtained from redshift
measurements of standard candles, which assumes that atomic absorption and emission
lines are the same at distant redshifts as in the laboratory.
By definition, this is the case in the Jordan frame but not in the Einstein frame
(where particle masses are actually time dependent).
On the other hand, these standard candles must not depend on the local gravity,
because in the Jordan frame Newton's constant becomes time dependent, so that these
candles are no longer standard (i.e., similar to those at $z=0$).
This rules out supernovae (which involve both local gravity and electrodynamics, within the
star) but allows one to use geometric candles such as baryon acoustic oscillations
\cite{Eisenstein1998a,Eisenstein2005} or the Alcock-Paczynski test \cite{Alcock1979}.

Using Eq.(\ref{conserv-back-nopress}), the matter and radiation densities now evolve as
\beq
\frac{\dd\bar{\rho}}{\dd t} = - 3 H \bar{\rho} , \;\;\;
\frac{\dd\bar{\rho}_{\rm (r)}}{\dd t} = - 4 H \bar{\rho}_{\rm (r)} .
\label{conserv-Jordan}
\eeq
Thus, we recover the usual conservation equations for matter and radiation in the Jordan
frame, whence
\beq
\bar{\rho}(t) = \frac{\bar{\rho}_0}{a^3} , \;\;\;
\bar{\rho}_{\rm (r)}(t) = \frac{\bar{\rho}_{\rm (r)0}}{a^4} , \;\;\;
\mbox{with} \;\;\; a_0 =1 ,
\label{rho-a3-a4-Jordan}
\eeq
where $\bar{\rho}_0$ are the mean Jordan-frame energy densities today, at $z=0$,
and we normalized the Jordan-frame scale factor by $a_0=1$.

From Eq.(\ref{rho-phi-no-conserv}), with $\bar{\tilde{p}}=0$, and Eq.(\ref{rho-de-Jordan}),
the Jordan-frame dark-energy density evolves as
\beq
\frac{\dd\bar{\rho}_{\rm de}}{\dd t} = - 3 H \left( \bar{\rho}_{\rm de}
+ \bar{p}_{\rm de} \right ) ,
\label{conserv-de-Jordan}
\eeq
where we defined the Jordan-frame dark-energy pressure as
\beqa
\bar{p}_{\rm de} & = & \bar{p}_{\varphi} +
\frac{\epsilon_2}{1-\epsilon_2} ( \bar{p}_{\rm (r)} + \bar{p}_{\varphi} )
+ \left( \epsilon_2 - \frac{2}{1-\epsilon_2} \frac{\dd\epsilon_2}{\dd\ln a} \right)
\nonumber \\
&& \times \frac{\bar{\rho} + \bar{\rho}_{\rm (r)} + \bar{\rho}_{\varphi}}{3 (1-\epsilon_2)^2} .
\label{p-de-Jordan}
\eeqa
On the other hand, from Eq.(\ref{KG-back-nopress}) the Klein-Gordon equation reads as
\beq
\frac{\dd}{\dd t}  \left[ \bar{A}^{-2} a^3 \frac{\dd\bar\varphi}{\dd t}
\bar{K}' \right] = - a^3 \bar{\rho} \frac{\dd \ln{\bar A}}{\dd\bar\varphi} .
\label{KG-Jordan}
\eeq
This can be integrated as
\beq
\frac{\dd\bar\varphi}{\dd t} \bar{K}' = - \bar{A}^2 a^{-3} \int_0^t \dd t' \,
\frac{\bar{\rho}_0 \beta(t')}{\tilde{M}_{\rm Pl}} ,
\label{KG-Jordan-int}
\eeq
because the integration constant must vanish to recover a realistic early-time cosmology
\cite{Brax:2014a}.

Finally, we define the Jordan-frame cosmological parameters as
\beq
\Omega_{\rm m} = \frac{\bar{\rho}}{\rho_{\rm crit}} , \;\;
\Omega_{\rm (r)} = \frac{\bar{\rho}_{\rm (r)}}{\rho_{\rm crit}} , \;\;
\Omega_{\rm de} = \frac{\bar{\rho}_{\rm de}}{\rho_{\rm crit}} ,
\label{Omega-Jordan-def}
\eeq
where $\rho_{\rm crit} = 3 M_{\rm Pl}^2 H^2 = \bar{A}^{-4}
(1-\epsilon_2)^{-2} \tilde{\rho}_{\rm crit}$ is the Jordan-frame critical density.
This gives
\beq
\Omega_{\rm m} = (1-\epsilon_2)^2 \tilde{\Omega}_{\rm m} , \;\;\;
\Omega_{\rm (r)} = (1-\epsilon_2)^2 \tilde{\Omega}_{\rm (r)} ,
\label{Omega-m-rad}
\eeq
and
\beq
\Omega_{\rm de} = \tilde{\Omega}_{\varphi} + (2\epsilon_2-\epsilon_2^2)
(\tilde{\Omega}_{\rm m} + \tilde{\Omega}_{\rm (r)} ) .
\eeq
We can check that
$\Omega_{\rm m}+\Omega_{\rm (r)}+\Omega_{\rm de} =
\tilde{\Omega}_{\rm m}+\tilde{\Omega}_{\rm (r)}+\tilde{\Omega}_{\varphi} =1$.
The effective dark-energy equation of state in the Jordan frame is simply defined as
\beq
w_{\rm de} = \bar{p}_{\rm de} / \bar{\rho}_{\rm de} .
\label{w-Jordan-def}
\eeq

\subsection{Perturbations}
\label{sec:Perturbations-Jordan}

The dynamics of large-scale perturbations in the Jordan frame are also obtained from the
equations derived in the Einstein frame in Sec.~\ref{sec:Einstein-frame}.

In the Einstein frame, the Newtonian gauge metric (\ref{Newtonian-gauge-Einstein}) reads as
$\dd\tilde{s}^2 = \tilde{a}^2 [ -(1+2\tilde{\Psi}_{\rm N}) \dd \tau^2 + (1-2\tilde{\Psi}_{\rm N})
\dd \vx^2 ]$, where we used Eq.(\ref{Psi-Phi}).
In the Jordan frame, we write
\beq
\dd s^2 = a^2 [ -(1+2\Phi) \dd \tau^2 + (1-2\Psi) \dd \vx^2 ] .
\label{Newtonian-gauge-Jordan}
\eeq
Then, using $\dd s^2 = A^2 \dd\tilde{s}^2$ and $a=\bar{A}\tilde{a}$, we obtain, up to
first order in $\delta A=A-\bar{A}$,
\beq
\Phi = \Psi_{\rm N} + \frac{\delta A}{\bar{A}} , \;\;\;
\Psi = \Psi_{\rm N} - \frac{\delta A}{\bar{A}} ,
\label{Phi-Psi-Jordan}
\eeq
where we introduced the Jordan-frame Newtonian potential, given by
\beq
\frac{1}{a^2} \nabla^2 \Psi_{\rm N} = \frac{1}{2M_{\rm Pl}^2} \delta\rho , \;\;\;
\mbox{whence} \;\;\; \Psi_{\rm N} = \tilde{\Psi}_{\rm N} .
\label{Poisson-small-scale-Jordan-1}
\eeq
The last equality follows from Eqs.(\ref{Poisson-small-scale-E}) and
(\ref{MPl-Jordan}).
Therefore, in the Jordan frame the two metric potentials are no longer equal, but their sum
remains equal to $2\Psi_{\rm N}$. This is related to the fact that photons do not feel the
effect of the fifth force, see also Eq.(\ref{conserv-Trad-E-J}).
Therefore, weak lensing statistics show the same dependence on the matter density
fluctuations as in GR and the impact of the modified gravity only arises through the
different evolution of the density field and the time-dependent Newton constant,
see also Sec.\ref{sec:dyn-wl} below.

The Klein-Gordon equation (\ref{KG-small-scale-E}) reads as
\beq
\frac{1}{a^2} \nabla^2 \varphi = \frac{\beta\bar{A}}{\bar{K}' M_{\rm Pl}} \delta\rho ,
\label{KG-Jordan-small-scale}
\eeq
and, up to first order over $\delta \varphi$, we obtain
\beq
\frac{\delta A}{\bar{A}} = \frac{\beta}{M_{\rm Pl}\bar{A}} \delta\varphi , \;\;\;
\frac{1}{a^2} \nabla^2 \frac{\delta A}{\bar{A}} = \frac{\beta^2}{M_{\rm Pl}^2 \bar{K}'}
\delta\rho .
\label{dA-dphi-drho}
\eeq
This also gives
\beq
\frac{1}{a^2} \nabla^2 \Phi =
\frac{1+\epsilon_1(t)}{2M_{\rm Pl}^2(t)} \delta\rho \;\;\; \mbox{and} \;\;\;
\Phi = (1+\epsilon_1) \Psi_{\rm N} ,
\label{Poisson-small-scale-Jordan-2}
\eeq
where $\epsilon_1$ is defined by
\beq
\epsilon_1(t) = \frac{2\beta^2}{\bar{K}'} ,
\label{tilde-epsilon1-def}
\eeq
and we recover the same factor as in the first relation (\ref{epsilon-def}).

The continuity equation (\ref{continuity-small-scale-E}) and the Euler equation
(\ref{Euler-small-scale-E}) become
\beq
\frac{\pl\rho}{\pl\tau} + \nabla \cdot (\rho\vv) + 3 {\cal H} \rho = 0
\label{continuity-small-scale-Jordan}
\eeq
and
\beq
\frac{\pl\vv}{\pl\tau} + (\vv\cdot\nabla)\vv + {\cal H} \vv
= - \nabla \Phi - \frac{\nabla p}{\rho} .
\label{Euler-small-scale-Jordan}
\eeq
Therefore, in contrast with the Einstein frame, in the Jordan frame the continuity and Euler
equations take the same form as in $\Lambda$-CDM, and the coupling to the scalar field
$\varphi$ only gives rise to the modified Poisson equation
(\ref{Poisson-small-scale-Jordan-2}), in terms of the formation of large-scale structures.
There is no longer a non-conservation term in the continuity equation nor an additional
friction term in the Euler equation.
However, in contrast with the Einstein frame and the $\Lambda$-CDM cosmology, the two
gravitational potentials $\Phi$ and $\Psi$ that enter the Newtonian gauge
metric are now different.

\subsection{Formation of large-scale structures}
\label{sec:Formation-Jordan}

Introducing the Jordan-frame matter density contrast,
\beq
\delta = \delta\rho / \bar{\rho} ,
\label{delta-Jordan-def}
\eeq
the continuity equation (\ref{continuity-small-scale-Jordan}) also writes as
\beq
\frac{\pl\delta}{\pl\tau} + \nabla \cdot [ (1+\delta) \vv] = 0 .
\label{continuity-delta-small-scale-Jordan}
\eeq
This is the same equation as (\ref{continuity-delta-small-scale-E}) and we have
\beq
\delta = \hat{\delta} = \tilde{\delta} .
\label{delta-equal}
\eeq
Indeed the matter density contrasts in the Einstein frame, whether we consider the density
$\tilde{\rho}$ or the ``conserved'' density $\hat\rho$ of Eq.(\ref{rho-hat-def})
(in case of zero pressure), and in the Jordan frame, are equal within the approximations
described in the previous sections.

On perturbative scales, we again set the pressure term to zero and we introduce the
two-component vector
\beq
\psi \equiv \left(\bea{c} \psi_1 \\ \psi_2 \ea \right) \equiv
\left( \bea{c} \delta \\ -(\nabla\cdot\vv)/(\dd a/\dd t) \ea \right) .
\label{psi-Jordan-def}
\eeq
Equations (\ref{continuity-2}) and (\ref{Euler-2}) become
\beqa
\frac{\pl\psi_1}{\pl\ln a} - \psi_2 & = & \int \dd\vk_1\dd\vk_2 \;
\delta_D(\vk_1\!+\!\vk_2\!-\!\vk) \hat{\alpha}(\vk_1,\vk_2) \nonumber \\
&& \times \; \psi_2(\vk_1) \psi_1(\vk_2) ,
\label{continuity-2-Jordan}
\eeqa
\beqa
\frac{\pl\psi_2}{\pl\ln a} - \frac{3}{2} \Omega_{\rm m}
(1+\epsilon_1) \psi_1 + \left( 2 + \frac{1}{H^2} \frac{\dd H}{\dd t} \right) \psi_2 & =  &
\nonumber \\
&& \hspace{-7.cm} \int\!\! \dd\vk_1\dd\vk_2 \; \delta_D(\vk_1\!+\!\vk_2\!-\!\vk)
\hat{\beta}(\vk_1,\vk_2) \psi_2(\vk_1) \psi_2(\vk_2) . \;\;\;\;\;\;\;\;\;\;
\label{Euler-2-Jordan}
\eeqa
We recover the same form as for the $\Lambda$-CDM cosmology, except for the factor
$\epsilon_1$ that corresponds to a time-dependent amplification of Newton's gravity,
from the modified Poisson equation (\ref{Poisson-small-scale-Jordan-2}).

On large scales or at early times, we can again linearize the equations of motion
and the evolution equation (\ref{D-linear}) for the linear modes becomes
\beq
\frac{\dd^2 D}{\dd(\ln a)^2} + \left( 2 + \frac{1}{H^2} \frac{\dd H}{\dd t} \right)
\frac{\dd D}{\dd\ln a} - \frac{3}{2} \Omega_{\rm m} (1+\epsilon_1) D = 0 .
\label{D-linear-Jordan}
\eeq
Again, as compared with the Einstein-frame Eq.(\ref{D-linear}) we find that the
coefficient $\epsilon_2$ has disappeared and the only difference from the $\Lambda$-CDM
case is the time-dependent amplification of the gravitational term by
$(1+\epsilon_1)$.

As in Galileon models, but in contrast with $f(R)$, Dilaton and Symmetron models,
the linear growing modes remain scale independent as in the $\Lambda$-CDM cosmology.
This is because we did not include a potential $V(\varphi)$ in the scalar-field Lagrangian
(\ref{K-def}) and the field is massless. Thus, the amplification of gravity extends up to
the Hubble scale and is only damped on galactic and smaller scales by the nonlinear
K-mouflage mechanism. See Sec.~\ref{sec:Comparison} for a discussion and comparison
with other modified-gravity theories.

\subsection{Spherical collapse dynamics}
\label{sec:Spherical-collapse-Jordan}

As can be derived from Eq.(\ref{Euler-small-scale-Jordan}), on large scales where the
pressure is negligible, the particle trajectories $\vr(t)$ read as
\beq
\frac{\dd^2 \vr}{\dd t^2}  - \frac{1}{a} \frac{\dd^2 a}{\dd t^2} \vr = - \nabla_{\vr} \Phi ,
\label{trajectory-Jordan}
\eeq
where $\vr=a\vx$ is the physical coordinate and $\nabla_{\vr}=\nabla/a$ the gradient
operator in physical coordinates.
To study the spherical collapse before shell crossing, it is convenient to label each shell
by its Lagrangian radius $q$ or enclosed mass $M$, and to introduce its
normalized radius $y(t)$ by
\beq
y(t) = \frac{r(t)}{a(t) q} \;\;\; \mbox{with} \;\;\;
q = \left(\frac{3M}{4\pi\bar\rho_0}\right)^{1/3} , \;\;\; y(t=0) = 1 .
\label{y-def-Jordan}
\eeq
In particular, the matter density contrast within radius $r(t)$ reads as
\beq
1+ \delta(<r) = y(t)^{-3} .
\label{deltaR-def}
\eeq
Since the Poisson equation (\ref{tilde-epsilon1-def}) is only modified by the time-dependent
prefactor $1+\epsilon_1(t)$ and the time dependence of Newton's constant,
for a spherical system the gravitational force is still set by the total mass within radius $r$,
\beq
\frac{\dd\Phi}{\dd r} = (1+\epsilon_1) \frac{\cG \delta M}{r^2} ,
\eeq
where $\delta M=4\pi \delta(<r) \bar{\rho} r^3/3$.
Then, Eq.(\ref{trajectory-Jordan}) gives for the evolution of the normalized radius $y$, or
density contrast $\delta(<r)=y^{-3}-1$,
\beq
\frac{\dd^2 y}{\dd(\ln a)^2} + \left( 2+\frac{1}{H^2} \frac{\dd H}{\dd t} \right)
\frac{\dd y}{\dd\ln a} + \frac{\Omega_{\rm m}}{2} (1+\epsilon_1) (y^{-3}-1) y = 0.
\label{y-lna-Jordan}
\eeq
Again, as in the $\Lambda$-CDM cosmology but in contrast with $f(R)$, Dilaton and Symmetron
models, the spherical collapse is scale invariant so that the dynamics of different mass shells
decouple.
This applies to the unscreened regime, from clusters of galaxies up to the Hubble radius.

\subsection{Halo mass function}
\label{sec:mass-function}

As usual, we can write the halo mass function $n(M) \dd M/M$ as
\beq
n(M) \frac{\dd M}{M} = \frac{\bar\rho_0}{M} f(\nu) \frac{\dd\nu}{\nu} ,
\;\;\; \mbox{with} \;\;\; \nu = \frac{\delta_L}{\sigma(M)} ,
\label{nM-def}
\eeq
where we used the fact that the linear growing modes are scale independent
[so that $\delta_L/\sigma(M)=\delta_{L\rm i}/\sigma_{\rm i}(M)$, where
the subscript ``i'' refers to the high redshift $z_{\rm i}$ where the Gaussian initial
conditions are defined, far before the dark-energy era].
Here $\sigma(M)$ is the root mean square of the linear density contrast
at scale $M$ and $\delta_L$ is the linear density contrast associated with
the nonlinear density threshold $\Delta_{\rm m}$ that defines the virialized halos.
The mapping $\delta_L \rightarrow \delta$ is obtained by solving the spherical
collapse dynamics (\ref{y-lna-Jordan}), with the initial condition
$y_{\rm i} = 1 - \delta_{L\rm i}/3$ at a very high redshift $z_{\rm i}$.
Inverting this relation gives the linear density threshold $\delta_L$ that is associated
with a given nonlinear density threshold $\delta=\Delta_{\rm m}$, where the subscript ``m''
denotes that $\delta=y^{-3}-1$ is the density contrast with respect to the mean density
of the Universe.

The scaling variable $\nu$ directly measures the probability of
density fluctuations in the Gaussian initial conditions.
Then, we take for the scaling function $f(\nu)$ the fit to $\Lambda$-CDM simulations
obtained in \cite{Valageas2009}, which obeys the exponential tail
$f(\nu) \sim e^{-\nu^2/2}$ at large $\nu$.
This means that the mass function (\ref{nM-def}) shows the correct large-mass
tail, which is set by the Gaussian initial conditions.

\subsection{Planck masses}
\label{sec:Planck-mass}

It is interesting to note that, depending on the physical process that is considered,
one can define several effective Planck masses. In other words, if we assume General
Relativity and measure the reduced Planck mass or Newton's constant from different
sets of observations, we would obtain different values. This could be used as
a signature of the modified-gravity theory.

From Eq.(\ref{Friedmann-1-Jordan}), the effective Planck mass that would be read
from the Friedmann equation, at the background level, is
\beq
M_{\rm Pl (Friedmann)}^2(t) = \frac{\tilde{M}_{\rm Pl}^2(1-\epsilon_2(t))^2}{\bar{A}^2(t)} .
\label{MPlanck-eff-Friedman}
\eeq
On the other hand, with respect to large-scale density fluctuations in the cosmological
unscreened regime, where the Klein-Gordon equation can be linearized over the scalar
field, the effective Planck mass that would be read from the modified Poisson equation
(\ref{Poisson-small-scale-Jordan-2}) is
\beq
M_{\rm Pl (unscreened)}^2(t) = \frac{\tilde{M}_{\rm Pl}^2}{\bar{A}^2(t)(1+\epsilon_1(t))}
\label{MPlanck-eff-Poisson}
\eeq
On small astrophysical scales, within the highly nonlinear screened regime,
the effective Planck mass is instead the one defined in Eq.(\ref{MPl-Jordan}),
\beq
M^2_{\rm Pl(screened)}(t) = \frac{\tilde{M}^2_{\rm Pl}}{\bar{A}(t)^2} .
\label{MPl-screened-Jordan}
\eeq
The difference between these various definitions is a signature of the modification of gravity
associated with the scalar-field models (\ref{S-def}), as seen from the Jordan frame.

\subsection{Symmetry of large-scale gravitational clustering}
\label{sec:symmetry-clustering}

On large scales, where we neglect shell crossing and pressure, the dynamics of
gravitational clustering is given by Eqs.(\ref{continuity-delta-small-scale-Jordan}),
(\ref{Euler-small-scale-Jordan}) (with $\tilde{p}=0$), and
(\ref{Poisson-small-scale-Jordan-2}). It is convenient to define the rescaled
velocity and metric potential by
\beq
\vv = \frac{\dd a}{\dd t} f \, \vu , \;\;\; \Phi = \left( \frac{\dd a}{\dd t} f \right)^{\!2} \phi ,
\label{symmetry-rescaling}
\eeq
where we introduced the linear growth rate
\beq
f = \frac{\dd\ln D_+}{\dd\ln a} ,
\label{tilde-f-def}
\eeq
and to change the time variable from $t$ to $\ln D_+$.
Then, the equations of motion write as
\beqa
\frac{\pl\delta}{\pl\ln D_+} + \nabla \cdot [ (1+\delta) \vu] & = & 0 , \label{continuity-symm} \\
\frac{\pl\vu}{\pl\ln D_+} + (\vu\cdot\nabla)\vu + ( \kappa-1 ) \vu & = & - \nabla \phi , \label{Euler-symm} \\
\nabla^2 \phi & = & \kappa \, \delta ,
\eeqa
where we introduced the time-dependent factor $\kappa(t)$, defined by
\beq
\kappa(t) = \frac{4\pi{\cal G} \bar{\rho} (1+\epsilon_1)}{\left(\frac{\dd\ln D_+}{\dd t}
\right)^2} = \frac{3\Omega_{\rm m}}{2 f^2} (1+\epsilon_1) .
\label{kappa-def}
\eeq
Therefore, after the change of time coordinate $t \rightarrow \ln D_+$ and the rescaling
(\ref{symmetry-rescaling}), the only dependence on cosmology that is left in large-scale
gravitational clustering is encapsulated in the function $\kappa(t)$.
This remains valid beyond shell crossing but it breaks down on small scales where
baryonic effects become important and introduce new characteristic scales,
which cannot be absorbed by the change of variables (\ref{symmetry-rescaling}).
Nevertheless, on large scales where gravity is the dominant process,
this symmetry means that all cosmological scenarios with the same function
$\kappa(D_+)$ show the same density and velocity fields $\{\delta,\vu\}$.
In particular, this means that quintessence models, where only the background dynamics
is modified (i.e., the Hubble expansion rate $H(z)$), and modified-gravity models
or dark-energy models (with dark-energy fluctuations) that only give rise to a modification
of Poisson equation by a time-dependent Newton's constant, are equivalent with respect
to gravitational clustering, if they show the same function $\kappa(D_+)$.
In the context of $\Lambda$-CDM cosmology, this property has also been used to derive
approximate consistency relations satisfied by the matter correlation functions
[valid at the nonlinear level within the approximation where the dependence
on cosmology of $\kappa(D_+)$ can be neglected]
\cite{Valageas2014,Nishimichi2014,Kehagias2014}.

In the case of the K-mouflage scenarios, this symmetry only holds on large scales
(down to cluster scales) where the Klein-Gordon equation can be linearized over
$\varphi$, as in Eq.(\ref{KG-Jordan-small-scale}). On smaller scales (galactic scales
and below), higher-order terms over $\varphi$ become important and the nonlinear
K-mouflage screening mechanism comes into play. Then, the modified Poisson equation
no longer takes the linear form (\ref{Poisson-small-scale-Jordan-2}) and the
symmetry (\ref{kappa-def}) breaks down.
On even smaller scales we actually recover General Relativity as
$\Phi \simeq \Psi_{\rm N}$, because the fifth force is screened.
In hierarchical scenarios, where smaller scales collapse first, larger scales are not very
sensitive to the details of the clustering on smaller scales, while small collapsed
scales are sensitive to the clustering up to the largest scale that has turned nonlinear.
Therefore, we expect the symmetry (\ref{kappa-def}) to apply to scales that are greater
than the transition to the K-mouflage screening regime, and not to smaller scales
(even though we recover General Relativity far inside the nonlinear screening
regime). Besides, on such small scales non-gravitational baryonic effects come into play
(such as AGN feedback) and the symmetry no longer holds.

For our purposes in this paper, the formulation (\ref{continuity-symm})-(\ref{kappa-def})
explicitly shows that, from cluster scales up to the Hubble scale, K-mouflage
cosmologies belong to the same family as the $\Lambda$-CDM and quintessence scenarios,
with respect to matter clustering. The equations that govern the gravitational dynamics in these
rescaled variables take the same form, except for a time-dependent function $\kappa(D_+)$.
However, the shape of this function is similar for realistic scenarios
(we shall see in Fig.~\ref{fig_eps_z} that $\epsilon_1$ is about $2\%$). Therefore, we can
expect that gravitational clustering shows the same qualitative properties in these cosmologies
and only small quantitative deviations. In particular, semianalytical methods should work
equally well for all these cosmologies, and phenomenological observations, such as
the fact that virialized halos are well described by Navarro-Frenk-White (NFW)
profiles \cite{Navarro:1996} in $\Lambda$-CDM cosmology, should remain valid in other
cases.
This justifies our modelization of clusters, described in Sec.~\ref{sec:clusters} below,
where we treat $\Lambda$-CDM and K-mouflage cosmologies in the same manner.

\section{Numerical results for large-scale structures}
\label{sec:Numerical-large-scale}

In this paper we consider two simple models for $K(\tilde{\chi})$. The first one, which we
call the ``arctan model'' in the following, is defined by:
\beq
K_{\rm arctan}(\tilde{\chi}) = -1 + \tilde{\chi} +K_* [ \tilde{\chi} - \chi_* \arctan(\tilde{\chi} / \chi_*) ] ,
\label{K-arctan-def}
\eeq
with the low-$\tilde{\chi}$ expansion
\beq
\tilde{\chi} \rightarrow 0 : \;\;\; K_{\rm arctan}(\tilde{\chi}) = -1 + \tilde{\chi} + \frac{K_*}{3}
\frac{\tilde{\chi}^3}{\chi_*^2} - \frac{K_*}{5} \frac{\tilde{\chi}^5}{\chi_*^4} + ...
\label{low-chi-1}
\eeq
and the choice of parameters
\beq
K_*= 10^3 , \;\;\; \chi_*= 10^2 .
\eeq
This gives a K-mouflage model that is consistent with both Solar System and cosmological
constraints (with $\beta=0.1$).
For comparison we also consider the model used in \cite{Brax:2014a,Brax:2014b},
which we call the ``cubic model'' in the following,
in which $K(\tilde{\chi})$ is a low-order polynomial:
\beq
K_{\rm cubic}(\tilde{\chi}) = -1 + \tilde{\chi} + K_{0}\tilde{\chi}^{m} ,
\label{K-cub-def}
\eeq
and the choice of the parameters
\beq
m = 3, \;\;\; K_{0} = 1.
\eeq
Here Eq.(\ref{K-cub-def}) should not be understood as a perturbative expansion around
$\tilde\chi=0$. It is rather a simple model that interpolates between the low-$\tilde\chi$
behavior (\ref{K-chi=0}) and a large-$\tilde\chi$ power-law behavior $\propto \tilde\chi^m$.
The cubic model is consistent with cosmological constraints but its form at large negative
$\chi$, $-\chi \gg 1$, is not consistent with Solar System constraints. Therefore, this is
an effective model that applies to the semiaxis $\chi \geq -1$ while the large-negative
domain is left unspecified. This is sufficient for our purposes, since the cosmological
background and large-scale perturbations correspond to $\chi>0$ and clusters correspond
to $\chi>-1$ (more precisely, $| \chi | \ll 1$).

For both models we choose an exponential form for the coupling function,
\beq
A(\varphi) = e^{\beta\varphi/\tilde{M}_{\rm Pl}} , \;\;\; \mbox{with} \;\;\; \beta=0.1
\label{A-exp-def}
\eeq
We also consider a reference $\Lambda$-CDM model for comparison.

All the cosmological scenarios are normalized to the same background cosmological
parameters today,
$\{\Omega_{\rm m0},\Omega_{\rm (r)0},\Omega_{\rm de0},H_0\}$.
In addition, we normalize the Planck mass (\ref{MPl-Jordan}) to the same value
$M_{\rm Pl0}^2$ today, as measured by Solar System experiments.
This means that we renormalize the Einstein-frame Planck mass by a factor $\bar{A}_0^2$,
where $\bar{A}_0=\bar{A}(z=0)$,
\beq
M_{\rm Pl}^2(t) = M_{\rm Pl0}^2 \frac{\bar{A}_0^2}{\bar{A}(t)^2} , \;\;\;
\mbox{whence} \;\;\; \tilde{M}_{\rm Pl}^2 = M_{\rm Pl0}^2 \bar{A}_0^2 .
\label{MPlanck-renorm}
\eeq

On the other hand, the matter density power spectrum $P(k)$ is normalized
to the same value at high redshift, when dark energy is subdominant and both
cosmologies almost coincide. However, these different scenarios do not exactly converge
in terms of the background expansion rate at a given matter density, because of the
different high-redshift reduced Planck masses.
Therefore, the normalization to the same power spectrum for the matter density contrast
at high $z$ is somewhat arbitrary, since the K-mouflage and $\Lambda$-CDM models
do not coincide.
Nevertheless, for our purposes this is a convenient choice as it illustrates how the difference
in the gravitational clustering dynamics that appear at low $z$, because of the fifth force
mediated by the scalar field, affect the late-time density field.
[For other normalization choices, it would be difficult to distinguish the effects due to the
different normalizations at high $z$, before the dark-energy era and when the fifth force was
negligible, and to the late-time dynamics characterized by different growth rates.]
This normalization also corresponds to the case where we can measure the density contrast
field, i.e. the patterns of large-scale structures (e.g., the scale associated with the nonlinear
transition $\sigma^2=1$) at high $z$, independently of accurate measures of the
background density and expansion rate.
In practice, this normalization ambiguity does not appear because one compares each
cosmological scenario with the data, rather than comparing with a theoretical reference
cosmology (in particular, the best fits associated with different theories will typically have
slightly different cosmological parameters and expansion rates at both $z=0$ and
$z\rightarrow\infty$).

In the following, we present our results for the choice of cosmological parameters today
given by $\Omega_{\rm m0}=0.25$, $\Omega_{\rm de0}=0.75$, $h=0.70$ and
$\sigma_8=0.7$.

\subsection{Background dynamics}
\label{sec:Background}

\begin{figure}
\begin{center}
\epsfxsize=8.5 cm \epsfysize=5.8 cm {\epsfbox{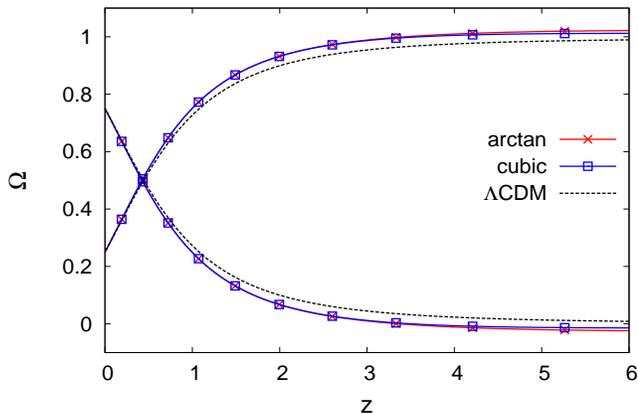}}
\end{center}
\caption{Evolution with redshift of the matter and dark-energy cosmological parameters
$\Omega_{\rm m}(z)$ and  $\Omega_{\rm de}(z)$.
We display the two K-mouflage models of Eq.(\ref{K-arctan-def}) (arctan model,
red crosses) and Eq.(\ref{K-cub-def}) (cubic model, blue squares), and the reference
$\Lambda$-CDM universe (black dashed lines).
The two scalar-field models almost coincide in this figure.}
\label{fig_Om_z}
\end{figure}

\begin{figure}
\begin{center}
\epsfxsize=8.5 cm \epsfysize=5.8 cm {\epsfbox{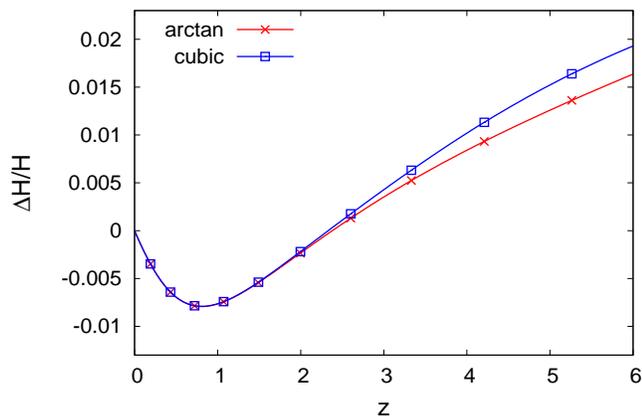}}
\end{center}
\caption{Relative deviation of the Hubble expansion rate with respect to the
$\Lambda$-CDM reference,
$\Delta H/H= H/H_{\Lambda\rm-CDM}-1$,
for the same K-mouflage models as in Fig.~\ref{fig_Om_z}.}
\label{fig_H_z}
\end{figure}

\begin{figure}
\begin{center}
\epsfxsize=8.5 cm \epsfysize=5.8 cm {\epsfbox{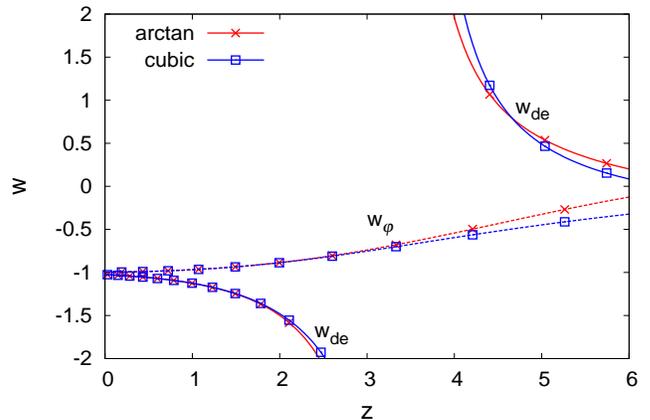}}
\end{center}
\caption{Effective equation of state parameters $w_{\rm de}(z)$ (solid lines with
a divergence and change of sign at $z\simeq 3$) and $w_{\varphi}(z)$
(dashed lines with a smooth behavior), for the same K-mouflage models as in
Fig.~\ref{fig_Om_z}.}
\label{fig_w_z}
\end{figure}

We consider the density parameters $\Omega_{\rm m}(z)$ and  $\Omega_{\rm de}(z)$ in
Fig.~\ref{fig_Om_z} for the two K-mouflage models defined in equations (\ref{K-arctan-def})
and (\ref{K-cub-def}) and for the reference $\Lambda$-CDM universe. Since we
normalize the density and dark-energy parameters to be equal to the ones
observed today, all models coincide at $z=0$ in terms of background quantities.
The deviations from the $\Lambda$-CDM scenario are slightly greater for the arctan model
(\ref{K-arctan-def}) than for the cubic model (\ref{K-cub-def}), in agreement with
\cite{Barreira2015}.
This is due to the fact that $K'(\tilde{\chi})$ is slightly smaller in the former case over
the range of redshifts of interest, $z \lesssim 6$, and that deviations from the
$\Lambda$-CDM scenario typically scale as $\beta^2/K'$, see for instance
Eq.(\ref{tilde-epsilon1-def}) and Ref.\cite{Brax:2014a}.

As for studies performed in the Einstein frame \cite{Brax:2014a} (where the cosmological
parameters are normalized by their Einstein-frame values today), we find that the 
dark-energy density becomes negative (and subdominant) at high redshift.
This gives $\Omega_{\rm m}>1$ at high $z$ for the two K-mouflage
models (\ref{K-arctan-def}) and (\ref{K-cub-def}) (but as in the $\Lambda$-CDM case
$\Omega_{\rm m}\rightarrow 1$ at high redshift).

In Fig.~\ref{fig_H_z} we consider the relative deviation, $H(z)/H_{\Lambda-\rm CDM}(z) - 1$,
of the Hubble rate with respect to the $\Lambda$-CDM reference. The deviation is slightly
larger for the cubic model (\ref{K-cub-def}) at $z \sim 6$, but this is only a transient effect
because at $z>12$ the deviation is slightly greater for the arctan model (\ref{K-arctan-def}),
as expected. At moderate redshifts the Hubble expansion rates differ by less than $2\%$
between the three cosmologies that we consider here.
This amplitude is mostly set by our choice of coupling constant $\beta=0.1$, because as
recalled above deviations from the $\Lambda$-CDM scenario scale as $\beta^2/\bar{K}'$
and at low $z$ we have $\bar{K}' \simeq 1$. Therefore $\beta=0.1$ typically leads to
percent deviations from the $\Lambda$-CDM scenario.
This value of $\beta$ (or lower values) is required to satisfy observational constraints on
cosmological and Solar System scales \cite{Barreira2015}, in particular from the expansion
rate at the time of Big Bang Nucleosynthesis and from the bounds on the current time
derivative of the gravitational coupling ${\cal G}$ provided by the Lunar Ranging experiment.

The deviation from the $\Lambda$-CDM reference does not vanish at high redshift
because the reduced Planck masses are different, see Eq.(\ref{MPlanck-renorm}).
Indeed, in the K-mouflage models $M_{\rm Pl}^2(t)$ becomes time dependent and we choose
to normalize all scenarios by their Planck mass today (when Solar System measurements and
laboratory experiments are performed).
Note that in studies performed in the Einstein frame \cite{Brax:2014a},
where the reduced Planck mass is constant, one can recover the $\Lambda$-CDM
expansion rate at both $z=0$ and at high redshift. However, this requires normalizing the
matter density today by the conserved density $\hat{\rho}$ of Eq.(\ref{rho-hat-def})
instead of the Einstein-frame density $\tilde{\rho}$.

In Fig.~\ref{fig_w_z} we display the effective equation of state parameter for the dark energy,
$w_{\rm de}=\bar{p}_{\rm de}/\bar{\rho}_{\rm de}$, evaluated using Eqs.(\ref{rho-de-Jordan})
and (\ref{p-de-Jordan}) for $\bar{\rho}_{\rm de}$ and $\bar{p}_{\rm de}$. For both models
$w_{\rm de} \rightarrow - 1$ at late times, mimicking the presence of a cosmological constant.
As in studies performed in the Einstein frame, the effective equation of state parameter
is beyond $-1$ at low $z$ and changes sign at a moderate redshift while going through
$\pm\infty$ (this does not correspond to a singularity in terms of the Hubble rate or 
dark-energy density but to the vanishing and change of sign of $\bar{\rho}_{\rm de}$).
We also show in Fig.~\ref{fig_w_z} the equation of state parameter $w_{\varphi}$ defined
as
\beq
w_{\varphi} = \frac{\bar{p}_{\varphi}}{\bar{\rho}_{\varphi}} = \frac{\bar{K}}{2\bar{\tilde\chi}
\bar{K}'-\bar{K}} ,
\label{wphi-def}
\eeq
where we used Eq.(\ref{rho-phi-def}).
In contrast to $w_{\rm de}$, $w_{\varphi}$ remains negative over $z \leq 6$ and shows
no divergence.
The difference between the behaviors of $w_{\rm de}$ and $w_{\varphi}$ shows the impact of
the coupling between the matter and scalar-field components. This makes the dark-energy
density and pressure significantly different from the bare scalar-field ones, see
Eqs.(\ref{rho-de-Jordan}) and (\ref{p-de-Jordan}), and can even make
$\bar{\rho}_{\rm de}$ and $\bar{\rho}_{\varphi}$ have different signs.

\subsection{Background scalar field}
\label{sec:background-scalar}

\begin{figure}
\begin{center}
\epsfxsize=8.5 cm \epsfysize=5.8 cm {\epsfbox{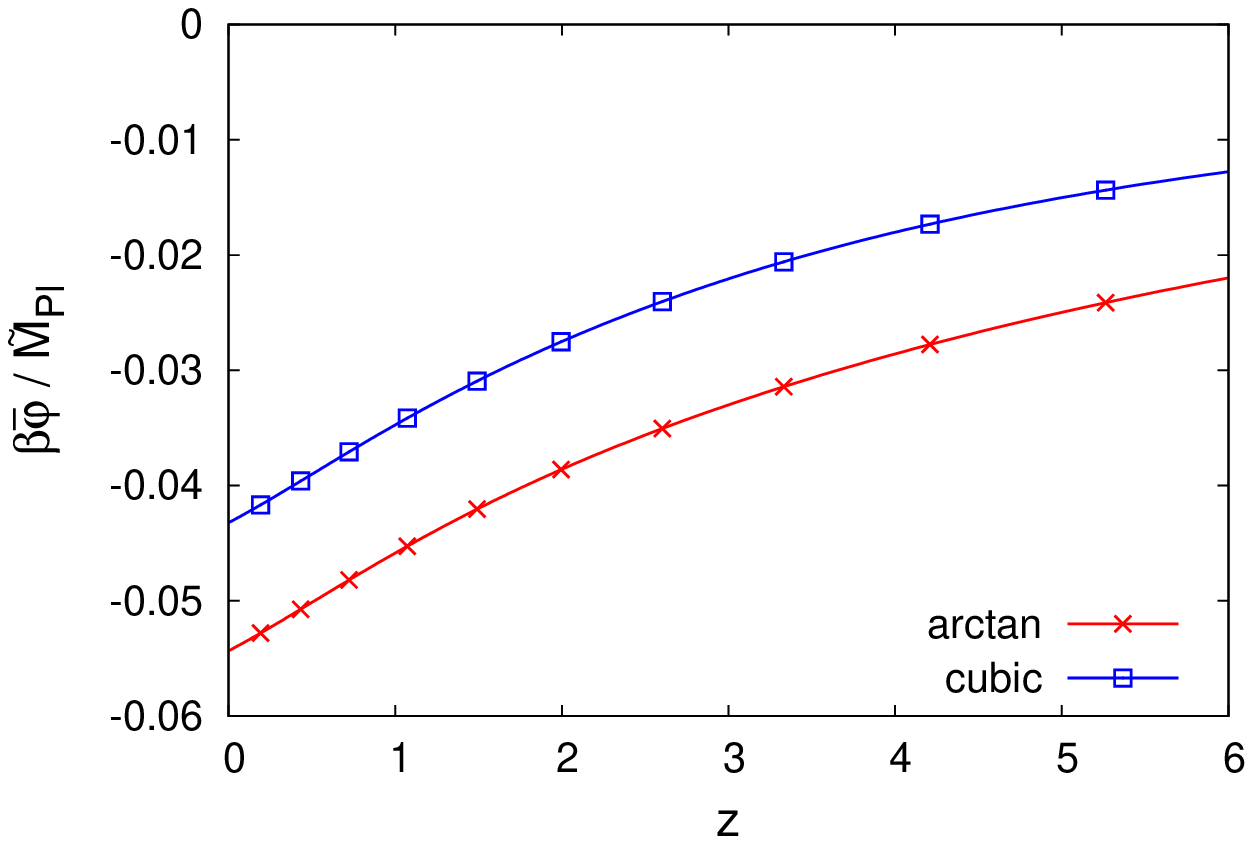}}\\
\epsfxsize=8.5 cm \epsfysize=5.8 cm {\epsfbox{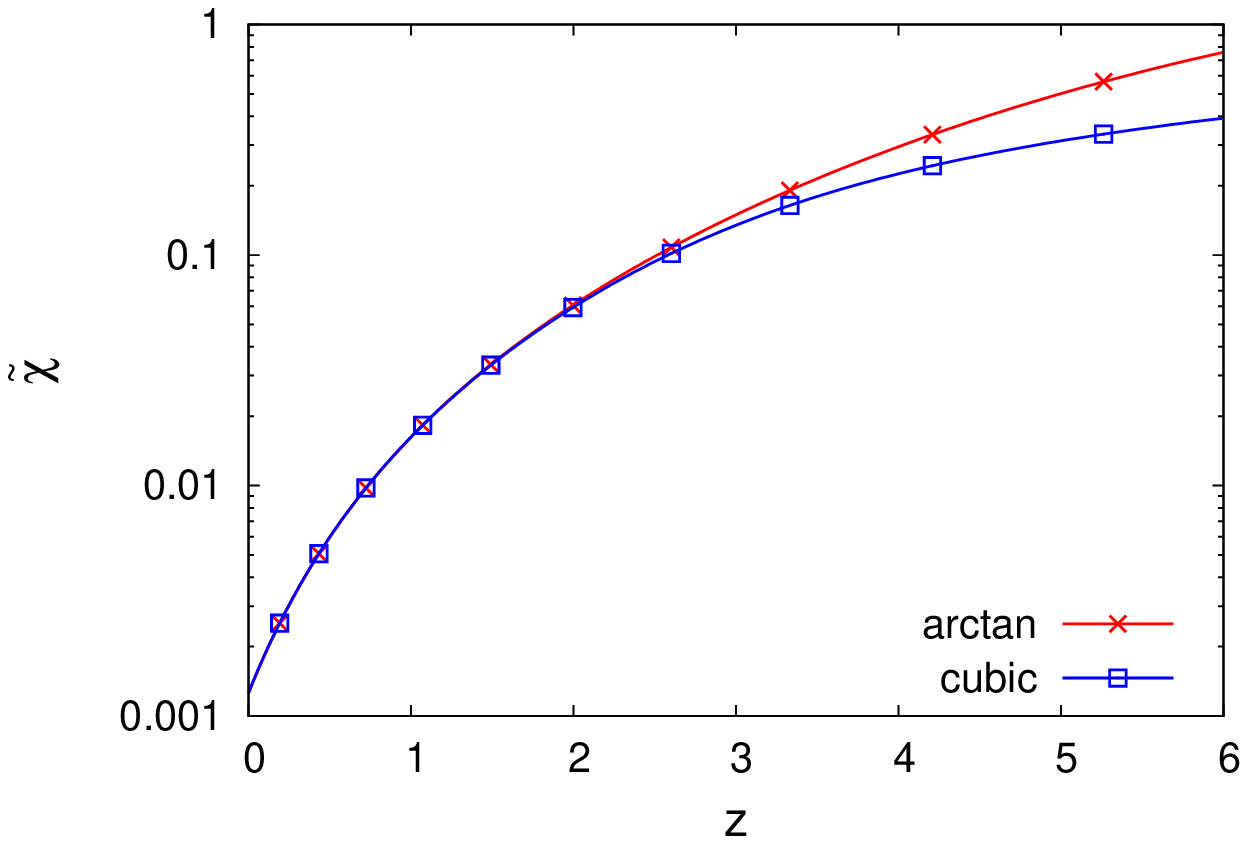}}
\end{center}
\caption{{\it Upper panel:} background scalar field $\bar{\varphi}$ as a function of redshift.
{\it Lower panel:} background kinetic term $\bar{\tilde\chi}$ as a function of redshift.}
\label{fig_chi_varphi_z}
\end{figure}

We show in Fig.~\ref{fig_chi_varphi_z} the background values $\bar\varphi$ and
$\bar{\tilde\chi}$ of the scalar field and of its kinetic term.
The scalar field $\bar\varphi$ is negative and its amplitude grows with redshift
(we chose the normalization $\bar\varphi=0$ at high redshift, $z\rightarrow\infty$).
We can see that $| \beta\bar\varphi/\tilde{M}_{\rm Pl} | \ll 1$ until $z=0$.
Thus, the coupling function $A(\varphi)$ is dominated by its low-order terms in the
expansion (\ref{A-phi=0}) and choosing for instance
$A(\varphi) = (1+\beta\varphi/n\tilde{M}_{\rm Pl})^n$, with $n>0$, would give
similar results to the exponential choice (\ref{A-exp-def}).

The kinetic term $\bar{\tilde\chi}$ decreases with time. It goes to infinity at high redshift,
$z\rightarrow\infty$, and we can check that at low $z$ it is significantly smaller than unity.
Then, the kinetic function $K(\tilde\chi)$ is dominated by its low-order terms in the expansion
(\ref{K-chi=0}). This explains why the two K-mouflage models converge at low $z$ in the
lower panel of Fig.~\ref{fig_chi_varphi_z}.

\subsection{Modified gravitational potentials, gravitational slip and effective Newton's constant}
\label{sec:potential}

\begin{figure}
\begin{center}
\epsfxsize=8.5 cm \epsfysize=5.8 cm {\epsfbox{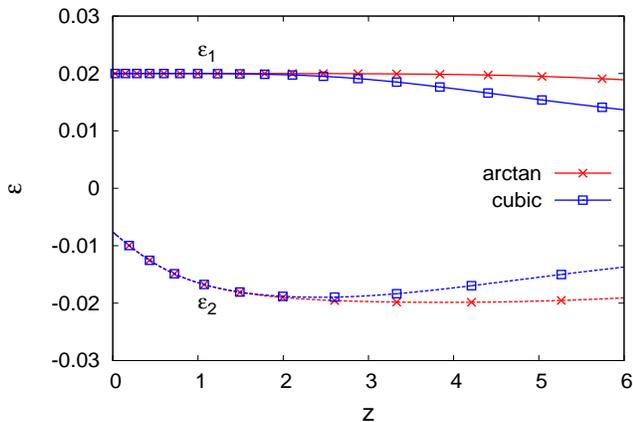}}
\end{center}
\caption{Coefficients $\epsilon_1$ and $\epsilon_2$, defined in Eqs.(\ref{tilde-epsilon1-def}) and
(\ref{tilde-epsilon2-def}) for the K-mouflage models, as functions of redshift.}
\label{fig_eps_z}
\end{figure}

As seen in Sec.~\ref{sec:Jordan-frame}, both for the background quantities and the
large-scale perturbative structures, the deviations from the $\Lambda$-CDM universe can be
measured by the two coefficients $\epsilon_1(t)$ and $\epsilon_2(t)$, defined in
Eqs.(\ref{tilde-epsilon1-def}) and (\ref{tilde-epsilon2-def}).
In particular, from Eqs.(\ref{Phi-Psi-Jordan}) and (\ref{dA-dphi-drho}) the two gravitational
potentials $\Phi$ and $\Psi$ of the Jordan-frame metric (\ref{Newtonian-gauge-Jordan})
read for large-scale unscreened structures as
\beq
\Phi = (1+\epsilon_1) \Psi_{\rm N} , \;\;\; \Psi= (1-\epsilon_1) \Psi_{\rm N} ,
\label{Phi-Psi-epsilon-1}
\eeq
and the normalized gravitational slip writes as
\beq
\eta \equiv \frac{\Psi-\Phi}{\Psi+\Phi}  = - \epsilon_1 .
\label{eta-def}
\eeq
We show both coefficients $\epsilon_1(t)$ and $\epsilon_2(t)$ in Fig.~\ref{fig_eps_z}.
We can see that they are of order $2\%$ at low $z$. Again, as can be seen from
Eq.(\ref{tilde-epsilon1-def}), this amplitude is set by our choice $\beta=0.1$ (to satisfy
observational constraints) as $\bar{K}'\simeq 1$ at low $z$ and deviations from the
$\Lambda$-CDM reference then scale as $\beta^2$.
This also sets the amplitude of the gravitational slip $\eta$.
At high $z$ the coefficients $\epsilon_i(t)$ go to zero, as $\bar{K}'$ goes to infinity
and we enter a cosmological nonlinear screening regime that also ensures that the 
dark-energy component becomes subdominant at early times.
This decrease of $\epsilon_i$ appears faster for the cubic model, because of its stronger
growth of $K'(\chi)$ at large positive $\chi$. As noticed above, this means that departures
from the $\Lambda$-CDM scenario are greater for the arctan model than for the cubic model
(with our choice of parameters).

The Jordan-frame coefficient $\epsilon_1$ is always positive and the gravitational slip
$\eta$ defined in Eq.(\ref{eta-def}) is negative.
From Eq.(\ref{tilde-epsilon2-def}) the coefficient $\epsilon_2$ also reads as
\beq
\epsilon_2(t) = \frac{\beta(t)}{\tilde{M}_{\rm Pl}} \frac{\dd\bar{\varphi}}{\dd\ln a}
= \frac{\beta(t)}{\tilde{M}_{\rm Pl} H(t)} \frac{\dd\bar{\varphi}}{\dd t} ,
\label{epsilon2-phi}
\eeq
which is negative from Eq.(\ref{KG-Jordan-int}) and of order $\beta^2/\bar{K}'$.

\begin{figure}
\begin{center}
\epsfxsize=8.5 cm \epsfysize=5.8 cm {\epsfbox{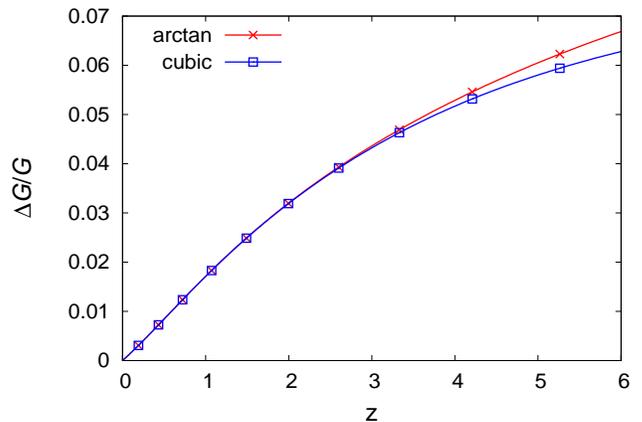}}
\end{center}
\caption{Relative drift with redshift of the effective Newton constant for the K-mouflage models.}
\label{fig_GNewt_z}
\end{figure}

In Fig.~\ref{fig_GNewt_z} we show the evolution with redshift of the effective Newton
constant, defined from Eq.(\ref{MPlanck-renorm}) as
\beq
\cG(t) = \cG_0 \frac{\bar{A}^2(t)}{\bar{A}_0^2} .
\label{GNewt-eff}
\eeq
Because of the dependence of the effective Newton coupling strength on the background
value of the scalar field, $\cG$ is a few percent higher at $z \sim 6$ than today.

\subsection{Linear theory}
\label{sec:Perturbation-theory}

\begin{figure}
\begin{center}
\epsfxsize=8.5 cm \epsfysize=5.8 cm {\epsfbox{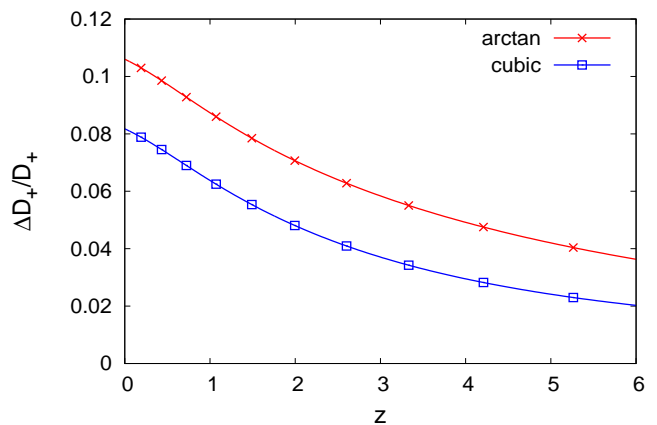}}
\end{center}
\caption{Relative deviation, $D_{+}/D_{+\Lambda-\rm CDM} - 1$, of the linear growing mode $D_+$
from the $\Lambda$-CDM reference.}
\label{fig_Dlin_z}
\end{figure}

\begin{figure}
\begin{center}
\epsfxsize=8.5 cm \epsfysize=5.8 cm {\epsfbox{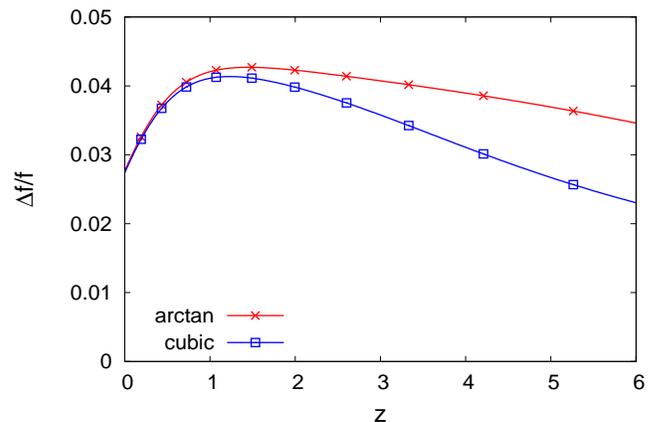}}
\end{center}
\caption{Relative deviation, $f/f_{\Lambda\rm CDM}-1$, of the linear growth rate,
$f(z)=\dd\ln D_{+}/\dd\ln a$, from the $\Lambda$-CDM reference.}
\label{fig_f_z}
\end{figure}

In Fig.~\ref{fig_Dlin_z} we show the relative deviation, $D_{+}/D_{+\Lambda-\rm CDM} - 1$,
of the linear growing mode, obtained by solving equation (\ref{D-linear-Jordan}),  from the
$\Lambda$-CDM reference case, and the linear growth rates $f(z)$. Again, the relative
deviation of the growing mode is greater for the arctan model (\ref{K-arctan-def}) than for
the cubic model (\ref{K-cub-def}), because of the lower value of $\bar{K}'$ over relevant
redshifts, see Eq.(\ref{tilde-epsilon1-def}) for the coefficient $\epsilon_1$ that modifies
the linear growing mode equation (\ref{D-linear-Jordan}).
All linear growing modes converge at high redshift, despite the slightly different Planck
masses and Hubble expansion rates. Indeed, far in the early matter-dominated era we
recover an Einstein-de Sitter cosmology and the Hubble term in the parenthesis in
Eq.(\ref{D-linear-Jordan}) converges to $H^{-2} \dd H/\dd t \rightarrow -3/2$.
Moreover, the factor $\epsilon_1$ goes to zero because of the nonlinear K-mouflage
screening mechanism, see Eq.(\ref{tilde-epsilon1-def}), as at high redshift $\tilde{\chi}$
and $\bar{K}'$ become large.
This large-$K'$ behavior is also required to ensure that the background dark-energy
density becomes subdominant.

We show the relative deviation of the linear growth rates $f(z)$ in Fig.~\ref{fig_f_z}.
Overall, $f(z)$ is greater for the K-mouflage scenarios, in agreement with the higher value of the linear
growing mode shown in Fig.~\ref{fig_Dlin_z}. The deviation is again of the order of a few percent.
The deviation for the linear modes $D_+$ shown in Fig.~\ref{fig_Dlin_z} could reach $10\%$ at $z=0$
because of the cumulative effect due to the integration over time.
The growth rates $f(z)$ converge to unity at high redshift but we can see that there remains a noticeable
difference between the K-mouflage models and the $\Lambda$-CDM reference up to $z \sim 6$.

This rather slow decrease of the deviations from the $\Lambda$-CDM reference at higher redshift
is a characteristic signature of K-mouflage models, as many other modified-gravity models,
such as $f(R)$ theories, lead to a faster convergence to the $\Lambda$-CDM scenario at $z \gtrsim 2$.
This is related to the fact that in the linear sub-horizon regime the K-mouflage effects are scale independent,
as the factors $\epsilon_1(t)$ and $\epsilon_2(t)$ only depend on time.
In contrast, in $f(R)$ theories or Dilaton models, the factor $\epsilon(k,t)$ that appears in the modified
Euler or Poisson equations, or in the evolution equation for the linear density modes, takes the form
$\epsilon(k,t) \propto \beta^2 k^2/(a^2m^2+k^2)$, with a characteristic physical scale $2\pi/m$ beyond
which the theory converges to General Relativity. At high redshift this scale typically goes to zero,
so that at a fixed physical (or also comoving) scale, deviations from the $\Lambda$-CDM scenario
vanish because the coupling $\beta$ decreases or one enters the unmodified regime beyond $2\pi/m$.
In the K-mouflage models that we consider in this paper, because there is no such characteristic scale
the convergence to General Relativity is only due to the vanishing of the effective coupling strength
$\beta^2/K'$, with $\beta$ being constant (in our case) and $K'$ increasing at high $z$ because of the
nonlinear K-mouflage mechanism itself.
However, this decrease of $\beta^2/K'$ at high $z$ is rather slow for generic kinetic functions $K(\chi)$,
as seen from the curve obtained for $\epsilon_1(t)$ in Fig.~\ref{fig_eps_z} for the simple cubic model.

\subsection{Halo mass function}
\label{sec:Numerical-mass-function}

\begin{figure}
\begin{center}
\epsfxsize=8.5 cm \epsfysize=5.8 cm {\epsfbox{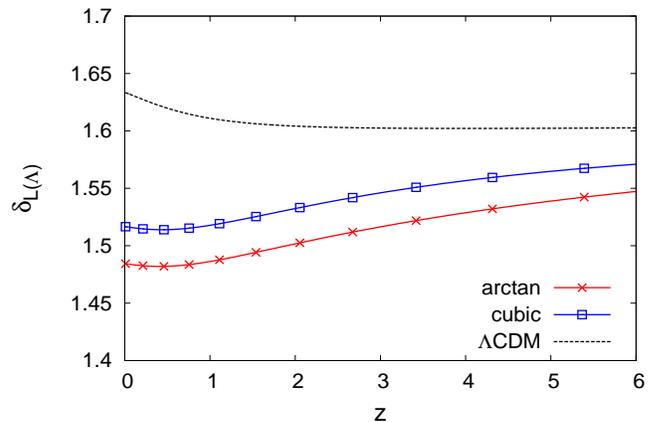}}
\end{center}
\caption{Linear density contrast threshold $\delta_{L(\Lambda)}$ associated with the nonlinear density
contrast $\Delta_c=200$ with respect to the critical density $\rho_{\rm crit}$, for the K-mouflage
models and the $\Lambda$-CDM reference.}
\label{fig_delta_L_z}
\end{figure}

\begin{figure}
\begin{center}
\epsfxsize=8.5 cm \epsfysize=5.8 cm {\epsfbox{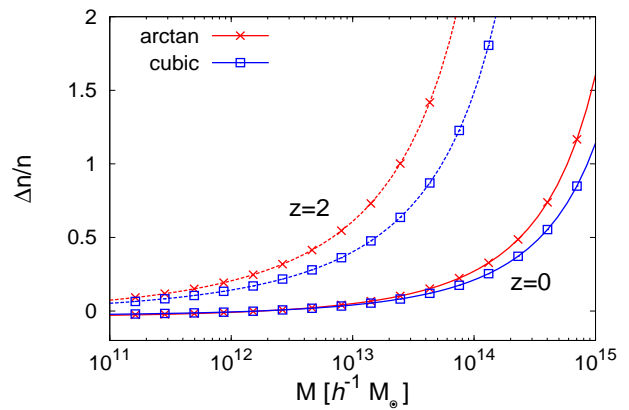}}
\end{center}
\caption{Relative deviation, $n(M)/n_{\Lambda\rm-CDM}(M) - 1$, of the halo mass function
of the K-mouflage models from the $\Lambda$-CDM reference, at $z=0$ (solid lines)
and $z=2$ (dashed lines). Halos are defined by
the density contrast $\Delta_c=200$ with respect to the critical density.}
\label{fig_dnM_z0}
\end{figure}

Solving the spherical collapse equation (\ref{y-lna-Jordan}), we can compute the linear density
contrast threshold $\delta_{L}(z)$ that corresponds to a nonlinear density contrast of
$\delta[\delta_L]=\Delta_{\rm m}$,
where $\Delta_{\rm m}$ is the nonlinear threshold that we choose to define halos.
As discussed in \cite{Brax:2014b}, we are not interested in  $\delta_{L}(z)$ at the observation redshift,
because it is not an observable quantity. Instead, we wish to evaluate the linear threshold
$\delta_{L_{\rm i}}$,  at a given high redshift $z_{\rm i}$, that is required to produce at later time
$z$ a nonlinear density contrast $\Delta_{\rm m}$. In other words, we want to estimate the
initial density fluctuation associated with a given nonlinear density contrast at the observed redshift.
To avoid the introduction of an arbitrary high redshift $z_{\rm i}$, following what it is done
in \cite{Brax:2014b} and usual practice, we translate all the initial thresholds $\delta_{L_{\rm i}}$
to redshift $z$ by multiplying them by $D_{+\Lambda-\rm CDM}(z)/D_{+\Lambda-\rm CDM}(z_{\rm i})$
[instead of using $D_{+}(z)/D_{+}(z_{\rm i})$,
i.e. the linear growing mode associated with each K-mouflage scenario],
and we denote this quantity the linear density contrast threshold $\delta_{L(\Lambda)}$.
This common multiplicative factor enables a meaningful comparison between the different scenarios.

In Fig.\ref{fig_delta_L_z} we display $\delta_{L(\Lambda)}$ when we define halos by a constant
density contrast threshold $\Delta_c=200$ with respect to the critical density $\rho_{\rm crit}$.
This corresponds to a density contrast with respect to the mean density $\bar{\rho}$ of
$\Delta_{\rm m}=\Delta_c/\Omega_{\rm m}(z)$.
We choose a constant $\Delta_c$ rather than $\Delta_{\rm m}$ because observational cluster surveys
usually define cluster halos by a constant overdensity with respect to the critical density $\rho_{\rm crit}$.
At high redshift both definitions coincide, as $\Omega_{\rm m}\rightarrow 1$, while at low redshift
or far in the dark-energy dominated era one can argue that $\Delta_c$ makes more physical sense.
Indeed, during an exponential acceleration phase the growth of large-scale structures freezes out and
one obtains isolated halos among increasingly large voids. Then, the mean universe density
$\bar{\rho}$ decreases as $a^{-3}$, following the dilution due to the expansion, and it does not correspond
to the typical density of halos (nor voids). In contrast, we can assume that the isolated halos keep
a roughly constant physical radius and density, like the critical density $\rho_{\rm crit}$ (in an exponential
phase where the Hubble rate is constant), so that it is more meaningful to express halo densities in terms
of $\rho_{\rm crit}$.

Both K-mouflage models accelerate the growth of large-scale structures as compared with the
$\Lambda$-CDM reference, as seen from the linear growing modes of Fig.~\ref{fig_Dlin_z}.
Therefore, we find in Fig.\ref{fig_delta_L_z} that a smaller linear density contrast is required to reach the
same nonlinear overdensity of $200$ (with respect to the critical density). Again, the deviation from the
$\Lambda$-CDM prediction is greater for the arctan model (\ref{K-arctan-def}).

From the linear threshold displayed in Fig.\ref{fig_delta_L_z} we obtain the halo mass function
as in Eq.(\ref{nM-def}) (note that
$\nu=\delta_L/\sigma=\delta_{L(\Lambda)}/\sigma_{(\Lambda)}=\delta_{L\rm i}/\sigma_{\rm i}$).
In Fig.\ref{fig_dnM_z0} we show the relative deviation $n(M)/n_{\Lambda-\rm CDM}(M) - 1$ of the halo
mass function from the $\Lambda$-CDM reference case, at $z=0$ and $z=2$.
As for the case of $\delta_{L(\Lambda)}$,
since the scalar field enhances gravitational clustering we find that the mass functions for the two K-mouflage
models are higher than the $\Lambda$-CDM reference in the high-mass tail, with the greater
deviation obtained for the arctan model (\ref{K-arctan-def}). As usual, the deviation increases at high mass
because the exponential falloff amplifies the sensitivity to slight departures of the growth of structures.
(The deviation becomes slightly negative at low mass because all mass functions obey the same
normalization to unity: there cannot be more matter in halos than the matter content of the universe.)

At fixed mass, $M \sim 10^{14} h^{-1} M_{\odot}$, the deviation from the $\Lambda$-CDM reference
is greater at $z=2$ than at $z=0$, despite the difference in linear density thresholds being lower,
as seen in Fig.~\ref{fig_delta_L_z}. This is because at fixed mass we are further into the rare
high-mass tail, which amplifies the dependence on the linear density threshold and more than
compensates the slow convergence between the K-mouflage and $\Lambda$-CDM thresholds.

\subsection{Matter density power spectrum and correlation function}
\label{sec:Numerical-Pk}

\begin{figure}
\begin{center}
\epsfxsize=8.5 cm \epsfysize=5.8 cm {\epsfbox{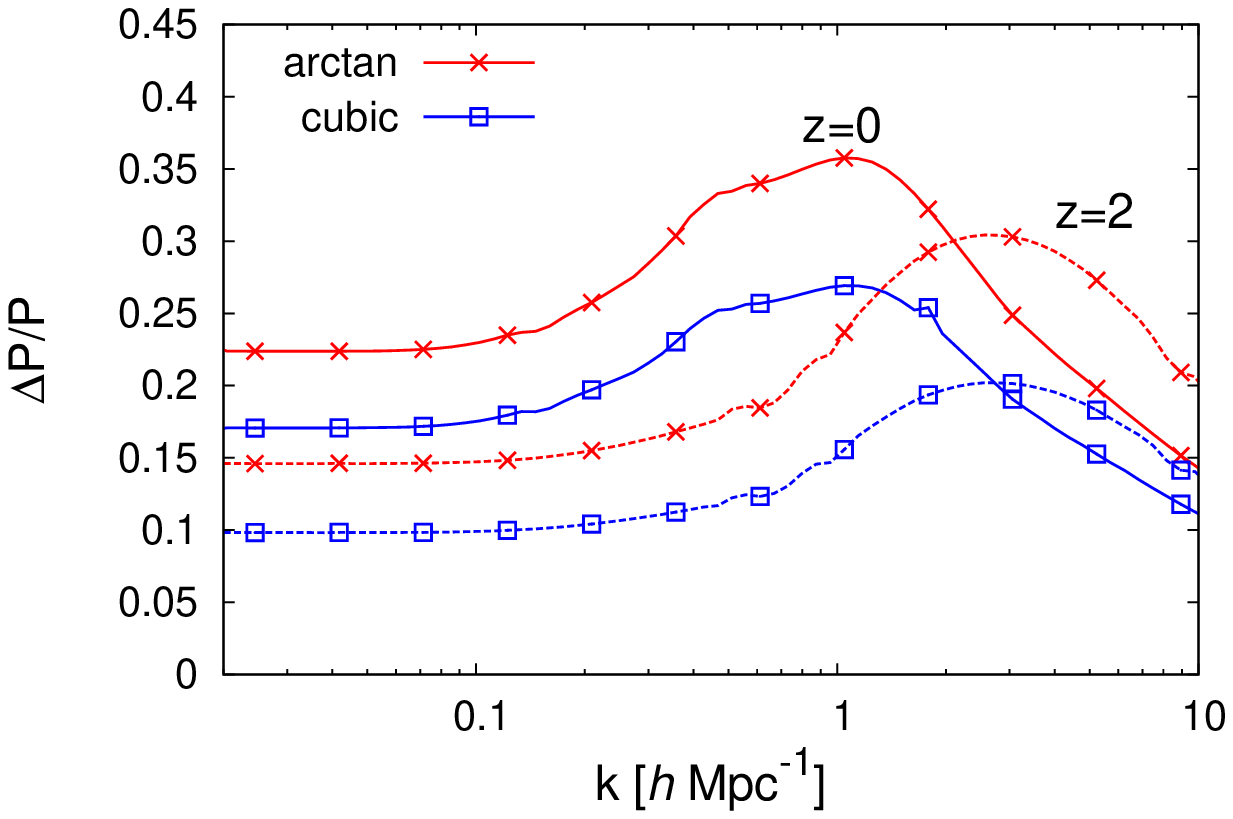}}\\
\epsfxsize=8.5 cm \epsfysize=5.8 cm {\epsfbox{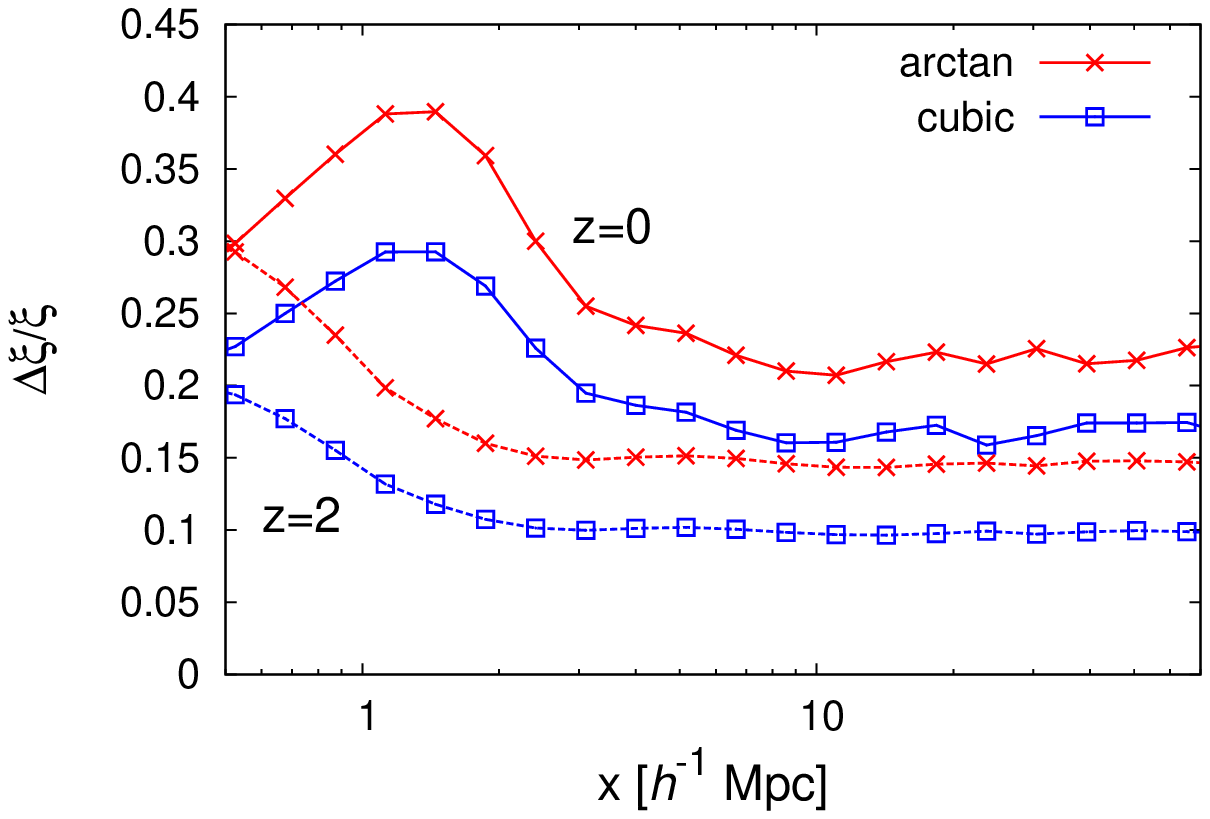}}\end{center}
\caption{{\it Upper panel:} relative deviation of the matter density power spectrum from the
$\Lambda$-CDM reference, at $z=0$ (solid lines) and $z=2$ (dashed lines), for the
K-mouflage models.
{\it Lower panel:} relative deviation of the matter density correlation function from the
$\Lambda$-CDM reference.}
\label{fig_dPk_z0}
\end{figure}

We show in Fig.~\ref{fig_dPk_z0} the matter density power spectra and correlation
functions at $z=0$ and $z=2$.
The computation of this power spectrum $P(k)$ combines perturbation theory up to one-loop order
with a halo model, as described in \cite{Brax:2014b} in the case of Einstein-frame studies
and following the approach introduced in \cite{Valageas2013}.
The two-point correlation function $\xi(x)$ is obtained from the Fourier transform of the
power spectrum.
On large scales we obtain the same constant relative deviation for the power spectra and
the correlation functions, as the linear growing modes $D_+(t)$ are scale independent in both
K-mouflage and $\Lambda$-CDM cosmologies (much below the horizon).
This deviation of $20\%$ is consistent with the deviation of $10\%$ obtained for the linear growing
modes in Fig.~\ref{fig_Dlin_z}.
The deviation from the $\Lambda$-CDM reference grows on mildly nonlinear scales,
as nonlinearities amplify the effects of the fifth force. This is related to the increase of the
large-mass tail of the halo mass function shown in Fig.~\ref{fig_dnM_z0}, because on these
scales the power spectrum and the correlation function probe the formation of massive halos,
as can be clearly seen in a halo model approach. The deviation decreases on smaller scales
because the power spectrum now probes the inner regions of halos and we assume
similar NFW profiles \cite{Navarro:1996} for all cosmologies
(but this regime shows a greater theoretical inaccuracy and numerical simulations would
be need to measure the impact of the modified gravity on small highly nonlinear scales
and halo profiles). However, in the nonlinear range shown in Fig.~\ref{fig_dPk_z0},
the impact of changes to the mass function is greater than that of halo profiles,
see also Ref.~\cite{Valageas2013a} for a detailed study of these various contributions.

The deviation from the $\Lambda$-CDM reference slowly decreases at high $z$, as the fifth
force mediated by the scalar field becomes negligible (as seen from the vanishing of the
key factor $\beta^2/\bar{K}'$ as $\bar{K}'\rightarrow\infty$).
This decrease of the deviations of large-scale clustering from the $\Lambda$-CDM reference
is slower than what is found in many other modified-gravity scenarios, such as the
$f(R)$ theories, and is characteristic of these K-mouflage models.

\section{Clusters of galaxies}
\label{sec:clusters}

To go beyond background quantities and the large-scale perturbative regime, we investigate in this
section the impact of K-mouflage scenarios on the largest collapsed structures that we observe,
that is, clusters of galaxies.
This provides another probe of modified-gravity models, which is complementary with background
and perturbative studies, as it corresponds to the nonlinear regime of the matter density field and
to the well-defined objects measured in actual surveys.

For our purposes, clusters present two advantages as compared with galaxies.
First, they are unscreened objects \cite{Brax:2014b}, so that the impact of the modification to gravity
is very simple and corresponds to a time-dependent effective Newton constant
(we shall check that this holds down to the cluster cores in Sec.~\ref{sec:unscreened}
below).
Therefore, one does not expect dramatic qualitative changes from the $\Lambda$-CDM case,
and the same semi-quantitative models can be applied to both K-mouflage and $\Lambda$-CDM
cosmologies.
This is also illustrated by the symmetry described in Sec.~\ref{sec:symmetry-clustering},
which shows that in this unscreened regime, from the point of view of nonlinear gravitational
clustering, the $\Lambda$-CDM cosmology, quintessence models, and K-mouflage scenarios,
belong to the same class. They obey the same equations of motion
(\ref{continuity-symm})-(\ref{Euler-symm}), with only slightly different time-dependent functions
$\kappa(t)$ from Eq.(\ref{kappa-def}).
Second, at first order clusters can be described by simple physical laws, such as hydrostatic
equilibrium for the gas profile and Bremsstrahlung emission for the X-ray luminosity, giving rise to
the so-called cluster ``scaling laws'' \cite{Kaiser1986}. This is especially true for the most massive
clusters that we focus on.

In contrast, galaxies probe the transition from the unscreened to screened regimes and also involve
many complex astrophysical phenomena, such as cooling processes, star formation, supernovae and
AGN winds and feedback. Therefore, although they would be very interesting probes they are much
more difficult to model, both for the modified-gravity sector and for the usual galaxy formation processes
that also appear in the $\Lambda$-CDM cosmology.

In this paper, our goal is to estimate the magnitude of the impact of K-mouflage models on clusters of
galaxies rather than building a very accurate description of clusters.
Therefore, we consider the simplest possible modelling of clusters with only few physical parameters.
This may not provide the highest-accuracy cluster model, but we can expect that it captures the main
physical processes and provides a robust estimate of the impact of modifications to gravity.
Moreover, we check that our predictions show a reasonable agreement with observations.

\subsection{Halo density profiles}
\label{sec:profiles}

\begin{figure}
\begin{center}
\epsfxsize=8.5 cm \epsfysize=5.8 cm {\epsfbox{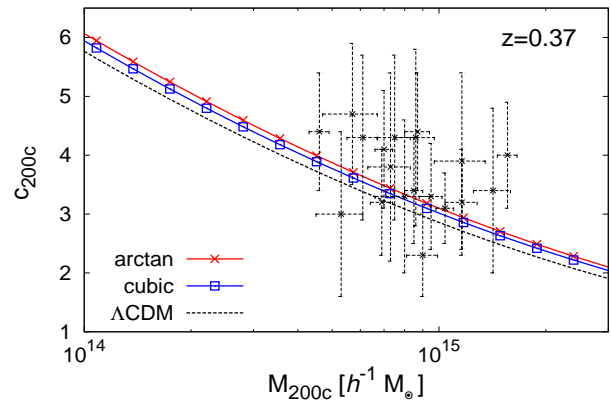}}
\end{center}
\caption{Mass-concentration relation for NFW halos, for the K-mouflage models and the
$\Lambda$-CDM reference, at $z=0.37$.
The black points (with their error bars) are observational measures taken from \cite{Merten2014}.}
\label{fig_cM_z0}
\end{figure}

To study the effects of K-mouflage scenarios on clusters of galaxies we need their
dark matter profile. Because in the unscreened regime gravitational clustering proceeds in the same
fashion in K-mouflage and $\Lambda$-CDM cosmologies, as described in
Sec.~\ref{sec:symmetry-clustering}, we assume in all cases NFW
profiles \cite{Navarro:1996},
\beq
\rho_{\rm DM}(r)=\frac{\rho_{s}}{(r/r_{s})(1+r/r_{s})^{2}} .
\label{NFW-profile}
\eeq
This profile is characterized by a scaling radius and density, respectively $r_{s}$ and $\rho_{s}$,
which can be expressed in terms of the concentration parameter $c=R_{\Delta_c}/r_{s}$.
Here $R_{\Delta_c}$ is the radius such that the mean density within $R_{\Delta_c}$ is $\Delta_c$
times the critical density, $\rho(<R_{\Delta_c})= \Delta_c \rho_{\rm crit}$, as we again define the
extension of the halos by an overdensity threshold with respect to the critical density.
From the definition of $c$, is possible to express $\rho_{s}$ as
\beq
\rho_{s}  = \rho_{\rm crit} \frac{\Delta_{c}}{3}\frac{c^{3}}{\ln(1+c) - c/(1+c)} ,
\label{rho-concentration-relation}
\eeq
which can be inverted to give $c$ as a function of $\rho_s$.

To consider the effects of the presence of the scalar field on the shape of the dark matter profile
we consider a simple model for the halo concentration. We assume that halos of mass $M$
typically form at a redshift $z_{\rm f}(M)$ determined by
\beq
\sigma(q,z_{\rm f}) = \sigma_{\rm f} ,
\label{halo-formation-condition}
\eeq
where $q=(3M/4\pi\bar{\rho}_0)^{1/3}$ is the halo Lagrangian radius and $\sigma_{\rm f}$
is a free parameter, and that the density of the newly-formed halo is proportional to
$\rho_{\rm crit}(z_{\rm f})$,
\beq
\rho_{s}(M) = \Delta_{\rm f} \, \rho_{\rm crit}(z_{\rm f}) ,
\label{rhos-densities-relation}
\eeq
with $\Delta_{\rm f}$ a second free parameter.
Equation (\ref{halo-formation-condition}) means that halos of a given mass typically form when
density fluctuations at this mass scale reach the nonlinear regime, while equation
(\ref{rhos-densities-relation}) assumes that the core of the cluster keeps a roughly constant density
after its formation, which is set by the critical density at the formation time.
As discussed in Sec.~\ref{sec:Numerical-mass-function}, we choose to rescale $\rho_s$ in terms
of the critical density rather than the mean density at redshift $z_{\rm f}$ because the former
is more physical at late times (whereas they coincide at high redshift) and it also corresponds to our
definition of halos.
Next, using equation (\ref{rho-concentration-relation}), we compute $c(M)$ and we define the
dark matter density profile using (\ref{NFW-profile}).

In Fig.~\ref{fig_cM_z0} we display the mass-concentration relation obtained with the choice of
parameters $\sigma_{\rm f}=0.2$ and $\Delta_{\rm f}=500$ (halos are again defined by
$\Delta_c=200$).
As is well known \cite{Navarro1997},
the concentration $c$ (and the scaling density $\rho_s$) is larger for smaller mass,
because in hierarchical scenarios smaller mass scales turned nonlinear at higher redshift when the
critical (and the mean) density of the Universe was greater.
This is of course consistent with our model
(\ref{halo-formation-condition})-(\ref{rhos-densities-relation}).
We compare these results with the mass-concentration relation obtained by \cite{Merten2014},
from the analysis of 19 X-ray selected galaxy clusters from the Cluster Lensing and
Supernova Survey with Hubble (CLASH), with a mean redshift $z \simeq 0.37$.
We can see that reasonable choices of the parameters $\sigma_{\rm f}$ and $\Delta_{\rm f}$
(we naturally expect $\sigma_{\rm f} \lesssim 1$ and $\Delta_{\rm f} \gtrsim 200$) allow us to obtain
a reasonable match
to observations. This suggests that this simple modelling captures the main features of the gravitational
formation of halos. Therefore, we do not consider here more sophisticated models, which involve
the past accretion history and merging trees of virialized halos. These could provide more accurate
modelling, at the price of additional complexity (and often additional parameters), but it is not
clear if their estimate of the dependence on the underlying gravity theory would be much more accurate.
Such studies are left for future works, where N-body simulations would be needed to check detailed models.

As expected, we find a small increase of the concentration $c(M)$ in the K-mouflage models, as compared
with the $\Lambda$-CDM reference. This is due to the faster growth of gravitational clustering, which implies
a slightly greater scaling density $\rho_s(M)$.
However, we can see that the effect is rather modest.

\subsection{Clusters are not screened}
\label{sec:unscreened}

\begin{figure}
\begin{center}
\epsfxsize=8.5 cm \epsfysize=5.8 cm {\epsfbox{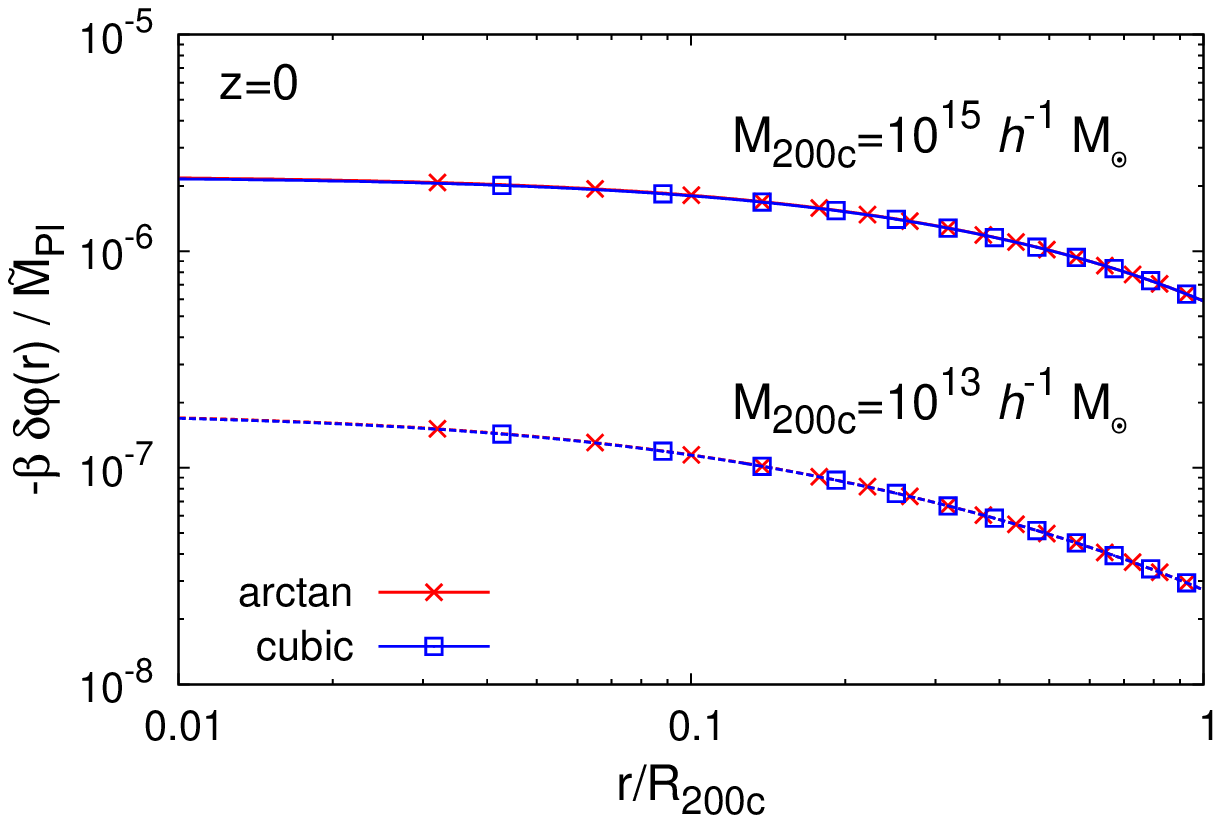}}\\
\epsfxsize=8.5 cm \epsfysize=5.8 cm {\epsfbox{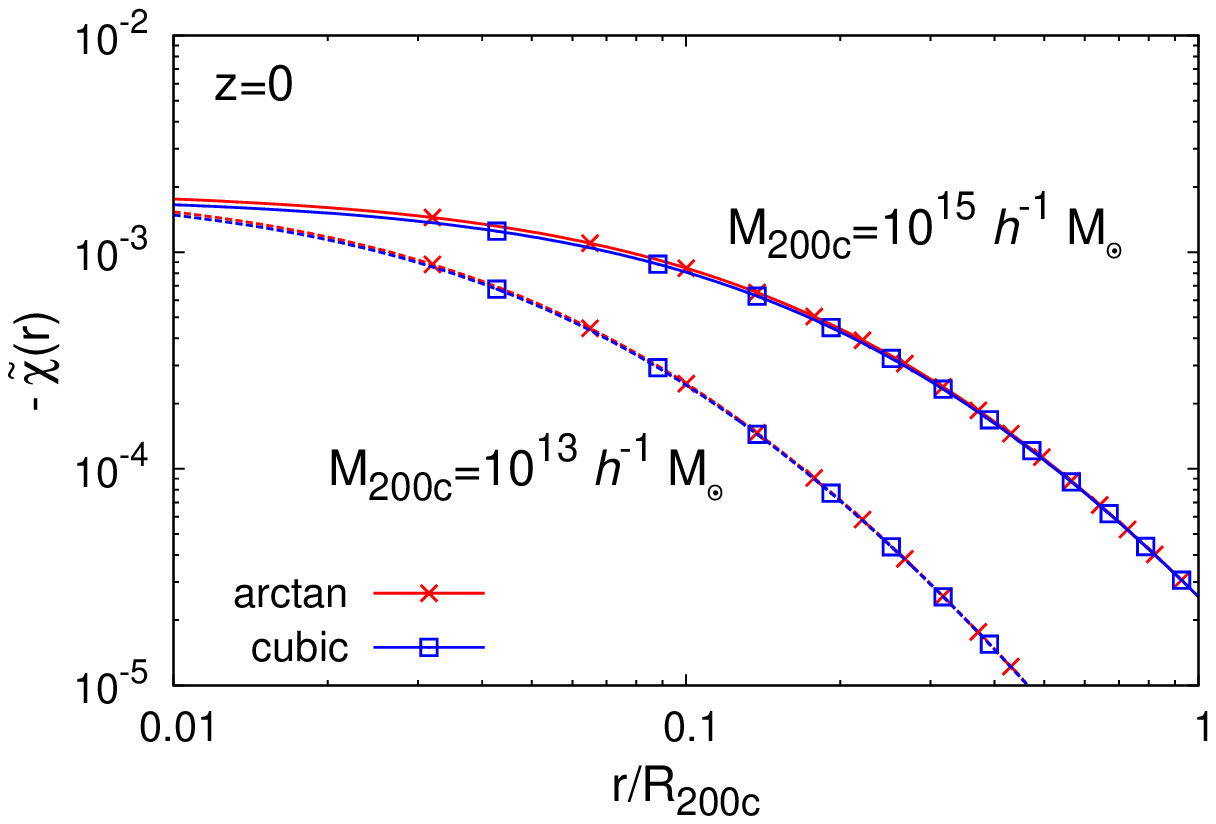}}
\end{center}
\caption{{\it Upper panel:} scalar-field radial profile, $\delta\varphi(r)=\varphi(r)-\bar\varphi$,
for halos of mass $10^{15} h^{-1} M_{\odot }$ (upper solid lines) and
$10^{13} h^{-1} M_{\odot }$ (lower dashed lines).
The scalar-field fluctuation is negative as the scalar field $\varphi$ is minimum at the center
of the halo.
{\it Lower panel:} radial profile of the ``kinetic energy'' $-\tilde{\chi}(r)$ of the scalar field,
for halos of mass $10^{15} h^{-1} M_{\odot }$ (upper solid lines) and
$10^{13} h^{-1} M_{\odot }$ (lower dashed lines).
Here $\tilde{\chi}<0$ because we consider the static limit, which is dominated by spatial
gradients.
The arctan and cubic K-mouflage models give almost identical results in this figure.}
\label{fig_chi_r_z0}
\end{figure}

As noticed in \cite{Brax:2014a,Brax:2014b}, clusters are unscreened, and the Klein-Gordon
equation (\ref{KG-small-scale-E}) can be kept at the linear level over the fluctuations of the
scalar field, as in Eq.(\ref{KG-Jordan-small-scale}). In this section we check that this property
extends far inside the cluster profile.

In the small-scale static limit, which corresponds for instance to high-density astrophysical
objects, the Klein-Gordon equation (\ref{KG-press-1}) becomes
\beq
\nabla_{\vr} \cdot (\nabla_{\vr} \varphi \; K' ) =
\frac{\beta\bar{A}}{c^2 M_{\rm Pl}} \rho ,
\label{KG-small-scale-J-static}
\eeq
where $\vr$ is the Jordan-frame physical coordinate and we assumed
$\delta\rho \simeq \rho$ (i.e., $\delta \gg 1$).
As compared with Eq.(\ref{KG-Jordan-small-scale}), here we do not make the approximation
$K' \simeq \bar{K}'$. Instead we consider the small-scale regime where
$\tilde{\chi} \simeq \delta\tilde{\chi}<0$.
For a spherically symmetric halo, using the Stokes theorem this gives
\beq
\frac{\dd\varphi}{\dd r} K' = \frac{\beta\bar{A} M(r)}{c^2M_{\rm Pl} 4\pi r^2} .
\label{KG-Jordan-static-spheric}
\eeq
As in \cite{Brax:2014a,Brax:2014b,Barreira2015} (but in Jordan-frame coordinates),
defining the ``K-mouflage screening radius'' $R_K$ by
\beq
R_K(M) = \left( \frac{\beta\bar{A}^2M}{4\pi c M_{\rm Pl}{\cal M}^2} \right)^{1/2} ,
\label{RK-def}
\eeq
where $M=M(R)$ is the total mass of the object of radius $R$, and introducing the
rescaled dimensionless variables $x=r/R_K$, $m(x)=M(<r)/M$, $\phi(x)=\varphi(r)/\varphi_K$,
with $\varphi_K={\cal M}^2R_K/c\bar{A}$, the integrated Klein-Gordon equation
(\ref{KG-Jordan-static-spheric}) reads as
\beq
\frac{\dd\phi}{\dd x} K' = \frac{m(x)}{x^2} , \;\;\; \mbox{with} \;\;\;
\tilde{\chi} = - \frac{1}{2} \left( \frac{\dd\phi}{\dd x} \right)^2 .
\label{KG-spherical-phi}
\eeq
As pointed out in \cite{Brax:2014a}, in the small-scale static regime we have
$\tilde{\chi}<0$ whereas the cosmological background value satisfies $\bar{\tilde\chi}>0$.
Using that ${\cal M}^4 \sim \bar{\rho}_{\rm de0}$ is roughly the dark-energy density today
and that $\bar{A} \sim 1$, we obtain
\beq
R_K(M) \simeq \sqrt{\frac{\beta M}{1 M_{\odot}}} 3470 \, {\rm a.u.} \simeq
\sqrt{\frac{\beta M}{10^{14} M_{\odot}}} 0.12 \, h^{-1} {\rm Mpc} .
\label{RK-num}
\eeq
The first equality shows that the Solar System is screened by the Sun, which allows these
K-mouflage scenarios to satisfy Solar System constraints \cite{Barreira2015}.
On the other hand, for $\beta=0.1$ the K-mouflage screening radius of a cluster of mass
$10^{14}M_{\odot}$ is $R_K \simeq 0.04 h^{-1}$Mpc.
This is much smaller than the radius of the cluster, which means that most of the cluster is
unscreened. Moreover, as we move inside the halo the enclosed mass $M(<r)$ decreases,
which further delays the onset of K-mouflage screening.
When $| \tilde{\chi} | \ll 1$ we have $K' \simeq 1$ and we obtain
\beq
| \tilde{\chi} | \ll 1 : \;\; \tilde{\chi}(r) \simeq - \frac{m(x)^2}{2 x^4} ,
\label{chi-cluster}
\eeq
which gives inside the halo
\beq
|\tilde{\chi}| \ll 1, \;\; r<R : \;\; \tilde{\chi}(r) \simeq - \frac{\rho_{\rm crit}}{6{\cal M}^4}
\left( \beta \bar{A}^2 \Delta_c(<r) \frac{H r}{c} \right)^{\!2} . \;\;
\label{chi-cluster-2}
\eeq
We show the radial profiles of $-\tilde{\chi}(r)$ in the lower panel in Fig.~\ref{fig_chi_r_z0},
for $M=10^{15} h^{-1} M_{\odot }$ and $10^{13} h^{-1} M_{\odot }$.
In both cases we can check that $|\tilde{\chi}| \ll 1$ over the full halo profile
[as seen from Eq.(\ref{chi-cluster}), $\tilde{\chi}(r)$ goes to a finite value at $r\rightarrow 0$
for NFW density profiles, because $\rho \propto 1/r$ and $m \propto x^2$ in the core].
This means that clusters are not screened
and that we can use the background value $\bar{K}'$ for the kinetic function.
In fact, at low $z$ we also have $\bar{\tilde\chi} \ll 1$ and $\bar{K}' \simeq 1$, so that
the kinetic function is dominated by the low-order terms in the expansion
(\ref{K-chi=0}) and the results are not very sensitive to the precise nonlinear form
of $K(\tilde\chi)$.
Then, the Klein-Gordon equation can be linearized in the scalar field as in
Eq.(\ref{KG-Jordan-small-scale}) and the gravitational potential $\Phi$ that governs the
dynamics of matter is again given by Eq.(\ref{Poisson-small-scale-Jordan-2}).

From the kinetic factor $\tilde\chi(r)$ we can obtain the radial profile of the scalar
field, $\varphi(r)$, by integrating $\dd\phi/\dd x=\sqrt{-2\tilde\chi}$ and
using $\varphi=\varphi_K \phi$. The boundary condition is $\varphi \rightarrow \bar\varphi$ at
infinity. We show the radial profile of the fluctuation $\delta\varphi=\varphi-\bar\varphi$
in the lower panel in Fig.~\ref{fig_chi_r_z0}.
We can check from the comparison with Fig.~\ref{fig_chi_varphi_z} that
$| \delta\varphi | \ll | \bar\varphi |$, in agreement with the scaling
$\delta\varphi/\bar\varphi \sim (aH/ck)^2 \delta\rho/\bar\rho$ obtained from
Eq.(\ref{dphi-estim-1}) and $\bar\varphi/\tilde{M}_{\rm Pl} \sim \beta/\bar{K}'$.
In particular, this explicitly shows that the coupling function $A(\varphi)$ remains dominated
by its low-order terms in the expansion (\ref{A-phi=0}), both for the background and
for large-scale structures such as clusters of galaxies.

The magnitude of $\delta\varphi$ can also be read from the modified Poisson equations
(\ref{Phi-Psi-Jordan}). In realistic models the fifth force should not have a magnitude greater
than the Newtonian force, which implies
$|\delta A/\bar{A}| \simeq |\beta\delta\varphi/\tilde{M}_{\rm Pl}| \lesssim |\Psi_{\rm N}|$,
whence $|\beta\delta\varphi/\tilde{M}_{\rm Pl}| \lesssim 10^{-5}$.

\subsection{Impact of nonlinear substructures}
\label{sec:substructures}

The Klein-Gordon equation (\ref{KG-small-scale-J-static}) that determines the
scalar field $\varphi$ is nonlinear, because of the factor $K'(\tilde\chi)$.
This means that substructures could have a strong impact on the solution $\varphi(\vr)$
as there is no longer a linear superposition property and the solution obtained for
the averaged halo profile is not identical to the average of the exact solutions obtained by
taking into account substructures.
In this section, we check that this nonlinearity does not play a significant role and
does not invalidate our approach described in Sec.~\ref{sec:unscreened}.

First, we note that for an object that is exactly spherically symmetric, the integrated
Klein-Gordon equation (\ref{KG-Jordan-static-spheric}) holds and the scalar-field profile
only depends on the integrated mass $M(r)$ within radius $r$. This smoothes out radial
substructures.
However, in practice clusters are not exactly spherically symmetric, and individual
cluster galaxies also break any overall spherical symmetry.
We have seen in Sec.~\ref{sec:unscreened} that clusters are unscreened as
the kinetic factor, $\tilde\chi_{\rm clus}(r)$, associated with the mean cluster density profile
(\ref{NFW-profile}), is much smaller than unity. Then, if galactic halos only form a small
fraction of the total cluster volume, throughout most of the cluster volume we can
linearize Eq.(\ref{KG-small-scale-J-static}), as in Eq.(\ref{KG-Jordan-small-scale}),
which gives
\beq
\mbox{unscreened region:} \;\;\;
\nabla_{\vr}^2 \varphi = \frac{\beta\bar{A}}{\bar{K}' c^2 M_{\rm Pl}} \rho ,
\label{KG-linear-cluster}
\eeq
where $\bar{K}'$ is the background value, with $\bar{K}' \simeq 1$ at low $z$.
This equation breaks down around each cluster galaxy, where the high matter
density, which is much greater than the NFW mean density $\rho_{\rm DM}(r)$
of Eq.(\ref{NFW-profile}) at that radius, makes the scalar field enter the nonlinear
screening regime.
Thus, around each galaxy ``i'', $i=1, .., N_{\rm gal}$, we must cut a patch $V_{Ki}$
where the equation (\ref{KG-linear-cluster}) must be replaced by the fully nonlinear
equation (\ref{KG-small-scale-J-static}).
By definition, the volume $V_{Ki}$ is given by the K-mouflage radius $R_{Ki}$ of the galaxy.
In practice, if $R_{Ki} \ll R_{\rm clus}$, we can build an approximate solution
by patching the solutions within each galaxy volume $V_{Ki}$ with the global solution
(\ref{KG-linear-cluster}) that holds in between galaxies. Around each galaxy,
using an approximate spherical symmetry around the galaxy center, we obtain
the local profile by solving Eq.(\ref{KG-Jordan-static-spheric}), where $M$ is replaced
by the galaxy mass $m_{\rm gal}(r)$, and the boundary condition at $R_{Ki}$ is
approximated as a constant obtained from the global solution (\ref{KG-linear-cluster}).

This scenario holds provided the regions $V_{Ki}$ do not extend far beyond the galaxy
volumes $V_i$ (defined for instance by their stellar content or by the region where matter
is gravitationally bound to the galaxy) and do not cover most of the cluster volume.
From Eq.(\ref{RK-def}) we have $R_K(m_{\rm gal}) \propto m_{\rm gal}^{1/2}$.
Defining the mass function $n(m) \dd m$ of the cluster galaxies, the total volume
built by the nonlinear regions $V_{Ki}$ reads as
\beq
V_{K\rm gal} = \int_0^{\infty} \dd m \, \frac{\dd n}{\dd m} V_K(m)
\propto \int_0^{\infty} \dd m \, \frac{\dd n}{\dd m} m^{3/2} .
\label{VK-gal-sum}
\eeq
The mass function of the cluster galaxies or of dark matter subhalos is typically
a power law at law mass with an exponential cutoff at high mass
\cite{Mo2010}.
In any case, the integral $\int \dd m (\dd n/\dd m) m = M_{\rm gal}$ is necessarily
finite and equal to the total mass associated with the galaxies, which is smaller than
the total cluster mass. Therefore, the integral (\ref{VK-gal-sum}) converges at low
mass and is dominated by the galaxies around the knee of the galaxy multiplicity function,
which typically corresponds to $M \sim 10^{12} M_{\odot}$
From Eq.(\ref{RK-num}) we have
$R_K(10^{12}M_{\odot}) \simeq 4 h^{-1}$kpc, with $\beta=0.1$,
which gives a volume fraction of the order of $(0.004)^3 \sim 5 \times 10^{-8}$.
Even if we have $\sim 20$ such galaxies in the cluster,
this only makes a fraction of order $10^{-6}$ of the cluster volume.

Moreover, we can see that $R_{Ki}$ is typically smaller than the actual radius $R_i$
of the galaxy (by a factor of a few). In the $\Lambda$-CDM cosmology itself, the
analysis of the hot gas that makes most of the intracluster medium and gives rise to the
cluster X-ray luminosity (based on hydrostatic equilibrium and scaling laws) only applies
outside of the cluster galaxies, where cooling and star formation processes play a major
role. Therefore, the nonlinearities of the
Klein-Gordon equation (\ref{KG-small-scale-J-static}) do not bring further restrictions
as compared with the standard $\Lambda$-CDM case, as they are ``hidden'' within the
galaxies, and the impact of the fifth force on the intracluster medium can be obtained
from the linearized equation (\ref{KG-linear-cluster}) associated with the
unscreened regime.
They do not modify global properties either, such as the cluster correlation function,
as the dynamics and formation of the clusters remain governed by the linearized
Klein-Gordon equation (\ref{KG-linear-cluster}).

\subsection{Hydrostatic equilibrium}
\label{sec:Hydrostatic equilibrium}

\begin{figure}
\begin{center}
\epsfxsize=8.5 cm \epsfysize=5.8 cm {\epsfbox{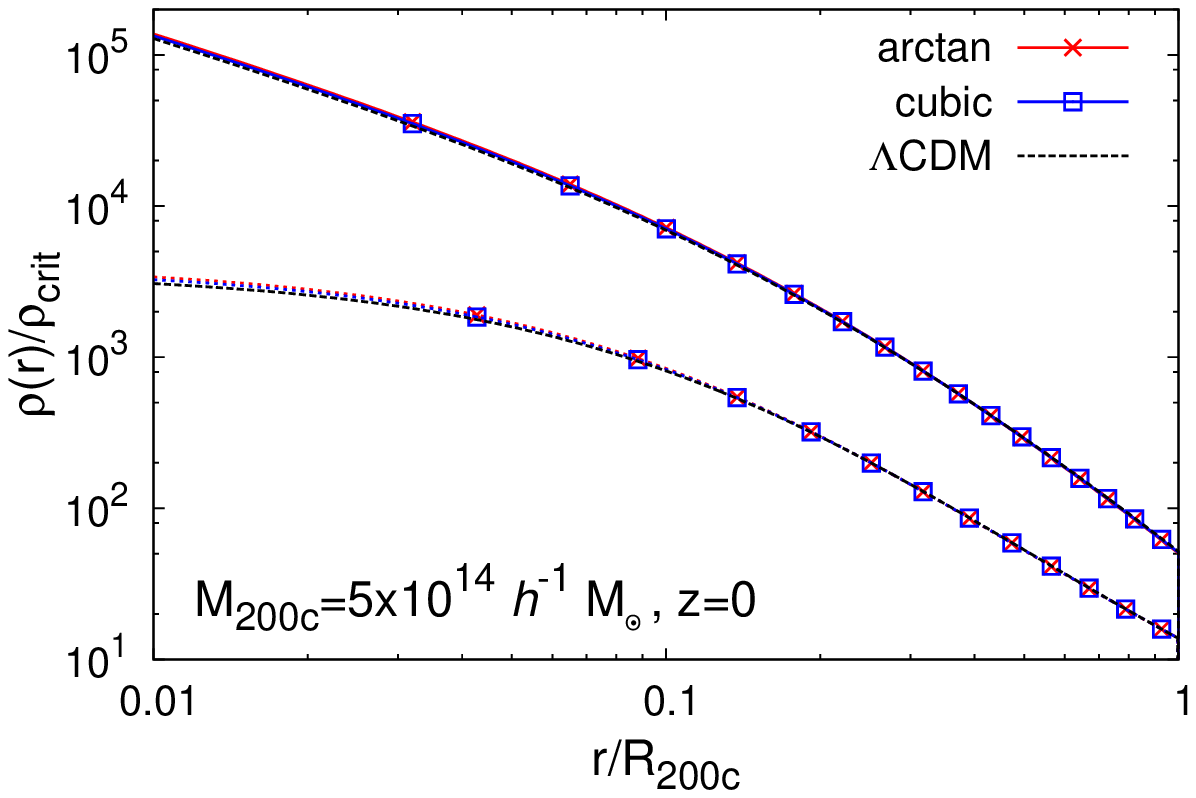}}\\
\epsfxsize=8.5 cm \epsfysize=5.8 cm {\epsfbox{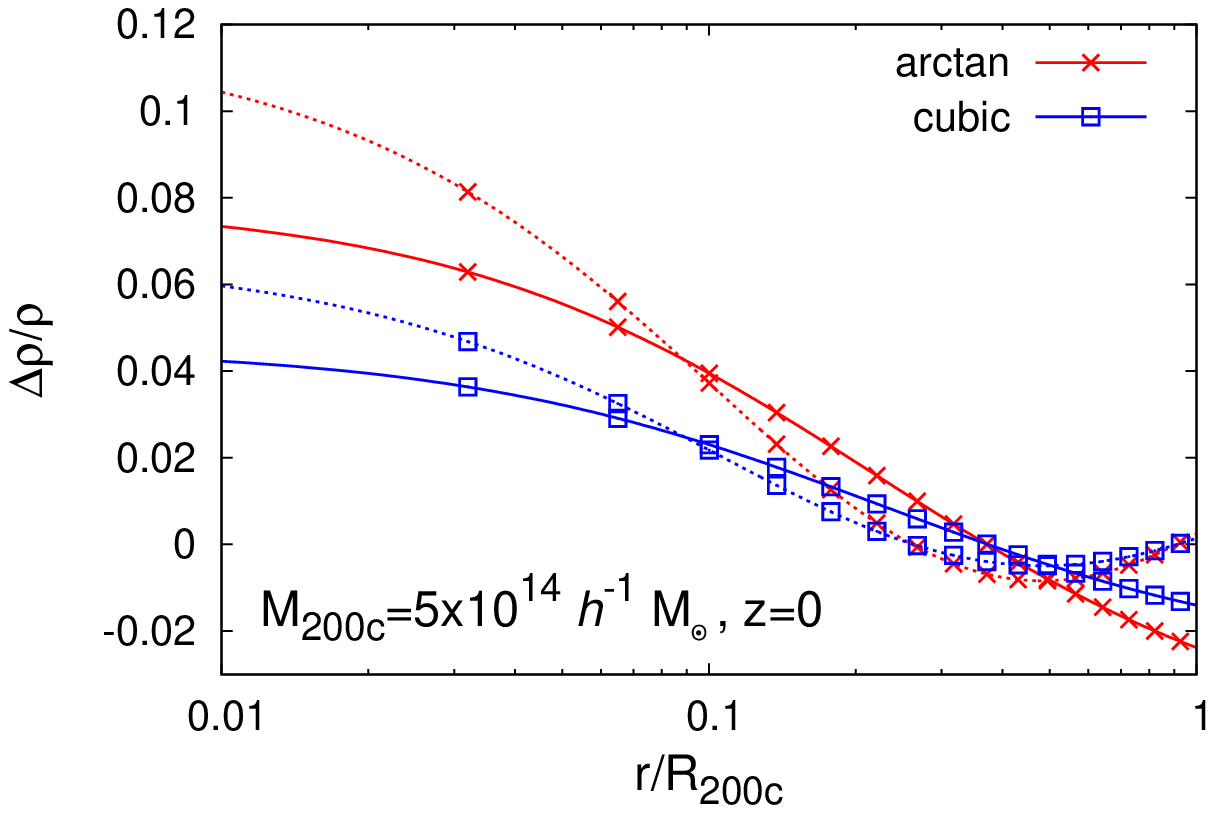}}
\end{center}
\caption{{\it Upper panel:} density profiles for a cluster of mass
$M_{200c}=5\times 10^{14} h^{-1} M_{\odot }$.
The upper solid lines refer to the dark matter density profiles and the lower dotted lines to the gas
density profiles. The K-mouflage models and the $\Lambda$-CDM reference cannot be distinguished
in this figure.
{\it Lower panel:} relative deviation from the $\Lambda$-CDM reference of the dark matter (solid
lines) and gas (dotted lines) density profiles.}
\label{fig_rhosca_z0}
\end{figure}

From equation (\ref{Euler-small-scale-Jordan}), the equation of hydrostatic equilibrium
for the gas density $\rho_g$ and pressure $p_g$ reads as
\beq
\nabla \Phi =  \nabla \left ( \Psi_{\rm N} + \frac{\beta c^2 \varphi}{\tilde{M}_{\rm Pl}}  \right ) =
 - \frac{\nabla p_g}{\rho_{g}} ,
\label{hydro-equ}
\eeq
where we used Eqs.(\ref{Phi-Psi-Jordan}) and (\ref{dA-dphi-drho}).
This explicitly shows how the pressure gradient is amplified, at fixed density profile,
by the fifth force. Assuming spherical symmetry this leads to
\beq
\frac{\dd p_g}{\dd r} = - \rho_{g} ( 1+\epsilon_1 ) \frac{\dd\Psi_{\rm N}}{\dd r} ,
\label{spherical-equ-rho}
\eeq
where we used the fact that clusters are unscreened, so that $K' \simeq \bar{K}'$
and the Klein-Gordon equation can again be linearized as in
Eq.(\ref{KG-linear-cluster}).
In the $\Lambda$-CDM cosmology we simply have $\epsilon_1=0$.
To obtain the gas profile from Eq.(\ref{spherical-equ-rho})
we also need an equation of state that gives the pressure as a function of the gas
density or temperature.
We consider an isothermal gas with $p_{g}=\rho_{g} k_B \bar{T}_{g}/(\mu m_{p})$,
where $k_B$ is Boltzmann's constant and $\mu m_p$ is the mean molecular weight of
the gas. This yields the gas density profile
\beq
\rho_{g}(r) \propto e^{-(1+\epsilon_1)\mu m_p \Psi_{\rm N}(r) /k_B \bar{T}_{g}} ,
\label{rhog-def}
\eeq
where the Newtonian potential $\Psi_{\rm N}$ is fixed by the dark matter profile.

To evaluate $\bar{T}_{g}$ we assume that the gas temperature is proportional to the mean
value of the dark matter ``temperature'', $T_{\rm DM}(r)$, which we define from
the velocity dispersion $\sigma_{\rm D}^{2}(r)$ as
\beq
k_B T_{\rm DM}(r) = \mu m_{p} \sigma_{\rm DM}^{2}(r) .
\label{Temperature-DM}
\eeq
The dark matter being collisionless it is not described by a thermodynamical
temperature. However, we can expect the virialization processes associated with
the formation of the halo to scale in the same fashion for the dark matter, as measured
by its velocity dispersion, and for the gas, as measured by its temperature.
In particular, the dark matter velocity dispersion obeys the Jeans equation,
which can be written at equilibrium as \cite{Binney1987}
\beq
\frac{1}{\rho_{\rm DM}} \frac{\dd [\rho_{\rm DM} \sigma_{\rm DM}^{2}]}{\dd r}
 = - \frac{\dd \Phi}{\dd r}  = - (1+\epsilon_1) \frac{\dd \Psi_{\rm N}}{\dd r} .
\label{Jeans-equation}
\eeq
For a given dark matter halo profile, set by the NFW profile (\ref{NFW-profile})
and concentration parameter $c(M)$, the Jeans equation (\ref{Jeans-equation})
determines the velocity dispersion profile $\sigma_{\rm DM}^2(r)$, whence
the effective dark matter temperature $T_{\rm DM}(r)$ defined in
Eq.(\ref{Temperature-DM}).
Then, we set the gas temperature $\bar{T}_g$ as
\beq
\bar{T}_{g} = \frac{1}{\beta_g} \bar{T}_{\rm DM} = \frac{1}{\beta_g} \;
\frac{\int \dd r \, 4\pi r^2 \rho_{\rm DM}(r) T_{\rm DM}(r)}
{\int \dd r \, 4\pi r^2 \rho_{\rm DM}(r)} ,
\label{T-gas}
\eeq
where $\beta_g$ is a free parameter that we fix to be equal to $0.6$, which is of the same
order as the values used in studies of clusters of galaxies \cite{Cavaliere1976}.
In other words, we assume that the kinetic and thermal energies of the dark matter
and the gas are proportional, because they are generated by the same process,
the formation and virialization of the halo.

Finally, to fully specify the gas density profile (\ref{rhog-def}) we normalize it as
\beq
M_{g} = \int_{0}^{R_{\Delta_c}} \dd r \, 4\pi r^{2} \rho_{g}(r) =
\frac{\Omega_{b}}{\Omega_{\rm DM}} M_{\rm DM} .
\label{rhogas-norma}
\eeq
Thus, we consider that the baryon and dark matter mass fractions in clusters
are given by the cosmological abundance.
This assumes that there is no significant redistribution and segregation of matter
on scales greater than cluster radii, which should be a reasonable approximation
for massive clusters.

Therefore, in terms of the intracluster medium, the differences between the K-mouflage and
$\Lambda$-CDM scenarios only arise through three effects, in our framework.
First, the dark matter profiles (\ref{NFW-profile}) are slightly different because of the small change
of the halo concentration shown in Fig.~\ref{fig_cM_z0}, which comes from the different growth rates
of large-scale structures.
Second, the equation of hydrostatic equilibrium (\ref{hydro-equ}) is modified by the factor
$\epsilon_1(t)$, which corresponds to the amplification of gravity by the fifth force in the
unscreened regime.
This implies slightly greater pressure gradients for the gas.
Third, the gas (and dark matter) temperature itself is also amplified by the same factor
$(1+\epsilon_1)$, at a fixed dark matter profile, because it also arises from the gravitational
collapse, see Eqs.(\ref{Jeans-equation}) and (\ref{T-gas}).
The second and third effects compensate in terms of the gas density profile, as the greater
potential depth is balanced by the greater gas temperature,
so that we can expect rather modest deviations between the different cosmological scenarios.

In Fig.~\ref{fig_rhosca_z0}, we show the dark matter and gas density profiles for
a cluster of mass $M=5\times 10^{14} h^{-1} M_{\odot }$ at $z=0$, for the K-mouflage models
and the $\Lambda$-CDM reference.
The presence of the scalar field makes the density profiles more compact, in agreement with
Fig.~\ref{fig_cM_z0}. As expected, the deviations from the $\Lambda$-CDM reference are of
the order of a few percent.

\subsection{Scaling laws}
\label{sec:Scaling laws}

From the gas density profile $\rho_g(r)$ and temperature $\bar{T}_{g}$, we
obtain the X-ray cluster luminosity within radius $R$ as \cite{Plionis2008}
\beq
L_{X}(<R) = 4\pi \epsilon_{X}(\bar{T}_g) \int_{0}^{R} \!\! n_{g}^{2}(r) r^{2} \dd r ,
\label{X-ray-luminosity}
\eeq
where $n_{g}(r)=\rho_g(r)/\mu m_{p}$ is the  cluster gas density and $\epsilon_{X}(T)$ is
the X-ray emissivity, which can be expressed in terms of the temperature as \cite{Mo2010}
\beq
\epsilon_{X}(T) = 4.836 \times 10^{-24} \frac{4-2Y}{(4-Y)^2} \left( \frac{T}{1 \rm keV}\right)^{1/2}
 \, \, \textrm{erg.s$^{-1}$.cm$^{3}$}
\label{X-ray-emissivity}
\eeq
Here $Y$ is the Helium mass fraction, and $\mu=2/(4-Y)$, $n_e/n_g=(2-Y)/(4-Y)$,
$(n_H+4n_{He})/n_g=2/(4-Y)$, where we assume complete ionization.
This applies to high temperatures of order 1 keV and above, where the X-ray emissivity
is dominated by Bremsstrahlung.
Equation (\ref{X-ray-luminosity}), with the emissivity (\ref{X-ray-emissivity}), gives
the total bolometric luminosity. In practice, one measures the radiation from
X-ray clusters within finite frequency bands.
Therefore, we also define the luminosity within frequency bands,
denoted for instance as bands ``$A=[\nu_1^A,\nu_2^A]$'', ``$B=[\nu_1^B,\nu_2^B]$'', ..,
by
\beq
L_{XA}(<R)= L_{X}(<R) \left ( e^{-h\nu_{1}^A/k_B\bar{T}_g} - e^{-h\nu_{2}^A /k_B\bar{T}_g} \right ) .
\label{band-X-ray-luminosity-function}
\eeq

\begin{figure}
\begin{center}
\epsfxsize=8.5 cm \epsfysize=5.8 cm {\epsfbox{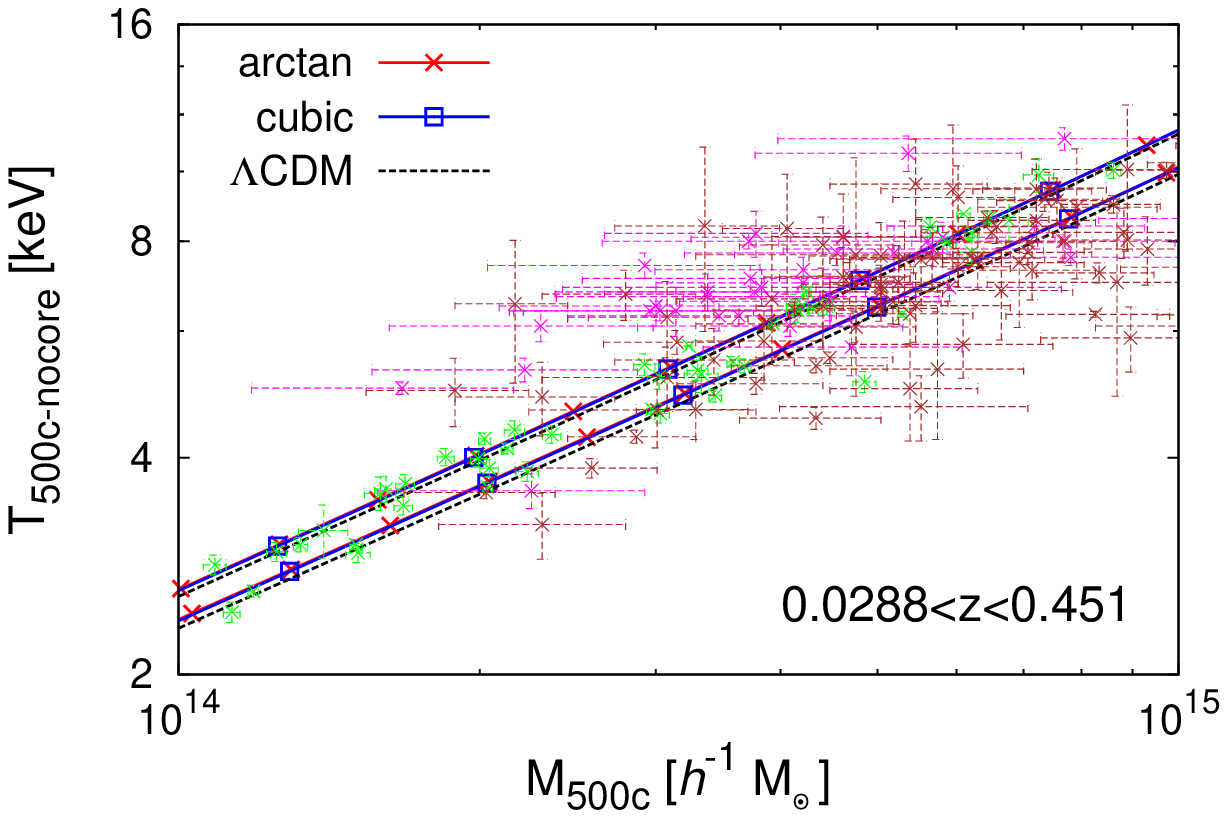}}\\
\epsfxsize=8.5 cm \epsfysize=5.8 cm {\epsfbox{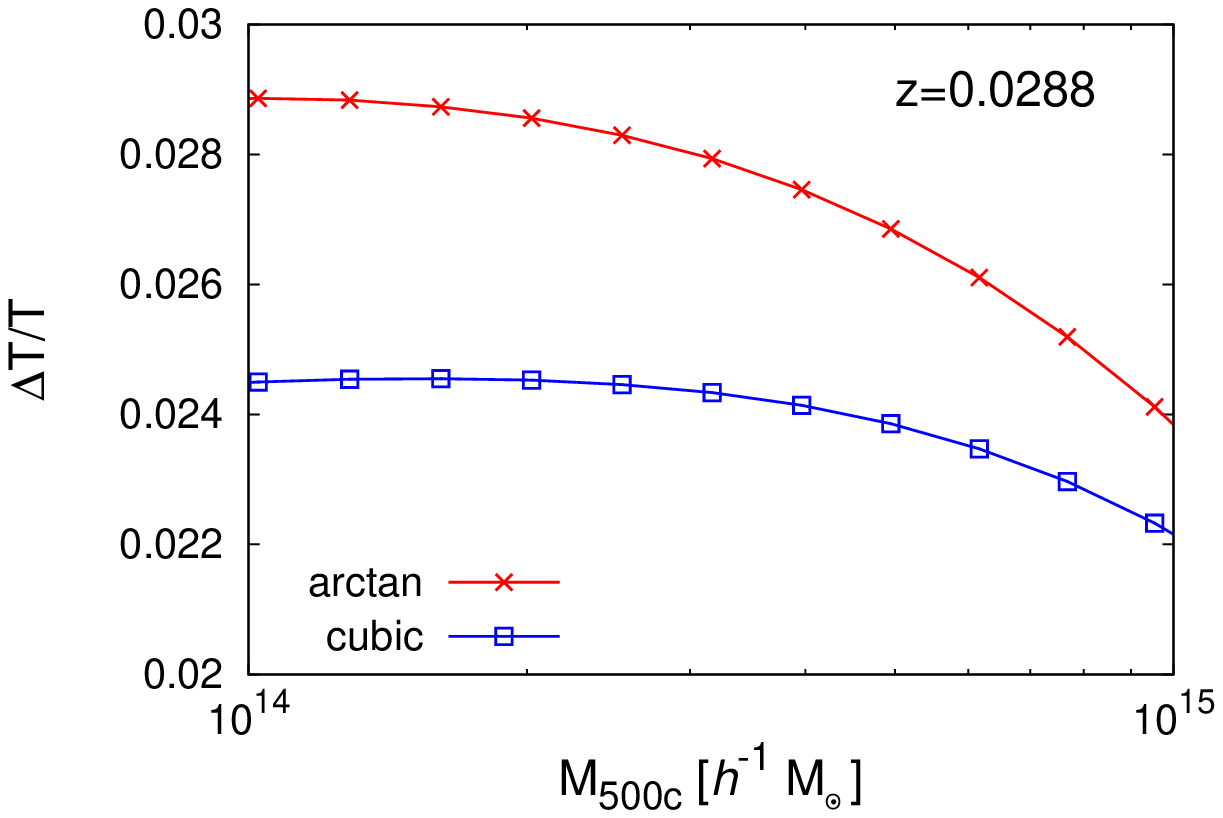}}
\end{center}
\caption{{\it Upper panel:} mass-temperature relation for the K-mouflage models and the
$\Lambda$-CDM reference, at $z=0.0288$ (lower curves) and $z=0.451$ (upper curves).
The data points are taken from observations made by
\cite{Vikhlinin2009b}  (in green), \cite{Zhang2008} (in magenta) and \cite{Mantz2010} (in brown),
with clusters in the redshift range $0.0288\leq z \leq 0.451$.
{\it Lower panel:} relative deviation of the cluster mass-temperature relation from the
$\Lambda$-CDM reference, at $z=0.0288$.}
\label{fig_M-T}
\end{figure}

\begin{figure}
\begin{center}
\epsfxsize=8.5 cm \epsfysize=5.8 cm {\epsfbox{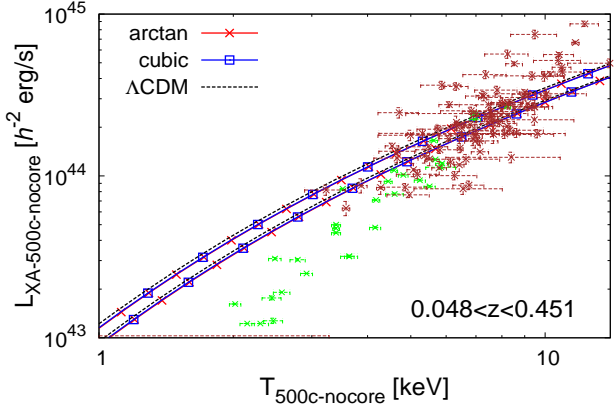}}\\
\epsfxsize=8.5 cm \epsfysize=5.8 cm {\epsfbox{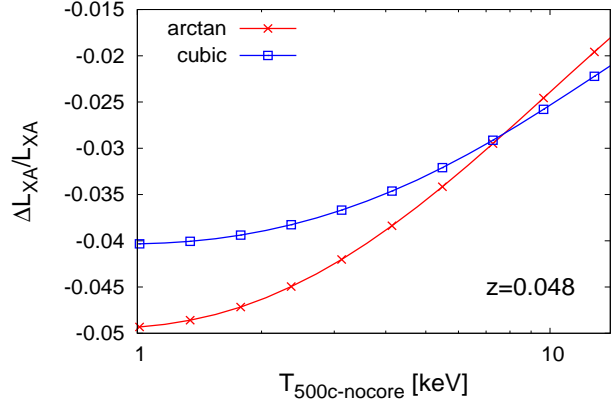}}
\end{center}
\caption{{\it Upper panel:} temperature-luminosity relation for the K-mouflage models and the
$\Lambda$-CDM reference, at $z=0.048$ (lower curves) and $z=0.451$ (upper curves).
The data points are taken from observations made by \cite{Pratt2009} (in green)
and \cite{Mantz2010} (in brown), with clusters in the redshift range $0.048\leq z \leq 0.451$.
{\it Lower panel:} relative deviation of the cluster temperature-luminosity relation from the
$\Lambda$-CDM reference, at $z=0.048$.}
\label{fig_LXA-T}
\end{figure}

Observational studies often measure the X-ray properties of galaxy clusters within a radius $R_X$
that is smaller than $R_{200c}$, because the luminosity scales as the squared density
[see Eq.(\ref{X-ray-luminosity})] so that inner high-density regions are easier to measure.
A popular choice is the radius set by the density threshold $\Delta_c=500$ with respect to the
critical density.
In the following, keeping our definition of halos by the threshold $\Delta_c=200$ as in
Figs.~\ref{fig_dnM_z0} and \ref{fig_cM_z0} - \ref{fig_rhosca_z0}, we use the density profile
obtained from Eq.(\ref{rhog-def}) and displayed in Fig.~\ref{fig_rhosca_z0} to compute
X-ray properties within $R_X$ defined by $\Delta_c=500$ (hence $R_X < R_{\rm halo}$).

To avoid the complications due to the internal structures of the clusters
(presence of massive galaxies in the center, importance of dissipative processes,
cooling cores,...) and also to follow the observational procedures, we define a core
radius $R_{\rm core}$ outside of which we evaluate the quantities of interest.
As in many observational analyses, we simply define $R_{\rm core}$ as a fixed fraction
of the cluster radius $R_X$ (as defined by the threshold $\Delta_c=500$ with respect to the
critical density), with $R_{\rm core} = f_{\rm core} R_X$ and typically $f_{\rm core} \sim 0.15$.
Then, we obtain for instance the luminosity in the outer cluster shells, between
$R_{\rm core} < r < R_X$, as
\beq
L_{XA \; \rm no-core} = L_{XA}(<R_X) - L_{XA}(<R_{\rm core}) .
\label{nocore-LXA}
\eeq

In Figs.~ \ref{fig_M-T} and \ref{fig_LXA-T}, we show, respectively, the
$M_{500c}-T_{\rm 500c-nocore}$ and $ T_{\rm 500c-nocore} - L_{XA-\rm500c-nocore}$ relations
compared to observations of clusters of galaxies in the X-ray, with the choice of
parameter $f_{\rm core}=0.15$ and the frequency ``A band'' [0.1 - 2.4] keV.
For the $M-T$ relation we obtain a good agreement with observations while our prediction for the
slope of the $T-L$ relation is too shallow. This is a well-known problem associated with a noticeable
breakdown of the naive ``scaling laws'' for the X-ray luminosity, especially for small clusters
\cite{Plionis2008}.
This is usually explained by a decrease of the gas fraction and a greater importance of nonthermal
effects, or departures from hydrostatic equilibrium, in small clusters.
However, because our goal is only to estimate the magnitude of the effects due to modifications of
gravity, we do not try to build a more accurate and more complex model in this paper. Moreover, our
simple model is sufficient to recover the typical X-ray luminosity in the range $4<T<15$ keV, which
corresponds to massive bright clusters.

At fixed mass, the temperature in the K-mouflage scenarios is greater than in the
$\Lambda$-CDM reference by about $2\%$. This is mostly set by the factor $\epsilon_1$,
which is about $2\%$ as seen in Fig.~\ref{fig_eps_z}.
Indeed, from Eq.(\ref{Jeans-equation}) we can see that at a fixed dark matter density profile
the fifth-force enhancement of gravity by the factor $(1+\epsilon_1)$ yields an increase
of the dark matter velocity dispersion and of the gas temperature by the same factor.
The small deviations from this $2\%$ value, which depend on mass, that appear
in Fig.~\ref{fig_M-T} correspond to the small changes of the dark matter profile through the
modification of the concentration parameter shown in Fig.~\ref{fig_cM_z0}.

At fixed temperature, the K-mouflage models give a slightly lower X-ray luminosity.
This is because at fixed mass K-mouflage models give a higher temperature, as seen
in Fig.~\ref{fig_M-T}. Therefore they give a lower mass at fixed temperature.
Since the X-ray luminosity scales as $L_X \sim \rho_s M \sqrt{T}$, a lower mass
implies a lower luminosity (disregarding the impact on $\rho_s$).
As expected, we find percent deviations as for the $M-T$ scaling law.

Thus, as for the quantities studied in previous sections, we obtain percent deviations from the
$\Lambda$-CDM scaling laws.
Unfortunately, this is probably too small to be used as a meaningful constraint on these
modified-gravity scenarios, in view of the observational and theoretical uncertainties.
Therefore, it is unlikely that cluster scaling laws can provide competitive constraints on such 
modified-gravity models, that must also pass very tight Solar-System bounds and satisfy larger-scale
cosmological constraints associated with the growth of large-scale structures or the evolution of the
Hubble expansion rate (e.g., constraints from BBN).

\subsection{Cluster Temperature function}
\label{sec:Luminosity function}

\begin{figure}
\begin{center}
\epsfxsize=8.5 cm \epsfysize=5.8 cm {\epsfbox{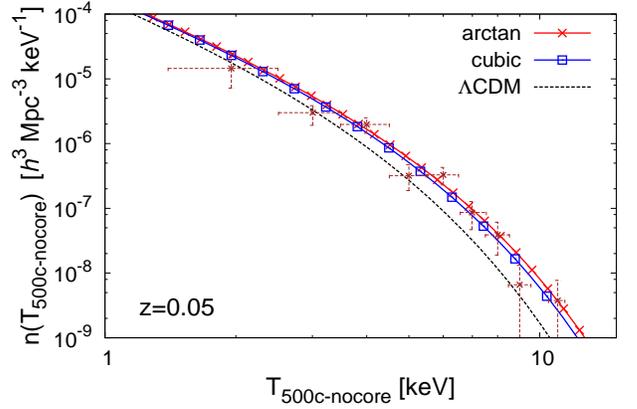}}\\
\epsfxsize=8.5 cm \epsfysize=5.8 cm {\epsfbox{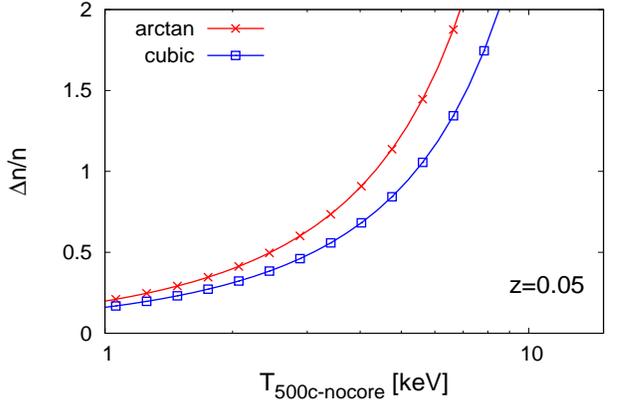}}
\end{center}
\caption{{\it Upper panel:} cluster temperature function for the K-mouflage models and the $\Lambda$-CDM
reference, at $z=0.05$. The data points are taken from observations made by \cite{Ikebe2002} from
a sample of clusters with $z \simeq 0.05$.
{\it Lower panel:} relative deviation of the cluster temperature function from the $\Lambda$-CDM reference
at $z=0.05$}
\label{fig_nMT}
\end{figure}

Neglecting the scatter of the mass-temperature relation, by combining the halo mass
function described in Sec.~\ref{sec:Numerical-mass-function} with the
mass-temperature relation obtained in Sec.~\ref{sec:Hydrostatic equilibrium}
and Fig.~\ref{fig_M-T}, we obtain the X-ray cluster temperature function
\beq
n(T) = n(M) \frac{\dd\ln{M}}{\dd\ln{T}} .
\label{cluster-temperature-function}
\eeq
In Fig.~\ref{fig_nMT} we show the temperature functions computed for the K-mouflage
models together with the $\Lambda$-CDM case,
evaluated at $z=0.05$ to compare them with the observations obtained by \cite{Ikebe2002}.

We obtain a reasonable agreement with observations. As is well known, this also means that the cluster
temperature is a rather robust quantity (as compared for instance with the X-ray luminosity) and that it is
not necessary to build very sophisticated models to recover the right order of magnitude.
As shown in the lower panel, we now obtain deviations for the cluster number counts that are of order
unity: the K-mouflage models can predict twice or three times more high-$T$ clusters than the
$\Lambda$-CDM reference.
As we have seen, this difference is not due to deviations in the cluster scaling laws, that is, in the
intracluster medium, which are quite small, but to the amplification of the high-mass tail of the halo
mass function already shown in Fig.~\ref{fig_dnM_z0}.
Therefore, this result should be rather robust as it is directly related to the faster growth of large-scale
structures in the K-mouflage scenarios.

\subsection{Sunyaev-Zel'dovich effect}
\label{sec:SZE}

\begin{figure}
\begin{center}
\epsfxsize=8.5 cm \epsfysize=5.8 cm {\epsfbox{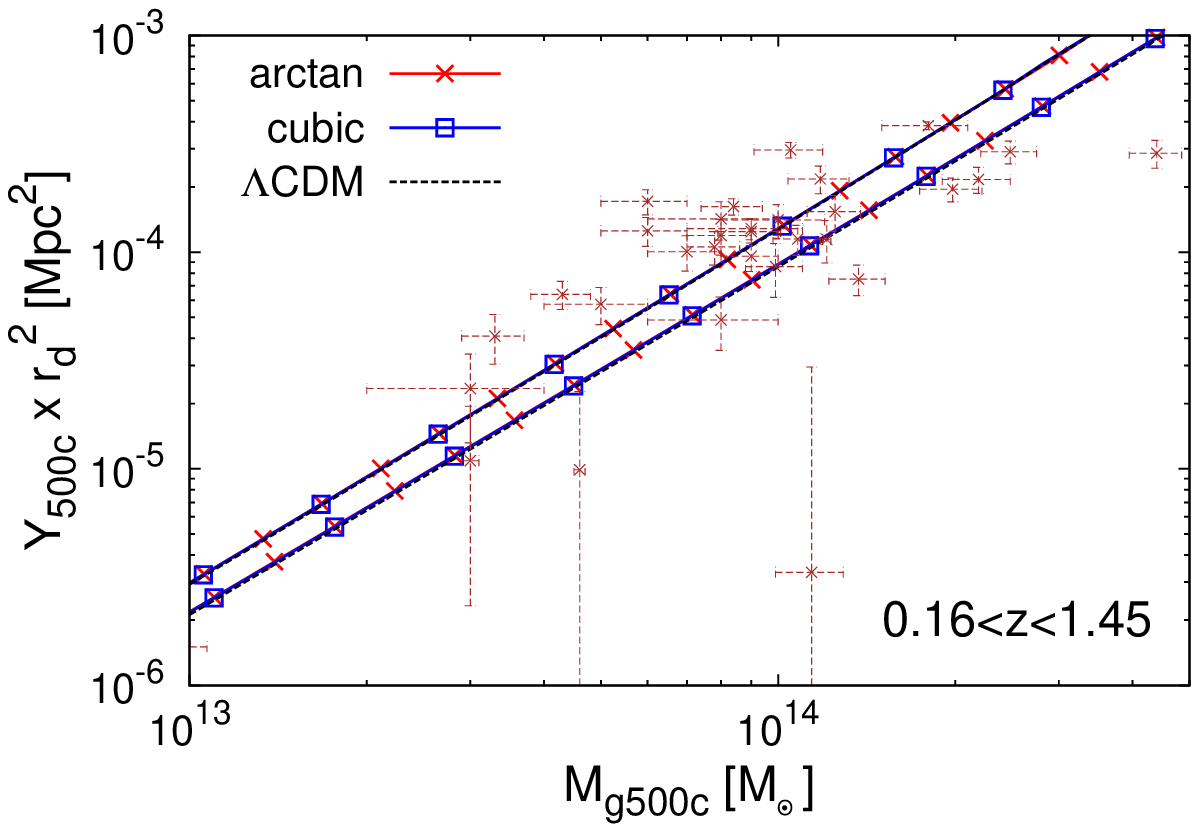}}\\
\epsfxsize=8.5 cm \epsfysize=5.8 cm {\epsfbox{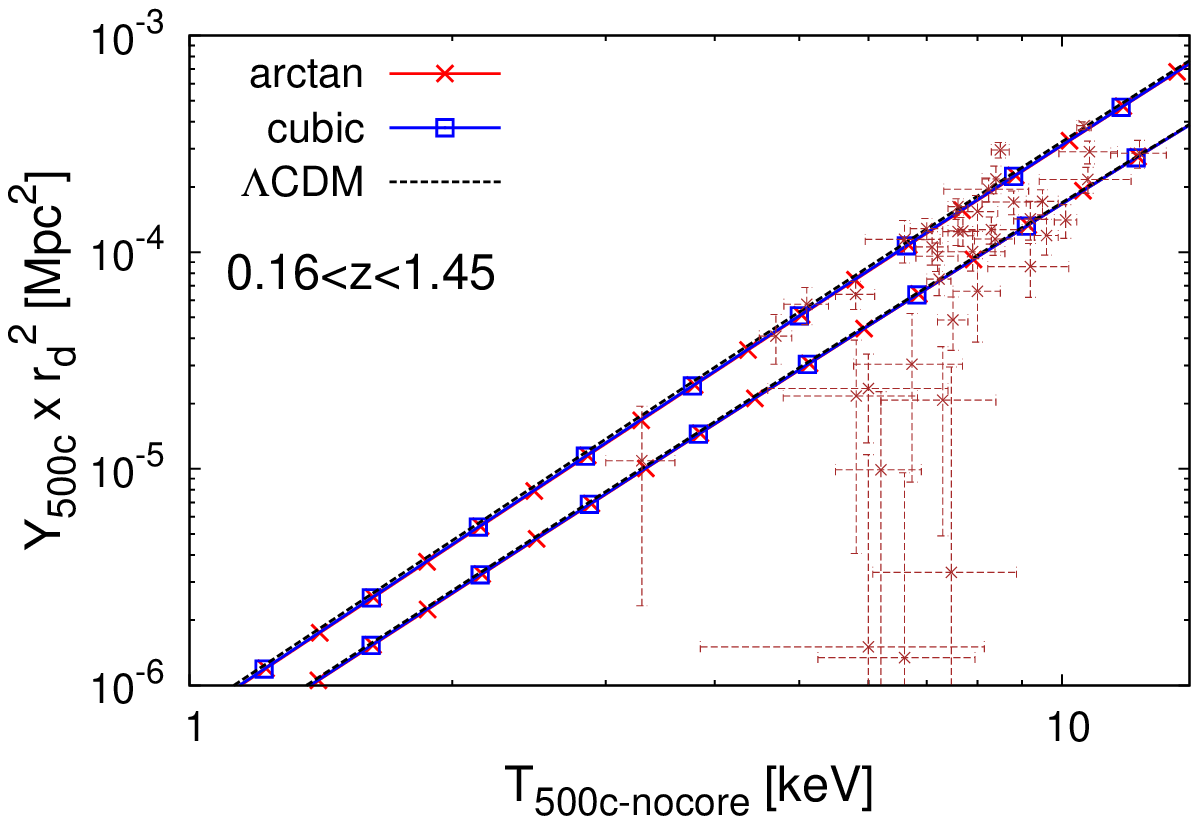}}
\end{center}
\caption{Integrated Comptonization within $R_{500c}$ as a function of the gas mass
(upper panel) and gas temperature (lower panel) for the K-mouflage models and the
$\Lambda$-CDM reference.
These different models cannot be distinguished in these figures.
We show our results for $z=0.16$ (lower curves in the upper panel and upper curves in the
lower panel) and $z=1.45$ (upper curves in the upper panel and lower curves in the lower panel).
The data points are measures from a sample of clusters in the range $0.16\leq z \leq 1.45$
\cite{Bender2014}.}
\label{fig_Y500M500}
\end{figure}

An indirect method to infer the properties of the clusters is to use the Sunyaev-Zel'dovich
effect (hereafter SZE) \cite{Sunyaev1972}. It occurs when photons from the CMB inverse
Thompson scatter in the intracluster medium. The measured CMB temperature is
then distorted with an amplitude proportional to the so-called Compton parameter
(see e.g. \cite{Bender2014}),
\beq
y  = \int n_{e} \sigma_{T} \frac{k_B T_{g}}{m_{e}c^2} \dd l ,
\label{Compton-parameter}
\eeq
where $n_{e}$ is the electron number density, $T_{g}$ is the gas temperature,
$m_{e}$ is the electron mass, $\sigma_{T}=6.65 \times 10^{-25}$ cm$^2$ is the
Thompson cross-section and $\dd l$ denotes the integration along the line of sight.
Following a common observational practice, by integrating over the angular area
of the cluster, defined for instance by the radius $R_{500c}$ associated with the
density contrast of $500$ with respect to the critical density, we define the
integral Compton parameter
\beq
Y_{500c} \equiv \int y \dd\Omega = r_{d}^{-2}(z) \int_{0}^{R_{500c}} 4 \pi r^2
n_{e}(r) \sigma_{T} \frac{k_B T_{g}}{m_{e}c^2} \dd r ,
\label{Compton-parameter-integrated}
\eeq
where $r_{d}(z)$ is the angular distance of the cluster located at redshift $z$.

In Fig.\ref{fig_Y500M500} we show the relations $M_{g\rm 500c} - Y_{\rm 500c} r_{d}^{2}(z)$
and $T_{\rm 500c-nocore}-Y_{\rm 500c} r_d^2(z)$ for the K-mouflage models and the
$\Lambda$-CDM reference, and we compare them to the observations made by
\cite{Bender2014}.
Again, we obtain a reasonable agreement with observations and a small deviation between the
different scenarios. The agreement is better for the $M_g - Y$ relation than for the $T - Y$
relation, but the latter shows a very large scatter and is probably contaminated by large
observational errors.

In any case, as in Sec.~\ref{sec:Scaling laws}, it appears that deviations of cluster scaling laws
associated with modified-gravity scenarios are too small as compared with observational error
bars and theoretical uncertainties to be competitive.
However, number counts, whether in terms of the cluster temperature or SZ parameter $Y$,
could provide useful constraints.

\subsection{Dynamical and weak lensing masses}
\label{sec:dyn-wl}

Finally, we briefly comment on the dynamical and weak lensing masses of clusters
and massive halos. In the unscreened regime, which applies to clusters and larger
scales, the dynamics of matter particles (dark matter and the gas) is governed by
the metric potential $\Phi$, as in Eqs.(\ref{Euler-small-scale-Jordan}) and
(\ref{trajectory-Jordan}).
This gravitational potential is related to the matter density through the modified
Poisson equation (\ref{Poisson-small-scale-Jordan-2}) and it is equal to the
standard Newtonian potential (but with a time-dependent Newton constant)
multiplied by the factor $1+\epsilon_1$. From observations of the dynamics
in clusters one would then measure the dynamical mass
\beq
M_{\rm dyn} \equiv [1+\epsilon_1(t)] \frac{{\cal G}(t)}{{\cal G}_0} M ,
\label{M-dyn}
\eeq
assuming General Relativity (GR) gravity with today's Newton's constant.
On the other hand, the weak lensing potential $\Phi_{\rm wl}$ that governs the
deflection of light rays by the perturbations of the metric is
\beq
\Phi_{\rm wl} = \frac{\Phi+\Psi}{2} = \Psi_{\rm N} ,
\label{Phi-wl}
\eeq
where we used Eq.(\ref{Phi-Psi-Jordan}).
Therefore, weak lensing observations of clusters would give the weak lensing mass
\beq
M_{\rm lens} \equiv \frac{{\cal G}(t)}{{\cal G}_0} M ,
\label{M-lens}
\eeq
and the ratio between the dynamical and weak lensing masses reads as
\beq
\frac{M_{\rm dyn}}{M_{\rm lens}} = 1 + \epsilon_1(t) .
\label{ratio-Mdyn-Mlens}
\eeq
We have shown the factor $\epsilon_1$ in Fig.~\ref{fig_eps_z}. Thus, we find that
the dynamical mass is greater than the lensing mass by about $2\%$.
As explained above, this is set by the value of $2\beta^2$, which is constrained to be
of order $2\%$ or below because of cosmological and Solar System constraints
\cite{Barreira2015}
(BBN constraint and bound on the time dependence of $\cal G$).
Therefore, this result on the ratio $M_{\rm dyn}/M_{\rm lens}$
gives an upper bound for its deviation from GR for all
realistic K-mouflage models.

\subsection{Cluster correlation function}
\label{sec:correlation}

\begin{figure}
\begin{center}
\epsfxsize=8.5 cm \epsfysize=5.8 cm {\epsfbox{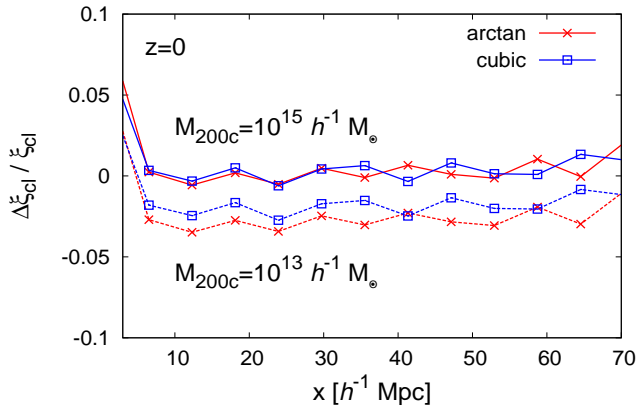}}
\end{center}
\caption{Relative deviation of the cluster correlation function from the $\Lambda$-CDM
reference for the K-mouflage models. We consider halos of mass
$10^{15} h^{-1} M_{\odot }$ (upper solid lines) and $10^{13} h^{-1} M_{\odot }$
(lower dashed lines).}
\label{fig_dxi_M_z0}
\end{figure}

In the previous sections we considered the internal and integrated properties of clusters,
as well as their abundance. Another probe of cosmology is provided by the cluster
correlation function. Following \cite{Cole1989,Mo1996}, we write the halo bias as
\beq
b(M) = 1 + \frac{\nu^2-1}{\delta_L} .
\label{bias-M}
\eeq
More accurate fitting formulas have been proposed for $\Lambda$-CDM cosmologies
\cite{Sheth2001d}, but they involve free parameters that might vary for different
modified-gravity scenarios. Moreover, numerical simulations find that the spherical
collapse model (\ref{bias-M}) provides reasonably good predictions that can fare
better than more sophisticated models for rare and massive halos
\cite{Tinker2010}, which we focus on here.
Therefore, Eq.(\ref{bias-M}) should be sufficient for our purposes and provide a simple
estimate of the impact of K-mouflage models. Note that because clusters are not
screened, the reasoning that leads to Eq.(\ref{bias-M}) in the $\Lambda$-CDM cosmology
remains valid for K-mouflage scenarios, as the only change is the time-dependent effective
Newton's constant as it would be defined from Eq.(\ref{MPlanck-eff-Poisson}).
This enters the bias (\ref{bias-M}) through the different values of $\nu(M)$ and
$\delta_L$, that we compute from the spherical collapse described in
Secs.~\ref{sec:Spherical-collapse-Jordan} and \ref{sec:Numerical-mass-function}.

Combining the halo bias (\ref{bias-M}) with the matter correlation function
$\xi(x)$ shown in Fig.~\ref{fig_dPk_z0}, we obtain the cluster correlation function
$\xi_{\rm cl}(x)$ displayed in Fig.~\ref{fig_dxi_M_z0}.
The comparison with Fig.~\ref{fig_dPk_z0} shows that the cluster correlation function
is much less affected by K-mouflage than the matter correlation and it can actually
be slightly lower than in the reference $\Lambda$-CDM cosmology.
This is because the amplification of gravity, associated with the greater
effective Newton's constant $(1+\epsilon_1(t)){\cal G}(t)/{\cal G}_0$, merely accelerates
the growth of large-scale structures. This amplifies the matter density power spectrum
and correlation function, as well as the large-mass tail of the halo mass function,
as seen in Figs.~\ref{fig_dPk_z0} and \ref{fig_dnM_z0}.
However, this same phenomenon also implies that, at a fixed mass $M$, massive halos
are less rare and have a smaller bias $b(M)$ [in particular, $\nu(M)$ becomes
smaller in Eq.(\ref{bias-M})].
This effect almost cancels the increase of the underlying matter density correlation
function.
Therefore, it appears that the matter correlation function, measured for instance
from weak lensing observations or galaxy surveys (using typical halos with a bias
of order unity that is not significantly changed by K-mouflage), is a better probe
of such modified-gravity scenarios than the cluster correlation function
(or more generally, the correlation of rare objects).

\section{Comparison with other modified-gravity models}
\label{sec:Comparison}

\subsection{Some other modified-gravity theories}
\label{sec:other-gravity-theory}

Before we conclude this study of K-mouflage models, it is interesting to compare
our results with other modified-gravity models that have been investigated in the
literature.
The main scenarios that have led to detailed analytical and numerical studies
are the $f(R)$ theories, scalar-field models such as Dilaton and Symmetron models,
and Galileon models \cite{Brax2012b,Khoury:2013tda,Brax:2013ida}.

The $f(R)$ models
\cite{Khoury:2003rn,Hu2007a,Brax:2008hh,Sotiriou:2008rp,Felice2010,Gannouji:2012iy},
can be recast as scalar-field models
with an Einstein-frame action of the form (\ref{S-def}), with a standard
kinetic term, an exponential coupling function $A(\varphi)$, and a scalar-field
potential $V(\varphi)$ \cite{Chiba2003}.
The Dilaton \cite{Gasperini:2001pc,Brax:2010gi,Brax:2011ja}
and Symmetron \cite{Hinterbichler:2010es,Brax:2012nk}
models are also scalar-tensor theories of this form, but with different coupling functions
$A(\varphi)$ and potentials $V(\varphi)$ (and standard kinetic terms).
Finally, Galileon models
\cite{Nicolis:2008in,Deffayet2009,Deffayet2009a,De-Felice2010a}
also involve a scalar field, with a non-standard kinetic
term (the scalar-field Lagrangian contains higher-order terms in
$\nabla\varphi$ and $\nabla^2\varphi$), but there is no coupling function
$A(\varphi)$ (i.e., $g_{\mu\nu}=\tilde{g}_{\mu\nu}$).
(Of course, it is possible to build more complex models that combine
these various ingredients.)

These different scenarios show different nonlinear screening mechanisms
that ensure convergence to GR in the Solar System,
Chameleon \cite{Khoury:2003aq} (for $f(R)$ models), Damour-Polyakov
\cite{Damour:1994zq} (for Dilaton and Symmetron models) and
Vainshtein \cite{Vainshtein:1972sx} (for Galileon models) mechanisms.
The theories that are closest to K-mouflage scenarios are the Galileon models,
as their screening mechanism also relies on the nonlinear derivative terms of the
scalar-field Lagrangian; but they also involve $\nabla^2\varphi$ instead of $\nabla\varphi$
only, which gives rise to different scaling exponents, for instance for the Vainshtein
and K-mouflage screening radii as a function of the mass $M$ of compact objects.

\subsection{Einstein and Jordan frames}
\label{sec:Einstein-Jordan}

A first important difference between the K-mouflage scenario and these
other modified-gravity models is the distinction between the Einstein and Jordan
frames. As recalled above, this distinction does not apply to the (simplest)
Galileon models, but the $f(R)$, Dilaton and Symmetron models also naturally
give rise to distinct Einstein and Jordan frames.
However, it turns out that in these scenarios the coupling function is constrained
to remain very close to unity.
Thus, $|A-1| \lesssim 10^{-4}$ for $f(R)$ models, because the ``mass'' of the scalar
field must be sufficiently large, $m \gtrsim 10^3 H/c$, to ensure an efficient
screening of the fifth force by a Chameleon mechanism in the Solar System.
For Dilaton and Symmetron models we have $|A-1| \lesssim 10^{-6}$ as the
coupling strength $\beta$ must vanish sufficiently fast in high-density regions to
screen the fifth force through a Damour-Polyakov mechanism.
This means that, in terms of background quantities (e.g., the Hubble expansion
rate and the scale factor) one can identify the Einstein and Jordan frames,
which also become almost identical to the $\Lambda$-CDM reference.
However, at the level of the metric perturbations $\Phi$ and $\Psi$, this is no longer
the case and the Einstein and Jordan gravitational potentials differ by terms
set by $\delta A$, and the dynamics of perturbations deviate from the
$\Lambda$-CDM reference because of the fifth force.

In the K-mouflage case, this identification already breaks down at the background
level. Indeed, $|A-1|$ can reach values of the order of a few percents
(see Fig.~\ref{fig_GNewt_z}) while being consistent
with Solar System and cosmological constraints \cite{Barreira2015}.
For the same reason, the background dynamics (in both Einstein and Jordan frames)
show percent deviations from the $\Lambda$-CDM reference.
Therefore, we must pay attention to the distinction between Einstein and Jordan
frames already at the background level. In particular, in this paper, as we study
clusters of galaxies that involve atomic or radiative processes (both for the
definition of their redshift, from atomic lines, and for their properties such as
X-ray emission), the Jordan frame is the one that is more directly connected to
observations and we work in this frame.
Another advantage of the Jordan frame is that the equations of motion take their
usual form, in particular matter is conserved, which permits a clear and simple
physical interpretation, and only gravity is modified. In contrast, in the Einstein
frame gravity takes a standard form but the equations of motion are modified
and the matter density is usually not conserved.

\subsection{Scale dependence and screening regime}
\label{sec:scale-dependence}

A second important difference between the K-mouflage scenario and some other
modified-gravity models is that the deviations from the $\Lambda$-CDM reference
are scale independent on perturbative scales (from cluster scales to the horizon).
This is most easily seen from the fact that the factor $\epsilon_1(t)$ that enters
the evolution equation (\ref{D-linear-Jordan}) of the linear growing mode
only depends on time, so that the linear growing mode $D_+(t)$ remains scale
independent as in the $\Lambda$-CDM cosmology. This is due to the fact that
in the scalar-field Lagrangian (\ref{K-def}) we focused on the nonstandard
kinetic term and neglected a possible potential term $V(\varphi)$.
Of course, in the highly nonlinear regime a new scale dependence appears,
as the fluctuations of the scalar field themselves become nonlinear and give rise
to the K-mouflage screening mechanism, which ensures the convergence to GR
in the Solar System. However, it happens that the screening transition appears
at galaxy scales, so that clusters remain unscreened and fully feel the effect
of the fifth force.

The same behavior is obtained in the Galileon models, where linear scales
below the horizon show scale-independent growing modes and the Vainshtein
screening mechanism applies to cluster scales and below \cite{De-Felice2012,Barreira2014};
but in the K-mouflage case the nonlinear screening only applies to galaxy scales and
below, as clusters remain unscreened.
In contrast, in $f(R)$ and Dilaton/Symmetron models, there is a characteristic
scale dependence, as we recover GR both on very large scales $x \gg 1 h^{-1}$Mpc
(because of the finite mass of the scalar field) and on very small scales
$x \ll 1 h^{-1}$Mpc (because of nonlinear screening mechanisms, here Chameleon
or Damour-Polyakov mechanisms \cite{Khoury:2003aq,Damour:1994zq}).
Then, the linear growing mode $D_+(k,t)$ shows a clear scale dependence on
quasilinear scales and nonlinear screening effects also add a further scale dependence
around $x \sim 1 h^{-1}$Mpc \cite{BraxPV2013}.
Thus, in these models clusters probe a scale-dependent regime and the
transition between the unscreened and screened regimes.

Therefore, clusters of galaxies are especially well-suited probes of K-mouflage
scenarios because they are unscreened (hence they feel the full amplitude of the
fifth force). Moreover, the modification of gravity is still scale independent on these
scales so that cluster properties should not be too difficult to model (the same modelizations
should apply equally well to the $\Lambda$-CDM and K-mouflage cosmologies).

\subsection{Clusters}
\label{sec:clusters-comp}

\subsubsection{Cluster profles}
\label{sec:profiles-comp}

The effect of the fifth force on the cluster matter and gas profiles within the context of
modified-gravity scenarios with chameleon mechanisms [mostly for $f(R)$
models] has been investigated in \cite{Terukina2012,Arnold2014,Terukina2014,Wilcox2015}.

As recalled above, the fifth force effect is somewhat different between $f(R)$,
Dilaton and Symmetron models, and K-mouflage scenarios.
In the former cases clusters are typically at the transition between the screened and
unscreened regimes. Then, massive clusters are screened by the Chameleon
or Damour-Polyakov mechanisms (the deviations from GR being most efficiently
suppressed in the Symmetron models) while for low-mass clusters only a small core
is efficiently screened. In particular, for $f(R)$ and Symmetron models the amplification
of gravity is localized in the outskirts of massive clusters.
This gives a distinct scale dependence of the modified-gravity effect in these theories
\cite{Lombriser2012,Clampitt2012},
but the efficient screening also decreases the overall deviation from GR.
See for instance
\cite{BraxPV2013} for a detailed analysis and comparison of these different models
and \cite{Schmidt2009,Lombriser2012,Moran2015,Gronke2015,Hammami2015}
for numerical simulations of various models.
In Galileon scenarios, depending on the model, clusters may be fully screened
(and their profiles are similar to those of a quintessence model with no fifth force
and the same expansion history) or only partly screened (which gives rise to
complex effects) \cite{Barreira2014}.
In the K-mouflage case, clusters still are in the unscreened regime. Therefore,
there is no characteristic scale dependence that can be used to distinguish them
from $\Lambda$-CDM cosmologies, but the amplitude of the smooth deviation is
greater (as compared with screened scenarios).

There have been no specific simulations of K-mouflage models so far, but we can
recall here some results from simulations of other modified-gravity models.

Ref.~\cite{Hammami2015} develops hydrodynamic N-body simulations to investigate
the impact on dark matter and gas profiles of $f(R)$ and Symmetron scenarios.
In agreement with the discussion above, the authors find that the dark matter density
is increased as compared with the $\Lambda$-CDM reference in the outskirts of
massive halos. This is because the fifth force applies to outer radii, which are
unscreened, and this also yields a greater velocity to the particles, which cannot
cluster as strongly within inner radii.
They also find a lower deviation from $\Lambda$-CDM for the gas density profile
than for the dark matter density profile. They note that this may be due to delays
in the collapse of the dark matter and the gas, with the screening of the halos taking
place after dark matter collapse and before gas collapse.
As explained above, for K-mouflage scenarios we do not expect such a localized
enhancement in the matter densities and different behaviors for the dark matter
and the gas, because clusters are unscreened and the dynamics remains similar
to the $\Lambda$-CDM cosmology, as illustrated by the explicit symmetry
described in Sec.~\ref{sec:symmetry-clustering}.
In fact, in Fig.~\ref{fig_rhosca_z0} we find that within our very simple model the
deviation from $\Lambda$-CDM is slightly greater for the gas than for the
dark matter [because of the small change in the concentration parameter and
the higher sensitivity for the gas that arises from the exponential equilibrium
distribution (\ref{rhog-def})].

We can note that N-body simulations of $f(R)$ models also find that in the
case $f_{R_0}=-10^{-4}$, where Chameleon screening is not efficient,
deviations from $\Lambda$-CDM are smooth and the velocity dispersion $\sigma_v^2$
and gas temperature $T_g$ are about $4/3$ times the GR value
\cite{Arnold2014,Moran2015}, due to the $4/3$ increase of the effective Newton's constant.
In our case, this corresponds to the $(1+\epsilon_1)$ enhancement in
Eq.(\ref{Jeans-equation}), but with $\epsilon_1 \simeq 2\%$ instead of $1/3$.
These simulations also find that dark matter halos remain well described by NFW profiles
\cite{Lombriser2012a}.
These results suggest that our approach for the density profiles and the gas temperature
described in Secs.~\ref{sec:profiles} and \ref{sec:Hydrostatic equilibrium} should
fare reasonably well.

For Chameleon scenarios, Ref.~\cite{Terukina2012} considers the effect of the fifth
force on the cluster gas profile, which becomes more compact for a given dark matter
profile [in our case, we also include the effect on the dark matter profile, through the
modification of the concentration parameter shown in Fig.~\ref{fig_cM_z0},
and we assume an isothermal gas (i.e., $\gamma=1$) instead of a polytropic
equation of state $p_g \propto \rho_g^{\gamma}$ with $\gamma\sim 1.2$].
However, they find that observational error bars are too large to give useful constraints
on $f(R)$ models. We reach the same conclusion for the K-mouflage models studied
in this paper.

Refs.~\cite{Terukina2014,Wilcox2015} combine X-ray observations (which probe the temperature
and electron number density profiles) and weak lensing signals (which probe the total
matter profile) to constrain deviations from General Relativity. Indeed, while
the gas profile is sensitive to the fifth force the lensing deflection of light rays
remains the same as in GR. This allows them to derive the upper bound
$|f_{R_0}| \leq 6\times 10^{-5}$.
The same behavior applies to the K-mouflage scenario. In this paper we found
a few percent deviations from GR, which should apply to all realistic K-mouflage models
that satisfy cosmological and Solar System constraints [as this is due to the constraint
on the coupling $\beta$, independently of the details of the kinetic function $K(\chi)$].
We leave a more general MCMC analysis of K-mouflage scenarios, combining different
probes, to future works.

\subsubsection{Cluster scaling relations}
\label{sec:scaling-comp}

The impact of $f(R)$ gravity on the cluster scaling relations has been studied in
Ref.\cite{Arnold2014} with numerical simulations. Again, we find similar behaviors
for the K-mouflage models as for the $f(R)$ model with $f_{R_0}=-10^{-4}$,
where clusters are not screened. In particular, the dark matter velocity dispersion and
gas temperature are increased, at fixed mass, and the X-ray luminosity is decreased,
at fixed temperature, as compared with the $\Lambda$-CDM reference.
However, whereas in $f(R)$ theories the deviations from $\Lambda$-CDM can reach
a factor $1/3$ in the unscreened regime, and deviations of order unity can also be
expected in Dilaton or Symmetron models, realistic K-mouflage models can only deviate
by a few percent at most.
Indeed, the magnitude of these deviations is set by the factor $\epsilon_1$, which itself is
set by $2\beta^2$ (at $z=0$) and the coupling strength must satisfy $\beta \leq 0.1$
because of cosmological and Solar System constraints \cite{Barreira2015}.

\subsubsection{Cluster lensing; Dynamical and lensing masses}
\label{sec:lensing-comp}

In $f(R)$, Dilaton and Symmetron scenarios the weak lensing potential
$\Phi_{\rm wl}$ that governs the deflection of light rays, given by
$\Phi_{\rm wl}=(\Phi+\Psi)/2$, is equal to the Newtonian potential $\Psi_{\rm N}$
[using $|f_{R_0}|\ll 1$ in the case of $f(R)$ theories]
\cite{Tsujikawa2008}.
Then, for weak lensing observations the only difference from the $\Lambda$-CDM scenario
comes from the different evolution of the matter density fields.
In contrast, the motions of matter particles or of small halos (e.g., satellite halos or small galaxies),
which fall towards massive clusters, feel the fifth force.
This gives rise to different lensing and dynamical masses.
This has been investigated in semianalytical models and N-body simulations
\cite{Lam2013,Moran2015}
and used in \cite{Terukina2014,Wilcox2015} in combination with X-ray observations,
for $f(R)$ theories, and in \cite{Zu2014}, using the galaxy infall kinematics onto
massive clusters, for $f(R)$ and Galileon models.

In the K-mouflage scenario, we again have $\Phi_{\rm wl}=\Psi_{\rm N}$,
as the factor $\delta A/\bar{A}$ cancels from the sum over the two metric potentials,
see Eq.(\ref{Phi-Psi-Jordan}). However, in contrast with the former theories the
effective Newton constant [which enters the Newtonian potential
(\ref{Poisson-small-scale-Jordan-1})] now depends on time, as in
Eq.(\ref{MPlanck-renorm}).
(This effect does not appear in the other scenarios
because they have $\bar{A}$ very close to unity, within $10^{-4}$ or better.)
On the other hand, this time-dependent prefactor cancels from the ratio between
the dynamical and lensing potentials or masses, see Eq.(\ref{ratio-Mdyn-Mlens}).
However, whereas in $f(R)$ theories this ratio can again deviate from unity by a factor
$1/3$ in the unscreened regime, and deviations of order unity can also be reached
in Dilaton or Symmetron models, in realistic K-mouflage models this ratio can only deviate
from unity by $2\%$ at most because of observational constraint on the scalar-field coupling,
$\beta \leq 0.1$.
Therefore, K-mouflage models cannot significantly decrease the tension between measures
of X-ray and lensing clusters masses.

\subsubsection{Cluster number counts}
\label{sec:counts-comp}

In most modified-gravity scenarios the growth of large-scale perturbations
differs from the GR evolution. This typically leads to a new scale dependence of the
linear growth rates (e.g., in $f(R)$, Dilaton and Symmetron models), as one goes
from the very large scales (beyond the Compton wavelength of the scalar field)
where GR is recovered to the quasi-linear scales where the fifth force is unscreened
and gives its maximum amplification of the gravitational interaction
(at smaller scales nonlinear screening leads again to a convergence to GR).
This amplification typically yields a faster growth of matter density perturbations
on scales $1 h^{-1}$Mpc to $10 h^{-1}$Mpc, whence a greater abundance of massive
halos and clusters as compared with the $\Lambda$-CDM cosmology,
see Refs.
\cite{Schmidt2009,Zhao:2010qy,Ferraro2011,Li2012,Lombriser2013,Barreira2014}
for numerical studies of various models.
As explained above, a similar enhancement is found in the K-mouflage scenarios,
with the important difference that all clusters are unscreened and that the
modification of gravity extends up to the horizon (so that the linear modes
grow faster than in $\Lambda$-CDM but remain scale independent).
In Galileon scenarios, the screening mechanism has a strong impact
and, depending on the models, the tail of the halo mass function
can be either increased or decreased, as compared with a quintessence scenario with
the same expansion history \cite{Barreira2014}.

The abundance of massive clusters has been used within $f(R)$ theories
to constrain $f_{R_0}$ \cite{Schmidt2009a,Cataneo2014,Boubekeur2014}.
In combination with CMB, BAO and SNeIa observations, one can obtain
an upper bound $|f_{R_0}| \lesssim 1.7\times 10^{-5}$
\cite{Cataneo2014,Boubekeur2014}, but most of the constraint comes from the
cluster data.
We can expect that for K-mouflage models similar results should be obtained,
especially since clusters are unscreened so that their abundance should provide
useful constraints. In this paper we presented the physics of K-mouflage scenarios
on cluster scales, highlighting the difference between the Einstein and Jordan frames
(which can be neglected in most other scenarios) and investigating both the modified
growth of structures and halo mass functions and the modified cluster scaling laws.
We leave a detailed MCMC analysis of the K-mouflage parameter space to future works.

\subsubsection{Cluster correlation function}
\label{sec:correlation-comp}

In modified-gravity scenarios, the correlation function and the power spectrum
of the matter density field are often increased as the growth of large-scale structures is
amplified by the fifth force. This also enhances the large-mass tail of the halo mass
function and decreases the bias of massive halos as they become less rare.
In the K-mouflage models that we considered in this paper, this decrease of the
cluster bias mostly cancels the increase of the underlying matter density correlation,
and the cluster correlation function is much closer to the $\Lambda$-CDM reference
than the matter correlation itself.
Therefore, the correlation of the matter density field, which can be measured
from weak lensing observations for instance, is a better probe of modified cosmology
than the clustering of massive halos.
The same effects, with a similar compensation between the smaller bias and the
higher matter correlation (but the compensation may only be partial, depending
on the model) appear in other modified-gravity scenarios, see for instance
Ref.\cite{Mak2012} for $f(R)$ theories.

\subsection{Other tests}
\label{sec:other-tests}

On slightly larger scales than those probed by clusters of galaxies, modified-gravity
models have also been tested from galaxy surveys, using redshift-space distortions
of the galaxy power spectrum \cite{Okada2013}, the clustering of Luminous
Red Galaxies \cite{Barreira2014}, or the shape of the galaxy correlation
function itself \cite{Bel2014}.
Similar studies could be performed for K-mouflage models, as they can lead to
$20\%$ deviations for the matter power spectrum and correlation function,
as seen in Fig.~\ref{fig_dPk_z0}.
This will be investigated in future works.

In between the cluster and cosmological scales (e.g., the formation of large-scale
structures and the background dynamics) and the Solar System scales
(where we must recover GR up to a very high accuracy), modified-gravity theories
have also been tested on intermediate galaxy scales.
In particular, within Chameleon scenarios, low-mass galaxies should be unscreened
so that the rotation curve of their diffuse gas component probes the fifth force,
while their stars, being compact high-density objects, are screened and move
as in GR \cite{Jain:2011ji}.
This can provide constraints as tight as $|f_{R_0}| \leq 10^{-6}$
for $f(R)$ theories \cite{Vikram2014}.
The same behaviors apply to K-mouflage models and we can expect that such
tests could also provide useful constraints on these scenarios. We leave this
task to future studies.

Numerical simulations have also been used to investigate the impact of $f(R)$
theories on Lyman-$\alpha$ forest observations (the transmitted flux fraction
and the flux power spectrum)  \cite{Arnold2015}. They find changes
that are too small as compared with observational error bars and do not provide
competitive constraints. Although this study does not distinguish the impact of screening
effects, screening can be expected to be rather inefficient for such moderate density
fluctuations. Therefore, it is likely that similar conclusions would be reached for
K-mouflage models, but we leave a detailed study for future works.

\section{Conclusions}
\label{sec:Conclusions}

We have extended previous works on K-mouflage models by deriving the equations of
motion in both the Einstein and Jordan frames for a fluid with pressure, and next
focusing on the usual case where the pressure arises from small-scale nonlinear processes.
In contrast with many modified-gravity scenarios, the Einstein and Jordan frames already
differ by a few percent at the background level, for K-mouflage models that are
consistent with cosmological and Solar System constraints.
Therefore, one must take into account these deviations and use the correct quantities when
comparing with observations.

We  focused on the Jordan frame, which is better suited to cosmological probes that
involve atomic processes, such as X-ray clusters.
We show that even though K-mouflage models only differ from the $\Lambda$-CDM reference
by a few percent at the background level, the linear growing modes can deviate by $10\%$
and the matter density power spectrum and correlation function by $30\%$.
The tail of the halo mass function is enhanced by factors of order unity for
$M \gtrsim 10^{14} h^{-1} M_{\odot}$ at $z=0$.

Galaxy clusters are not screened by the K-mouflage mechanism, contrary to what happens
for chameleon models like $f(R)$ in the large curvature limit
or Galileon models subject to the Vainshtein screening.
For this reason, we investigate the effects of the K-mouflage modification of gravity on
clusters. We take into account the impact
of the fifth force mediated by the K-mouflage scalar field on both the dark matter and gas
profiles, through the modifications to the growth of large-scale structures and to the
hydrostatic equilibrium.
We find that K-mouflage makes clusters slightly more compact. Similarly, the gas
temperature and the X-ray luminosity differ from their $\Lambda$-CDM counterparts
by a few percent, an order of magnitude which follows directly from the constraints on
K-mouflage (especially on the coupling strength $\beta$) due to Solar System tests.
The only deviation of noticeable order appears in the cluster temperature function,
as the number of clusters is larger than in the $\Lambda$-CDM scenario for K-mouflage
models, because of the faster growth of large-scale structures.
This appears to be large enough that one can hope that this will be within the reach of the
future large-scale surveys.
On the other hand, the cluster correlation function only deviates by a few percent
from the $\Lambda$-CDM case because the increase of the underlying matter density
correlation function is compensated by the lower cluster bias, as massive halos become
less rare at fixed mass, due to the enhanced structure formation.

In this paper we only considered two kinetic functions $K(\tilde{\chi})$ to illustrate the
K-mouflage scenarios and to estimate the amplitude of the effects that can be reached
by realistic models, which are consistent with cosmological and Solar System constraints.
We leave to future works a more detailed MCMC analysis of K-mouflage scenarios, which
would provide the parameter space of K-mouflage models that is allowed by cluster
observations and combinations with other observational probes.

\begin{acknowledgments}
This work is supported in part by the French Agence Nationale de la Recherche under Grant ANR-12-BS05-0002. P.B.
acknowledges partial support from the European Union FP7 ITN
INVISIBLES (Marie Curie Actions, PITN- GA-2011- 289442) and from the Agence Nationale de la Recherche under contract ANR 2010
BLANC 0413 01.
\end{acknowledgments}

\appendix

\section{Equations of motion in the Einstein frame}
\label{app:Einstein-frame}

In this appendix we derive the equations of motion of the scalar field and of the
matter component in the Einstein frame, for a cosmological fluid with a nonzero pressure.
The derivation is similar to the one presented in previous papers
\cite{Brax:2014a,Brax:2014b}, where we studied the background cosmological dynamics
and the formation of large-scale cosmological structures, but with the addition of the
pressure terms.

\subsection{Energy-momentum tensors and equations of motion}
\label{app:Energy-momentum-tensors}

We consider three components of the energy density of the Universe, a matter fluid with
nonzero pressure, radiation, and the scalar field. The Einstein-frame and Jordan-frame
matter energy-momentum tensors are given by Eq.(\ref{Tm-def}) and they
satisfy the relations (\ref{tildeT-T}).
Assuming a perfect fluid, the matter energy-momentum tensor writes as
\be
\tilde{T}^{\mu}_{\nu} = (\tilde{\rho}+\tilde{p}) \, \tilde{u}^\mu \tilde{u}_\nu
+ \tilde{p} \, \delta^{\mu}_{\nu} ,
\label{T-perfect-fluid-def}
\ee
where $\tilde{u}^\mu$ is the velocity 4-vector, normalized such that
$\tilde{u}^\mu \tilde{u}_\mu=-1$, and $\tilde{\rho}$ and $\tilde{p}$ are the Einstein-frame
density and pressure, which are related to their Jordan-frame counterparts by
Eq.(\ref{rho-p-E-J}).

For the radiation component, we neglect perturbations and only consider the contribution to
the background, with the mean density and pressure $\bar{\tilde\rho}_{\rm (r)}$ and
$\bar{\tilde{p}}_{\rm (r)}= \bar{\tilde\rho}_{\rm (r)}/3$.
Their Jordan-frame counterparts are again given as in Eq.(\ref{rho-p-E-J}).

The Einstein-frame energy-momentum tensor of the scalar field reads as
\beq
\tilde{T}_{\mu\nu{\rm (\varphi)}} = \frac{-2}{\sqrt{-\tilde{g}}}
\frac{\delta S_{\varphi}}{\delta \tilde{g}^{\mu\nu}}
= K' \tilde{\nabla}_{\mu}\varphi \tilde{\nabla}_{\nu}\varphi + {\cal M}^4 K \tilde{g}_{\mu\nu} .
\label{Tphi-def}
\eeq

The Klein-Gordon equation that governs the dynamics of the scalar field $\varphi$ is
obtained from the variation of the action (\ref{S-def}) with respect to $\varphi$.
This yields
\beq
\tilde{\nabla}_{\mu} \left[  \tilde{\nabla}^{\mu} \varphi
\; K' \right] =  (\tilde{\rho}-3\tilde{p}) \frac{\dd\ln A}{\dd\varphi} .
\label{KG-press-1}
\eeq
Here we used the fact that the trace of the matter energy-momentum tensor is
$\tilde{T}^{\mu}_{\mu} = - \tilde{\rho}+3\tilde{p}$, from Eq.(\ref{T-perfect-fluid-def}),
while it is zero for the radiation component as $\tilde{p}_{\rm (r)} = \tilde{\rho}_{\rm (r)}/3$.
Combining with Eq.(\ref{Tphi-def}), we find for the scalar field the ``non-conservation''
equation
\beq
\tilde{\nabla}_{\mu} \tilde{T}^{\mu}_{\nu{\rm (\varphi)}} = (\tilde{\rho}-3\tilde{p}) \,
\tilde{\nabla}_{\nu} \ln A .
\label{conserv-Tphi-E}
\eeq

The matter energy-momentum tensor is conserved in the Jordan frame,
\beq
\nabla_{\mu} T^{\mu}_{\nu} = 0 ,
\label{conserv-Tm-J}
\eeq
which gives in the Einstein frame the non-conservation equation
\beq
\tilde{\nabla}_{\mu} \tilde{T}^{\mu}_{\nu} = - (\tilde{\rho}-3\tilde{p}) \,
\tilde{\nabla}_{\nu} \ln A .
\label{conserv-Tm-E}
\eeq
On the other hand, the radiation energy-momentum tensor is conserved in both frames,
\beq
\nabla_{\mu} T^{\mu}_{\nu{\rm (r)}} = 0 , \;\;\;
\tilde{\nabla}_{\mu} \tilde{T}^{\mu}_{\nu{\rm (r)}} = 0 .
\label{conserv-Trad-E-J}
\eeq
The sum of all energy-momentum tensors is also conserved in both frames, and
$\tilde{\nabla}_{\mu}[\tilde{T}^{\mu}_{\nu}+\tilde{T}^{\mu}_{\nu{\rm (r)}}
+\tilde{T}^{\mu}_{\nu{\rm (\varphi)}}] = \nabla_{\mu}[T^{\mu}_{\nu}+T^{\mu}_{\nu{\rm (r)}}
+T^{\mu}_{\nu{\rm (\varphi)}}] = 0$.

Finally, in the Einstein frame the Einstein equations take the usual form,
$\tilde{M}_{\rm Pl}^2 \tilde{G}^{\mu}_{\nu} = \tilde{T}^{\mu}_{\nu}
+\tilde{T}^{\mu}_{\nu{\rm (r)}}+\tilde{T}^{\mu}_{\nu{\rm (\varphi)}}$.

\subsection{Background dynamics}
\label{app:background}

At the level of the cosmological background, the Friedmann equations take the
usual form in the Einstein frame,
\beqa
3 \tilde{M}_{\rm Pl}^2 \tilde{H}^2 & = & \bar{\tilde{\rho}} + \bar\tilde{\rho}_{\rm (r)}
+ \bar{\tilde{\rho}}_{\varphi} ,
\label{Friedmann-1} \\
-2 \tilde{M}_{\rm Pl}^2 \frac{\dd \tilde{H}}{\dd\tilde{t}} & = & \bar{\tilde{\rho}}
+ \bar{\tilde{p}} + \bar{\tilde{\rho}}_{\rm (r)} + \bar{\tilde{p}}_{\rm (r)}
+ \bar{\tilde{\rho}}_{\varphi} + \bar{\tilde{p}}_{\varphi} ,
\label{Friedmann-2}
\eeqa
where $\bar{\tilde{\rho}}_{\varphi}$ and $\bar{\tilde{p}}_{\varphi}$ are the background 
scalar-field energy density and pressure (in the Einstein frame), given by
Eq.(\ref{rho-phi-def}).

The Klein-Gordon equation (\ref{KG-press-1}) gives
\beq
\frac{\dd}{\dd \tilde{t}}  \left[ \tilde{a}^3 \frac{\dd\bar\varphi}{\dd\tilde{t}} \bar{K}'
\right] = - \tilde{a}^3 (\bar{\tilde\rho} - 3 \bar{\tilde{p}} )
\frac{\dd \ln{\bar A}}{\dd\bar\varphi} ,
\eeq
and the scalar-field energy density satisfies
\beq
\frac{\dd\bar{\tilde\rho}_{\varphi}}{\dd\tilde{t}} = - 3 \tilde{H} ( \bar{\tilde\rho}_{\varphi}
+ \bar{\tilde{p}}_{\varphi} ) - (\bar{\tilde\rho} - 3 \bar{\tilde{p}} )
\frac{\dd\ln\bar{A}}{\dd\tilde{t}} .
\label{rho-phi-no-conserv}
\eeq

The non-conservation equation (\ref{conserv-Tm-E}) gives for the matter density the evolution
equation
\beq
\frac{\dd\bar{\tilde\rho}}{\dd\tilde{t}} = - 3 \tilde{H} (\bar{\tilde\rho} + \bar{\tilde{p}} )
+ (\bar{\tilde\rho} - 3 \bar{\tilde{p}}) \frac{\dd\ln{\bar A}}{\dd \tilde{t}} .
\label{conserv-back-1}
\eeq
In particular, we have
$\dd(\bar{\tilde\rho}+\bar{\tilde\rho}_{\varphi})/\dd\tilde{t} = - 3 \tilde{H}
(\bar{\tilde\rho}+\bar{\tilde{p}}+\bar{\tilde\rho}_{\varphi}+\bar{\tilde{p}}_{\varphi})$.
When the pressure is zero, we can define a conserved density $\hat{\rho}$ by
\beq
\hat{\rho} = \tilde{\rho}/A , \;\;\; \hat{p} = \tilde{p}/A .
\label{rho-hat-def}
\eeq
Indeed, substituting into Eq.(\ref{conserv-back-1}) gives
\beq
\frac{\dd\bar{\hat{\rho}}}{\dd\tilde{t}} = -3 \tilde{H} ( \bar{\hat{\rho}}
+ \bar{\hat{p}} ) - 3 \bar{\hat{p}} \frac{\dd\ln\bar{A}}{\dd \tilde{t}} .
\eeq
Thus, if $\hat{p}=0$ we obtain a conservation of the standard form in the Einstein frame,
$\dd\bar{\hat{\rho}}/\dd\tilde{t}= -3 \tilde{H} \bar{\hat{\rho}}$. However,
if $\hat{p} \neq 0$ it is no longer possible to cancel the non-conservation term of
Eq.(\ref{conserv-back-1}) by such a simple redefinition of the density.

The background radiation density obeys the usual conservation equation,
\beq
\frac{\dd\bar{\tilde\rho}_{\rm (r)}}{\dd\tilde{t}} = - 3 \tilde{H} (\bar{\tilde\rho}_{\rm (r)}
+ \bar{\tilde{p}}_{\rm (r)} ) = - 4 \tilde{H} \bar{\tilde\rho}_{\rm (r)} ,
\label{conserv-rad-back-1}
\eeq
in agreement with the second Eq.(\ref{conserv-Trad-E-J}).

\subsection{Perturbations}
\label{app:perturbations}

In the conformal Newtonian gauge the Einstein-frame metric can be written as
\beq
\dd \tilde{s}^2 = \tilde{a}^2 [ -(1+2\tilde{\Phi}) \dd \tau^2 + (1-2\tilde{\Psi}) \dd \vx^2 ] ,
\label{Newtonian-gauge-Einstein}
\eeq
where $\tau = \int \dd \tilde{t}/\tilde{a}$ is the conformal time, and $\tilde{\Phi}$ and
$\tilde{\Psi}$ are the two metric gravitational potentials.
Throughout this paper, we consider the non-relativistic and weak-gravitational-fields
regime, with $\tilde{\Phi} \ll 1$, $\tilde{\Psi} \ll 1$,
$v^2 \ll 1$ (where $\vv=\dd\vx/\dd\tau$ is the peculiar velocity of the particles), and we
expand up to first order in $\{\tilde{\Phi},\tilde{\Psi},v^2\}$.
Then, in the final equations, we
only keep zeroth-order terms, $1+\tilde{\Psi} \simeq 1$ and $1+v^2\simeq 1$, except when
the potentials or the velocity arise with a gradient operator, as in Eqs.(\ref{Euler-E-1})
and (\ref{Poisson-E-1}).
In particular, we have for the matter velocity 4-vector,
\beq
\tilde{u}^{\mu} = \tilde{a}^{-1} (1-\tilde{\Phi}+v^2/2,v^i) , \;\;
\tilde{u}_{\mu} = - \tilde{a} (1+\tilde{\Phi}+v^2/2,-v_i) ,
\label{4-velocity-def}
\eeq
where we denote
\beq
v_i = v^i = \frac{\dd x^i}{\dd\tau}
\label{pec-v-def}
\eeq
 the peculiar velocity.

The non-conservation equation (\ref{conserv-Tm-E}) gives
 \beqa
&& (\dot {\tilde\rho} +\dot {\tilde p}) \tilde{u}_{\nu} + 3 \tilde{h} (\tilde{\rho}+ \tilde{p})
\tilde{u}_{\nu} + (\tilde{\rho}+\tilde{p})  \dot{\tilde{u}}_{\nu}
+ \tilde{\nabla}_{\nu} \tilde{p} = \nonumber \\
&& -  (\tilde{\rho} - 3 \tilde{p}) \tilde{\nabla}_{\nu}(\ln A) ,
\label{pert-matter-1}
\eeqa
where we have introduced
\beq
\dot{\tilde\rho} \equiv \tilde{u}^\mu \tilde{\nabla}_\mu \tilde{\rho} , \;\;
\dot{\tilde{u}}_{\nu} \equiv \tilde{u}^\mu \tilde{\nabla}_{\mu} \tilde{u}_\nu , \;\;
3 \tilde{h} \equiv \tilde{\nabla}_{\mu} \tilde{u}^{\mu} .
\eeq
Contracting  with $\tilde{u}^\nu$ and using $\tilde{u}^\nu \tilde{u}_\nu=-1$, we get
\be
\dot{\tilde\rho} + 3 \tilde{h} (\tilde{\rho}+\tilde{p}) = (\tilde{\rho}-3\tilde{p})
\frac{\dot  A}{A} .
\label{cr}
\ee
It is easy to see that
$\dot{\tilde\rho}= \tilde{a}^{-1} [\pl_{\tau}\tilde{\rho} + (\vv\cdot\nabla){\tilde\rho}]$,
where $\nabla\equiv\pl/\pl\vx$ is the standard 3D spatial gradient, and
$3 \tilde{h}= \tilde{a}^{-1} [3\tilde{\cal H}+(\nabla\cdot\vv)]$, where
$\tilde{\cal H}=\dd\ln\tilde{a}/\dd\tau$ is the conformal expansion rate in the Einstein
frame. Therefore, this is explicitly
\beqa
&& \frac{\pl{\tilde\rho}}{\pl\tau} + (\vv\cdot\nabla) \tilde{\rho} + (3 \tilde{\cal H}
+ \nabla\cdot\vv) (\tilde{\rho}+\tilde{p})  \nonumber \\
&& = (\tilde{\rho} - 3 \tilde{p}) \; \left[ \frac{\pl\ln A}{\pl\tau} + (\vv\cdot\nabla) \ln A \right]  .
\label{continuity-E-1}
\eeqa

Next, the non-conservation equation (\ref{pert-matter-1}) can be simplified by subtracting
Eq.(\ref{cr}) multiplied by $\tilde{u}_{\nu}$. This leads to
\beq
\dot{\tilde{u}}_{\nu} = - \frac{\tilde{\nabla}_{\nu} \tilde{p} + \tilde{u}_{\nu} \dot{\tilde{p}}}
{\tilde{\rho} + \tilde{p}} - \frac{\tilde{\rho}-3 \tilde{p}}{\tilde{\rho}+\tilde{p}}
\frac{\tilde{\nabla}_{\nu} A + \tilde{u}_{\nu} \dot{A}}{A} .
\eeq
This is the generalized geodesic equation. Specializing to $\mu=i$, we get the Euler
equation of K-mouflage hydrodynamics,
\beqa
\frac{\partial\vv}{\partial\tau} + (\vv\cdot\nabla) \vv + \left( \tilde{\cal H} +
\frac{1}{\tilde{\rho}+\tilde{p}} \frac{\pl\tilde{p}}{\pl\tau}
+ \frac{\tilde{\rho}-3\tilde{p}}{\tilde{\rho}+\tilde{p}} \frac{\pl\ln A}{\pl\tau}
\right) \vv = && \nonumber \\
- \nabla \tilde{\Phi} - \frac{\nabla \tilde{p}}{\tilde{\rho}+\tilde{p}}
- \frac{\tilde{\rho}-3\tilde{p}}{\tilde{\rho}+\tilde{p}} \nabla\ln A \hspace{1cm} &&
\label{Euler-E-1}
\eeqa

From Eq.(\ref{Euler-E-1}) we can see that
$(\vv\cdot\nabla)\ln A \sim \pl_{\tau}v^2 + v^2 (\nabla\cdot\vv)$,
hence this term can be neglected in the continuity equation (\ref{continuity-E-1}) in the
non-relativistic limit $v^2 \ll 1$. This simplifies as
\beq
\frac{\pl\tilde{\rho}}{\pl\tau} + (\vv\cdot\nabla) \tilde{\rho} + (3 \tilde{\cal H}
+ \nabla\cdot\vv) (\tilde{\rho}+\tilde{p})
 = (\tilde{\rho} - 3 \tilde{p})  \frac{\pl\ln A}{\pl\tau} .
\label{continuity-E-2}
\eeq

The Klein-Gordon equation (\ref{KG-press-1}) writes as
\beq
\frac{1}{\tilde{a}^3} \frac{\pl}{\pl\tilde{t}} \left( \tilde{a}^3 \frac{\pl \varphi}{\pl\tilde{t}}
\; K' \right) - \frac{1}{\tilde{a}^2} \nabla \cdot (\nabla\varphi \; K' ) =
- (\tilde{\rho} - 3 \tilde{p}) \frac{\dd\ln A}{\dd \varphi}  .
\label{KG-press-3}
\eeq

The $(0,0)$ component of Einstein equations,
$\tilde{M}_{\rm Pl}^2 \tilde{G}^0_0 = \tilde{T}^0_0+\tilde{T}^0_{0(\varphi)}$,
gives
\beq
\tilde{\Psi} =  \tilde{\Psi}_{\rm N} \;\;\; \mbox{with} \;\;\;
\frac{1}{\tilde{a}^2} \nabla^2 \tilde{\Psi}_{\rm N} = \frac{1}{2 \tilde{M}_{\rm Pl}^2}
( \delta{\tilde\rho} + \delta{\tilde\rho}_{\varphi} ),
\label{Poisson-E-1}
\eeq
where $\delta{\tilde\rho}=\tilde{\rho}-\bar{\tilde\rho}$ and
$\delta{\tilde\rho}_{\varphi}=\tilde{\rho}_{\varphi}-\bar{\tilde\rho}_{\varphi}$,
with $\tilde{\rho}_{\varphi} = - {\cal M}^4 K + K' (\pl\varphi/\pl\tilde{t})^2$.
Here we denoted $\tilde{\Psi}_{\rm N}$ the usual Newtonian potential.
The $(i,j)$ components of the Einstein equations give (focusing on the part that is not
proportional to $\delta^i_j$),
\beq
\frac{\tilde{M}_{\rm Pl}^2}{\tilde{a}^2} \pl_i \pl_j (\tilde{\Psi}-\tilde{\Phi}) =
(\tilde{\rho}+\tilde{p}) v_i v_j + \frac{K'}{\tilde{a}^2} \pl_i\varphi \pl_j\varphi .
\label{Poisson-ij-E-1}
\eeq
From Eq.(\ref{Poisson-E-1}) we can see that on the left-hand side in Eq.(\ref{Poisson-ij-E-1})
we have $\tilde{M}_{\rm Pl}^2 \tilde{a}^{-2} \pl_i\pl_j\tilde{\Psi} \sim \delta\tilde{\rho}
\sim \tilde{\rho}$.
Therefore, the first term on the right-hand side, $(\tilde{\rho}+\tilde{p}) v_i v_j$, is negligible
as $v^2 \ll 1$.
Next, from the Klein-Gordon equation (\ref{KG-press-3}) we obtain, on scales that are much
smaller than the horizon,
\beq
k / \tilde{a}\tilde{H}  \gg 1 : \;\;\; \delta\varphi \sim \frac{\beta}{\tilde{M}_{\rm Pl}}
\frac{\tilde{a}^2}{k^2} \frac{\delta{\tilde\rho}-3\delta\tilde{p}}{K'}
\label{dphi-estimate}
\eeq
where $k$ is the typical comoving wave number of interest. Then, the second term in the
right-hand side in Eq.(\ref{Poisson-ij-E-1}) is of order
\beq
\frac{K'}{\tilde{a}^2} \pl_i\varphi \pl_j\varphi \sim \delta\tilde{\rho}
\frac{\delta{\tilde\rho}}{\bar{\tilde\rho}} \frac{\beta^2}{K'}
\frac{\tilde{a}^2\tilde{H}^2}{k^2} \ll \delta\tilde{\rho} ,
\eeq
which is again negligible  compared to
$\tilde{M}_{\rm Pl}^2 \tilde{a}^{-2} \pl_i\pl_j\tilde{\Psi}$.
Therefore, the Einstein equations (\ref{Poisson-ij-E-1}) give
\beq
\tilde{\Phi} = \tilde{\Psi} = \tilde{\Psi}_{\rm N} ,
\label{Psi-Phi}
\eeq
within the approximations that we use in this paper.

To close the system formed by the equations of motion obtained above, we must specify
the pressure, for instance through an (effective) equation of state such as
$\tilde{p} = w \tilde{\rho}$ with some parameter $w$.

\subsection{Pressure due to small-scale nonlinear physics}
\label{app:small-pressure}

In the previous sections, we derived the equations of motion for a cosmological
fluid with a nonzero pressure, in the non-relativistic limit $v^2 \ll 1$ for the mean fluid
velocity and in the weak field regime $\tilde{\Psi}_{\rm N} \ll 1$.
We made no approximation for the pressure and the equations of motion also apply
to fluids with a pressure of the same order as the density, such as
$\tilde{p}=w \tilde{\rho}$ where $w$ is a parameter of order unity.
However, in the usual CDM context, the pressure is negligible on cosmological scales and
it is built on small scales by nonlinear processes, such as the collapse of gas clouds that
generate shocks or the virialization of dark matter halos (which generate an effective pressure
through the velocity dispersion of the particles).
Then, the pressure is of the order $\tilde{p} \sim \tilde{\rho} c_s^2$, where $c_s$ is the
speed of sound or the velocity dispersion, and $c_s^2 \sim \tilde{\Psi}_{\rm N}$ because it is
generated by the gravitational collapse (for instance, if we have hydrostatic equilibrium
we typically have $\nabla\tilde{\Psi}_{\rm N} \sim \nabla \tilde{p}/\tilde{\rho}$ as the
pressure balances gravity).

Then, in the regime $\tilde{p}/\tilde{\rho} \sim \tilde{\Psi}_{\rm N} \sim v^2 \ll 1$,
the background pressure is zero, $\bar{\tilde{p}}=0$, and we recover the cosmological
dynamics studied in \cite{Brax:2014a} for a pressureless fluid.
Thus, the Friedmann equations read as
Eqs.(\ref{Friedmann-1-nopress})-(\ref{Friedmann-2-nopress})
and the matter and radiation densities evolve as in Eq.(\ref{conserv-back-nopress}).
The Klein-Gordon equation becomes as in Eq.(\ref{KG-back-nopress}).

For the perturbations, the continuity and Euler equations (\ref{continuity-E-2}) and
(\ref{Euler-E-1}) simplify as
\beq
\frac{\pl\tilde{\rho}}{\pl\tau} + \nabla \cdot (\tilde{\rho} \vv) + 3 \tilde{\cal H} \tilde{\rho}
= \tilde{\rho} \frac{\pl\ln A}{\pl\tau} ,
\label{continuity-nopress}
\eeq
and
\beq
\frac{\pl\vv}{\pl\tau} + (\vv\cdot\nabla)\vv + \left( \tilde{\cal H} + \frac{\pl\ln A}{\pl\tau}
\right) \vv = - \nabla ( \tilde{\Psi}_{\rm N} + \ln A ) - \frac{\nabla\tilde{p}}{\tilde\rho} ,
\label{Euler-nopress}
\eeq
while the Poisson equation remains identical to Eq.(\ref{Poisson-E-1}) and the
Klein-Gordon equation (\ref{KG-press-3}) becomes
\beq
\frac{1}{\tilde{a}^3} \frac{\pl}{\pl\tilde{t}} \left( \tilde{a}^3 \frac{\pl \varphi}{\pl\tilde{t}}
\; K' \right) - \frac{1}{\tilde{a}^2} \nabla \cdot (\nabla\varphi \; K' ) = - \tilde{\rho}
\frac{\dd\ln A}{\dd \varphi}  .
\label{KG-nopress}
\eeq
Therefore, in this regime the only effect of the pressure is to add the usual
pressure term in the Euler equation, without mixed terms involving the coupling function
$A(\varphi)$.

\subsection{Sub-horizon regime}
\label{app:Sub-horizon}

To simplify the Einstein equations (\ref{Poisson-ij-E-1}) we already used the small-scale limit,
$k/\tilde{a}\tilde{H} \gg 1$, which corresponds to scales that are much below the Hubble
scale $\tilde{r}_{\rm H} = 1/\tilde{H}$.
This is the regime that is relevant for the formation of cosmological large-scale structures,
such as clusters of galaxies.
Then, the continuity and Euler equations (\ref{continuity-nopress})-(\ref{Euler-nopress})
and the Poisson equation (\ref{Poisson-E-1}) can be further simplified.
Indeed, as in Eq.(\ref{dphi-estimate}), we obtain the estimates
\beqa
\hspace{-1cm} k/\tilde{a}\tilde{H}  \gg 1 & : & \frac{\delta\varphi}{\tilde{M}_{\rm Pl}}
\sim \frac{\beta}{K'} \frac{\tilde{a}^2 \tilde{H}^2}{k^2} \frac{\delta{\tilde\rho}}{\bar{\tilde\rho}}
\ll \frac{\delta{\tilde\rho}}{\bar{\tilde\rho}} ,
\label{dphi-estim-1} \\
&& \hspace{-1.5cm} \delta A \sim \frac{\beta \delta\varphi}{\tilde{M}_{\rm Pl}} \sim
\frac{\beta^2}{K'} \frac{\tilde{a}^2 \tilde{H}^2}{k^2}
\frac{\delta{\tilde\rho}}{\bar{\tilde\rho}} \ll \frac{\delta{\tilde\rho}}{\bar{\tilde\rho}} ,
\label{dA-estim} \\
&& \hspace{-1.5cm} \delta{\tilde\chi} \simeq - \frac{(\nabla\delta\varphi)^2}
{2{\cal M}^4 \tilde{a}^2} \sim \frac{\beta^2}{K'^2} \frac{\tilde{a}^2 \tilde{H}^2}{k^2}
\frac{(\delta{\tilde\rho})^2}{\bar{\tilde\rho}^2}
\ll \frac{(\delta{\tilde\rho})^2}{\bar{\tilde\rho}^2} ,
\label{dchi-estim} \\
&& \hspace{-1.5cm} \frac{\delta{\tilde\rho}_{\varphi}}{\delta{\tilde\rho}} \sim
\frac{\beta^2}{K'} \frac{\tilde{a}^2 \tilde{H}^2}{k^2}
\left(1+\frac{\delta{\tilde\rho}}{\bar{\tilde\rho}}\right) \ll 1
\label{drhophi-estim} .
\eeqa
Then, in the continuity and Euler equations (\ref{continuity-nopress})-(\ref{Euler-nopress}),
we can write $\pl\ln A/\pl\tau \simeq \dd\ln\bar{A}/\dd\tau$, which leads to
Eqs.(\ref{continuity-small-scale-E}) and (\ref{Euler-small-scale-E}).
In the Poisson equation (\ref{Poisson-E-1}), we can neglect $\delta{\tilde\rho}_{\varphi}$,
which gives Eq.(\ref{Poisson-small-scale-E}).
In the Klein-Gordon equation (\ref{KG-nopress}), we can neglect the fluctuations of $A$ and
only keep the spatial gradients. This leads to Eq.(\ref{KG-small-scale-E}),
which also corresponds to the quasi-static approximation.

\subsection{Formation of large-scale structures}
\label{app:Formation}

Introducing the Einstein-frame matter density contrast,
\beq
\tilde{\delta} = \delta{\tilde\rho} / \bar{\tilde\rho} ,
\label{delta-E-def}
\eeq
the continuity equation (\ref{continuity-small-scale-E}) also writes as
\beq
\frac{\pl\tilde\delta}{\pl\tau} + \nabla \cdot [ (1+\tilde\delta) \vv] = 0 .
\label{continuity-delta-small-scale-E}
\eeq
Thus, in terms of the density contrast we recover the usual continuity equation, without
any $A$-term left. This is related to the fact that $\tilde{\delta}=\hat{\delta}$, where
$\hat{\delta}=\delta\hat\rho/{\bar{\hat\rho}}$ is the conserved matter density introduced
in Eq.(\ref{rho-hat-def}), within our set of approximations
($\tilde{p} \ll \tilde{\rho}$ and $A\simeq \bar{A}$, so that the factor $\bar{A}$ cancels
out in the ratio $\delta\hat\rho/{\bar{\hat\rho}}$).

On perturbative scales, we set the pressure term to zero, as in standard perturbation
theory, because it is generated by non-perturbative effects such as shell crossing and
virialization (shocks). Then, the formation of large-scale structures can be tackled through
a perturbative approach, as in the usual $\Lambda$-CDM case.
Introducing the two-component vector $\tilde{\psi}$,
\beq
\tilde{\psi} \equiv \left(\bea{c} \tilde{\psi}_1 \\ \tilde{\psi}_2 \ea \right) \equiv
\left( \bea{c} \tilde{\delta} \\ -(\nabla\cdot\vv)/(\dd\tilde{a}/\dd\tilde{t}) \ea \right) ,
\label{psi-E-def}
\eeq
equations (\ref{continuity-delta-small-scale-E}) and (\ref{Euler-small-scale-E}) read in
Fourier space as
\beqa
\frac{\pl\tilde{\psi}_1}{\pl\ln\tilde{a}} - \tilde{\psi}_2 & = & \int \dd\vk_1\dd\vk_2 \;
\delta_D(\vk_1\!+\!\vk_2\!-\!\vk) \hat{\alpha}(\vk_1,\vk_2) \nonumber \\
&& \times \; \tilde{\psi}_2(\vk_1) \tilde{\psi}_1(\vk_2) ,
\label{continuity-2}
\eeqa
\beqa
\frac{\pl\tilde{\psi}_2}{\pl\ln\tilde{a}} - \frac{3}{2} \tilde{\Omega}_{\rm m}
(1+\tilde{\epsilon}_1) \tilde{\psi}_1 + \left( 2 + \frac{1}{\tilde{H}^2}
\frac{\dd\tilde{H}}{\dd\tilde{t}} + \tilde{\epsilon}_2 \right) \tilde{\psi}_2 & =  &
\nonumber \\
&& \hspace{-8cm} \int\!\! \dd\vk_1\dd\vk_2 \; \delta_D(\vk_1\!+\!\vk_2\!-\!\vk)
\hat{\beta}(\vk_1,\vk_2) \tilde{\psi}_2(\vk_1) \tilde{\psi}_2(\vk_2) ,
\label{Euler-2}
\eeqa
with
\beq
\hat{\alpha}(\vk_1,\vk_2)= \frac{(\vk_1\!+\!\vk_2)\cdot\vk_1}{k_1^2} ,
\hat{\beta}(\vk_1,\vk_2)= \frac{|\vk_1\!+\!\vk_2|^2(\vk_1\!\cdot\!\vk_2)}{2k_1^2k_2^2} .
\label{alpha-beta-def}
\eeq
The two differences from the equations of motion obtained in the $\Lambda$-CDM cosmology
are the two time-dependent factors $\tilde{\epsilon}_i(t)$, defined by
\beq
\tilde{\epsilon}_1(\tilde{t}) \equiv \frac{2\beta^2}{\bar{K}'} , \;\;\;
\tilde{\epsilon}_2(\tilde{t}) \equiv  \frac{\dd \ln \bar{A}}{\dd\ln\tilde{a}}
= \frac{\beta}{\tilde{M}_{\rm Pl}} \frac{\dd\bar\varphi}{\dd\ln\tilde{a}} .
\label{epsilon-def}
\eeq
In Eq.(\ref{Euler-2}) the factor $\tilde{\Omega}_{\rm m} (1+\tilde{\epsilon}_1)$ can also
be written as $\hat{\Omega}_{\rm m} (1+\hat{\epsilon}_1)$, where $\hat{\Omega}_{\rm m}$
is the cosmological parameter associated with the conserved density $\hat{\rho}$ 
defined in Eq.(\ref{rho-hat-def}),
[$\hat{\Omega}_{\rm m} = \bar{\hat\rho}/\tilde{\rho}_{\rm crit}=\tilde{\Omega}_{\rm m}/\bar{A}$,
with $\tilde{\rho}_{\rm crit} = 3 \tilde{M}_{\rm Pl}^2 \tilde{H}^2$ the Einstein-frame
critical density], and $\hat{\epsilon}_1=\bar{A}(1+\tilde{\epsilon}_1)-1=\bar{A}-1+2\bar{A}\beta^2/\bar{K}'$.

On large scales or at early times, we can linearize the equations of motion
(\ref{continuity-2})-(\ref{Euler-2}). This gives for the linear growing and decaying modes
$\tilde{D}_{\pm}(\tilde{t})$ the evolution equation
\beq
\frac{\dd^2\tilde{D}}{\dd(\ln\tilde{a})^2} + \left( 2 + \frac{1}{\tilde{H}^2}
\frac{\dd\tilde{H}}{\dd\tilde{t}} + \tilde{\epsilon}_2 \right)
\frac{\dd\tilde{D}}{\dd\ln\tilde{a}} - \frac{3}{2} \tilde{\Omega}_{\rm m}
(1+\tilde{\epsilon}_1) \tilde{D} = 0 .
\label{D-linear}
\eeq

\section{Comparison of Einstein-frame and Jordan-frame backgrounds}
\label{sec:E-J-backgrounds}

\begin{figure}
\begin{center}
\epsfxsize=8.5 cm \epsfysize=5.8 cm {\epsfbox{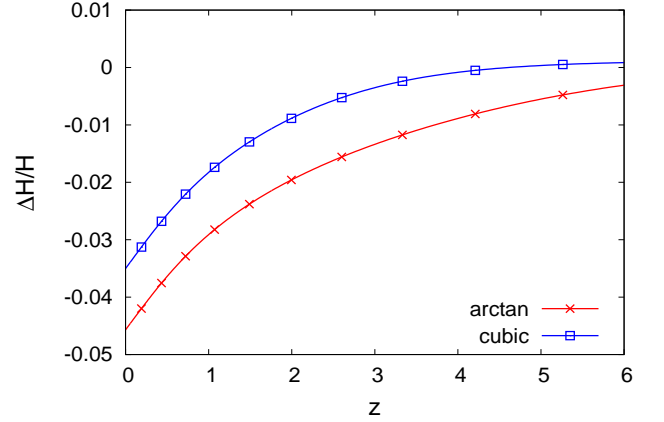}}
\end{center}
\caption{Relative deviation, $\tilde{H}/H-1$, of the Einstein-frame Hubble rate from the Jordan-frame
Hubble rate, as a function of the Jordan-frame redshift.}
\label{fig_H_E}
\end{figure}

\begin{figure}
\begin{center}
\epsfxsize=8.5 cm \epsfysize=5.8 cm {\epsfbox{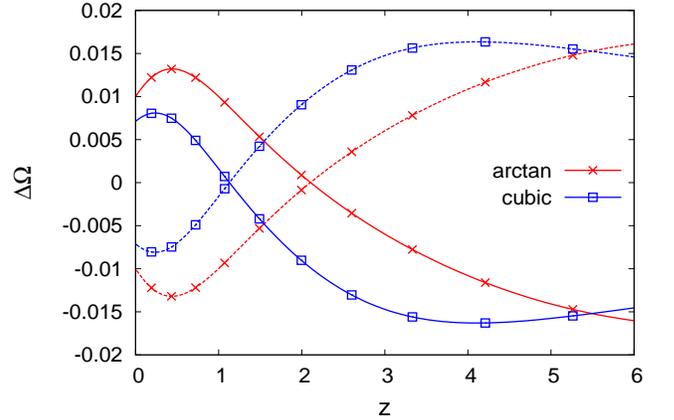}}
\end{center}
\caption{Deviation of the Einstein-frame cosmological parameters from their Jordan-frame
counterparts, $\hat{\Omega}-\Omega$, as a function of the Jordan-frame redshift.
We show the matter density parameters (solid lines) and the dark-energy density
parameters (dashed lines). In the Einstein frame we consider the effective matter and dark-energy
densities as given by Eqs.(\ref{hat-Omega_m-Omega_m}) and (\ref{hat-Omega-de}).}
\label{fig_Om_E}
\end{figure}

In this appendix we compare the Einstein-frame background quantities with their Jordan-frame
counterparts.
We show in Fig.~\ref{fig_H_E} the relative deviation of the Einstein-frame Hubble rate from the
Jordan-frame Hubble rate. From Eq.(\ref{tH-H-def}) this is given by
\beq
\frac{\tilde{H}-H}{H} = \bar{A} (1-\epsilon_2) - 1 .
\label{tH-H-comp}
\eeq
We can see that at low redshift the Einstein-frame Hubble rate is typically lower than its Jordan-frame
counterpart, at a fixed Jordan-frame redshift.
The comparison with Fig.~\ref{fig_H_z} shows that at $z \lesssim 4$ the deviation between these
two expansion rates is greater than the deviation between the Jordan-frame expansion and its
$\Lambda$-CDM reference.
As expected, this deviation is again of the order of a few percent, set by the value of $\beta^2$.
This clearly shows that for the K-mouflage scenario one cannot treat both frames as approximately
identical, contrary to what happens in many modified-gravity theories such as $f(R)$ models
or dilaton models.
At high $z$ the relative difference between both Hubble rates vanishes, as can be seen from
Eq.(\ref{tH-H-def}) as $\bar{A}\rightarrow 1$ and $\epsilon_2\rightarrow 0$.

We show in Fig.~\ref{fig_Om_E} the deviation between the Einstein-frame and Jordan-frame
density cosmological parameters.
More precisely, for the Einstein frame we consider the cosmological ``conserved'' matter density
$\hat{\rho}$ introduced in Eq.(\ref{rho-hat-def}). Using Eq.(\ref{Omega-m-rad}) this is given by
\beq
\hat{\Omega}_{\rm m} = \frac{\tilde{\Omega}_{\rm m}}{\bar{A}}
= \frac{\Omega_{\rm m}}{\bar{A} (1-\epsilon_2)^2} .
\label{hat-Omega_m-Omega_m}
\eeq
On the other hand, the radiation density parameter in the Einstein frame is
$\tilde{\Omega}_{(r)} = \Omega_{(r)}/(1-\epsilon_2)^2$, from Eq.(\ref{Omega-m-rad}), and the
Einstein-frame dark-energy density is then
\beq
\hat{\Omega}_{\rm de} \equiv 1 - \hat{\Omega}_{\rm m} - \tilde{\Omega}_{(r)}
= \frac{\bar{A}-1}{\bar{A}} \tilde{\Omega}_{\rm m} + \tilde{\Omega}_{\varphi} .
\label{hat-Omega-de}
\eeq
Again, we find in Fig.~\ref{fig_Om_E} that the differences between the Einstein-frame and Jordan-frame
cosmological parameters are of the order of $1\%$.

\section{Unitarity constraints}
\label{sec:Unitarity}

The K-mouflage models involve higher-order operators in the derivatives of $\varphi$ and a coupling
of the scalar field to matter $\beta$. This coupling induces a Yukawa interaction of the scalar field
with fermions,
\begin{equation}
{\cal L}_I= \beta \frac{m_f}{M_{\rm Pl}} \, \bar \psi \psi \, \delta\varphi ,
\end{equation}
where $\delta\varphi=\varphi-\bar{\varphi}$ are the fluctuations around a background $\bar{\varphi}$.
Interaction terms in the scalar Lagrangian of the type ${\cal M}^4 \tilde\chi^n$ in a background field
configuration $\bar{\tilde\chi}$ imply the existence of the two-body scattering processes
$f\bar f\to \varphi\varphi$ at tree level, with the exchange of one scalar field $\varphi$.
In quantum mechanics, unitarity of the scattering matrix requires that the scattering amplitude
${\cal M}_{f\bar f \to \varphi\varphi}$ for this process should satisfy
${\cal M}_{f\bar f \to \varphi\varphi}\le 16\pi$. In the clusters that we consider in the main body of the paper,
the background value of $\bar{\tilde\chi}\lesssim 10^{-3}$ is small, and the background value of
$\bar K' \simeq 1$ implies that clusters are unscreened. We focus on processes that can happen inside the
hot gas of the clusters and consider the two-body scattering processes involving either electrons or nuclei.
For temperatures of the gas less than $T_g\lesssim 10$ keV, the particles are non-relativistic.
We only consider K-mouflage functions $K(\tilde\chi)$ whose small $\tilde\chi$ expansion starts at
the cubic order. In this case, the three-point self-interaction of the scalar field is of order
\be
{\cal L}_3 \sim {\cal M}^4 \bar{\tilde\chi} \frac{(\partial_{\vr} \bar\varphi) (\partial_{\vr} \delta\varphi)}{{\cal M}^4}
\frac{(\partial\delta\varphi)^2}{{\cal M}^4} ,
\ee
where we consider a quasi-static background configuration.
The energies of the two outgoing scalars are $E_{1,2}\sim m_f$ whereas the spatial 3-momentum of the
particles is $p_r\sim \sqrt{m_f T_g}$.
The scalar propagator gives a factor of $1/m_f^2$ in the non-relativistic limit and finally we get that
the scattering amplitude can be estimated as
\begin{equation}
{\cal M}_{f\bar f \to \varphi\varphi} \sim \beta \frac{m_f^2}{M_{\rm Pl} {\cal M}^2} \bar{\tilde\chi}^{3/2}
\sqrt{m_f T_g}.
\end{equation}
For protons and neutrons at temperatures $T_g \lesssim 10$ keV for $\beta=0.1$ and using
$\bar{\tilde\chi} \lesssim 10^{-3}$, we find that ${\cal M}_{f\bar f \to \varphi\varphi}\lesssim 10^{-2}$,
implying that unitarity is respected in the two-body processes.

Terms of higher order in $\tilde\chi$ can lead to processes such as $f\bar f \to n\varphi$ involving $n$ scalars
in the final state. The scattering cross section grows fast with the number of outgoing particles and
can exceed the Froissart bound on the $\sigma_{\rm total} \lesssim \frac{1}{m_f^2}$.
This violation is relaxed by classicalization \cite{Dvali:2010jz,Keltner:2015xda} in the same fashion
as in Galileon models \cite{Brax:2015a}, where a classical lump sourced by the incoming energy of the
two fermions in the center-of-mass frame is created. For K-mouflage models, this classical configuration
has a typical size given by the K-mouflage radius $R_K$ \cite{Brax:2014c} and the scattering cross section
becomes equal to the geometrical cross section proportional to $R_K^2$. This process is analogous to the
creation of black holes in high-energy collisions.

K-mouflage models also satisfy a non-renormalization theorem analogous to the one for
Galileons \cite{deRham:2012az}. The quantum corrections going beyond the K-mouflage Lagrangian
are negligible when $r\gtrsim {\cal M}_K^{-1}$, where
${\cal M}_K= \bar K'^{1/4} {\cal M}$. Inside galaxy clusters, this is a short scale around 1 mm
and we can completely neglect quantum corrections on cluster scales.

\bibliography{ref1}   

\begin{thebibliography}{96}
\expandafter\ifx\csname natexlab\endcsname\relax\def\natexlab#1{#1}\fi
\expandafter\ifx\csname bibnamefont\endcsname\relax
  \def\bibnamefont#1{#1}\fi
\expandafter\ifx\csname bibfnamefont\endcsname\relax
  \def\bibfnamefont#1{#1}\fi
\expandafter\ifx\csname citenamefont\endcsname\relax
  \def\citenamefont#1{#1}\fi
\expandafter\ifx\csname url\endcsname\relax
  \def\url#1{\texttt{#1}}\fi
\expandafter\ifx\csname urlprefix\endcsname\relax\def\urlprefix{URL }\fi
\providecommand{\bibinfo}[2]{#2}
\providecommand{\eprint}[2][]{\url{#2}}

\bibitem[{\citenamefont{Babichev et~al.}(2009)\citenamefont{Babichev, Deffayet,
  and Ziour}}]{Babichev:2009ee}
\bibinfo{author}{\bibfnamefont{E.}~\bibnamefont{Babichev}},
  \bibinfo{author}{\bibfnamefont{C.}~\bibnamefont{Deffayet}}, \bibnamefont{and}
  \bibinfo{author}{\bibfnamefont{R.}~\bibnamefont{Ziour}},
  \bibinfo{journal}{Int.J.Mod.Phys.} \textbf{\bibinfo{volume}{D18}},
  \bibinfo{pages}{2147} (\bibinfo{year}{2009}), \eprint{0905.2943}.

\bibitem[{\citenamefont{Brax et~al.}(2013)\citenamefont{Brax, Burrage, and
  Davis}}]{Brax:2012jr}
\bibinfo{author}{\bibfnamefont{P.}~\bibnamefont{Brax}},
  \bibinfo{author}{\bibfnamefont{C.}~\bibnamefont{Burrage}}, \bibnamefont{and}
  \bibinfo{author}{\bibfnamefont{A.-C.} \bibnamefont{Davis}},
  \bibinfo{journal}{JCAP} \textbf{\bibinfo{volume}{1301}}, \bibinfo{pages}{020}
  (\bibinfo{year}{2013}), \eprint{1209.1293}.

\bibitem[{\citenamefont{Khoury and
  Weltman}(2004{\natexlab{a}})}]{Khoury:2003aq}
\bibinfo{author}{\bibfnamefont{J.}~\bibnamefont{Khoury}} \bibnamefont{and}
  \bibinfo{author}{\bibfnamefont{A.}~\bibnamefont{Weltman}},
  \bibinfo{journal}{Phys.Rev.Lett.} \textbf{\bibinfo{volume}{93}},
  \bibinfo{pages}{171104} (\bibinfo{year}{2004}{\natexlab{a}}),
  \eprint{astro-ph/0309300}.

\bibitem[{\citenamefont{Khoury and
  Weltman}(2004{\natexlab{b}})}]{Khoury:2003rn}
\bibinfo{author}{\bibfnamefont{J.}~\bibnamefont{Khoury}} \bibnamefont{and}
  \bibinfo{author}{\bibfnamefont{A.}~\bibnamefont{Weltman}},
  \bibinfo{journal}{Phys. Rev.} \textbf{\bibinfo{volume}{D69}},
  \bibinfo{pages}{044026} (\bibinfo{year}{2004}{\natexlab{b}}),
  \eprint{astro-ph/0309411}.

\bibitem[{\citenamefont{Damour and Polyakov}(1994)}]{Damour:1994zq}
\bibinfo{author}{\bibfnamefont{T.}~\bibnamefont{Damour}} \bibnamefont{and}
  \bibinfo{author}{\bibfnamefont{A.~M.} \bibnamefont{Polyakov}},
  \bibinfo{journal}{Nucl. Phys.} \textbf{\bibinfo{volume}{B423}},
  \bibinfo{pages}{532} (\bibinfo{year}{1994}), \eprint{hep-th/9401069}.

\bibitem[{\citenamefont{Vainshtein}(1972)}]{Vainshtein:1972sx}
\bibinfo{author}{\bibfnamefont{A.}~\bibnamefont{Vainshtein}},
  \bibinfo{journal}{Phys.Lett.} \textbf{\bibinfo{volume}{B39}},
  \bibinfo{pages}{393} (\bibinfo{year}{1972}).

\bibitem[{\citenamefont{{Brax} and
  {Valageas}}(2014{\natexlab{a}})}]{Brax:2014a}
\bibinfo{author}{\bibfnamefont{P.}~\bibnamefont{{Brax}}} \bibnamefont{and}
  \bibinfo{author}{\bibfnamefont{P.}~\bibnamefont{{Valageas}}},
  \bibinfo{journal}{\prd} \textbf{\bibinfo{volume}{90}}, \bibinfo{eid}{023507}
  (\bibinfo{year}{2014}{\natexlab{a}}), \eprint{1403.5420}.

\bibitem[{\citenamefont{{Brax} and
  {Valageas}}(2014{\natexlab{b}})}]{Brax:2014b}
\bibinfo{author}{\bibfnamefont{P.}~\bibnamefont{{Brax}}} \bibnamefont{and}
  \bibinfo{author}{\bibfnamefont{P.}~\bibnamefont{{Valageas}}},
  \bibinfo{journal}{\prd} \textbf{\bibinfo{volume}{90}}, \bibinfo{eid}{023508}
  (\bibinfo{year}{2014}{\natexlab{b}}), \eprint{1403.5424}.

\bibitem[{\citenamefont{{Barreira}
  et~al.}(2014{\natexlab{a}})\citenamefont{{Barreira}, {Brax}, {Clesse}, {Li},
  and {Valageas}}}]{Barreira:2014a}
\bibinfo{author}{\bibfnamefont{A.}~\bibnamefont{{Barreira}}},
  \bibinfo{author}{\bibfnamefont{P.}~\bibnamefont{{Brax}}},
  \bibinfo{author}{\bibfnamefont{S.}~\bibnamefont{{Clesse}}},
  \bibinfo{author}{\bibfnamefont{B.}~\bibnamefont{{Li}}}, \bibnamefont{and}
  \bibinfo{author}{\bibfnamefont{P.}~\bibnamefont{{Valageas}}},
  \bibinfo{journal}{ArXiv e-prints}  (\bibinfo{year}{2014}{\natexlab{a}}),
  \eprint{1411.5965}.

\bibitem[{\citenamefont{{Sawicki} et~al.}(2013)\citenamefont{{Sawicki},
  {Saltas}, {Amendola}, and {Kunz}}}]{Sawicki:2013}
\bibinfo{author}{\bibfnamefont{I.}~\bibnamefont{{Sawicki}}},
  \bibinfo{author}{\bibfnamefont{I.~D.} \bibnamefont{{Saltas}}},
  \bibinfo{author}{\bibfnamefont{L.}~\bibnamefont{{Amendola}}},
  \bibnamefont{and} \bibinfo{author}{\bibfnamefont{M.}~\bibnamefont{{Kunz}}},
  \bibinfo{journal}{\jcap} \textbf{\bibinfo{volume}{1}}, \bibinfo{eid}{004}
  (\bibinfo{year}{2013}), \eprint{1208.4855}.

\bibitem[{\citenamefont{{Brax} and
  {Valageas}}(2014{\natexlab{c}})}]{Brax:2014c}
\bibinfo{author}{\bibfnamefont{P.}~\bibnamefont{{Brax}}} \bibnamefont{and}
  \bibinfo{author}{\bibfnamefont{P.}~\bibnamefont{{Valageas}}},
  \bibinfo{journal}{\prd} \textbf{\bibinfo{volume}{90}}, \bibinfo{eid}{123521}
  (\bibinfo{year}{2014}{\natexlab{c}}), \eprint{1408.0969}.

\bibitem[{\citenamefont{{Barreira} et~al.}(2015)\citenamefont{{Barreira},
  {Brax}, {Clesse}, {Li}, and {Valageas}}}]{Barreira2015}
\bibinfo{author}{\bibfnamefont{A.}~\bibnamefont{{Barreira}}},
  \bibinfo{author}{\bibfnamefont{P.}~\bibnamefont{{Brax}}},
  \bibinfo{author}{\bibfnamefont{S.}~\bibnamefont{{Clesse}}},
  \bibinfo{author}{\bibfnamefont{B.}~\bibnamefont{{Li}}}, \bibnamefont{and}
  \bibinfo{author}{\bibfnamefont{P.}~\bibnamefont{{Valageas}}},
  \bibinfo{journal}{ArXiv e-prints}  (\bibinfo{year}{2015}),
  \eprint{1504.01493}.

\bibitem[{\citenamefont{{Sunyaev} and {Zeldovich}}(1972)}]{Sunyaev1972}
\bibinfo{author}{\bibfnamefont{R.~A.} \bibnamefont{{Sunyaev}}}
  \bibnamefont{and} \bibinfo{author}{\bibfnamefont{Y.~B.}
  \bibnamefont{{Zeldovich}}}, \bibinfo{journal}{Comments on Astrophysics and
  Space Physics} \textbf{\bibinfo{volume}{4}}, \bibinfo{pages}{173}
  (\bibinfo{year}{1972}).

\bibitem[{\citenamefont{{Caldwell} et~al.}(1998)\citenamefont{{Caldwell},
  {Dave}, and {Steinhardt}}}]{Caldwel1998}
\bibinfo{author}{\bibfnamefont{R.~R.} \bibnamefont{{Caldwell}}},
  \bibinfo{author}{\bibfnamefont{R.}~\bibnamefont{{Dave}}}, \bibnamefont{and}
  \bibinfo{author}{\bibfnamefont{P.~J.} \bibnamefont{{Steinhardt}}},
  \bibinfo{journal}{Physical Review Letters} \textbf{\bibinfo{volume}{80}},
  \bibinfo{pages}{1582} (\bibinfo{year}{1998}), \eprint{astro-ph/9708069}.

\bibitem[{\citenamefont{{Steinhardt} et~al.}(1999)\citenamefont{{Steinhardt},
  {Wang}, and {Zlatev}}}]{Steinhardt1999}
\bibinfo{author}{\bibfnamefont{P.~J.} \bibnamefont{{Steinhardt}}},
  \bibinfo{author}{\bibfnamefont{L.}~\bibnamefont{{Wang}}}, \bibnamefont{and}
  \bibinfo{author}{\bibfnamefont{I.}~\bibnamefont{{Zlatev}}},
  \bibinfo{journal}{\prd} \textbf{\bibinfo{volume}{59}}, \bibinfo{eid}{123504}
  (\bibinfo{year}{1999}), \eprint{astro-ph/9812313}.

\bibitem[{\citenamefont{{Esposito-Far{\`e}se} and
  {Polarski}}(2001)}]{Esposito2001}
\bibinfo{author}{\bibfnamefont{G.}~\bibnamefont{{Esposito-Far{\`e}se}}}
  \bibnamefont{and}
  \bibinfo{author}{\bibfnamefont{D.}~\bibnamefont{{Polarski}}},
  \bibinfo{journal}{\prd} \textbf{\bibinfo{volume}{63}}, \bibinfo{eid}{063504}
  (\bibinfo{year}{2001}), \eprint{gr-qc/0009034}.

\bibitem[{\citenamefont{{Brax} and {Burrage}}(2014)}]{Brax:2014d}
\bibinfo{author}{\bibfnamefont{P.}~\bibnamefont{{Brax}}} \bibnamefont{and}
  \bibinfo{author}{\bibfnamefont{C.}~\bibnamefont{{Burrage}}},
  \bibinfo{journal}{\prd} \textbf{\bibinfo{volume}{90}}, \bibinfo{eid}{104009}
  (\bibinfo{year}{2014}), \eprint{1407.1861}.

\bibitem[{\citenamefont{{Brax} and {Burrage}}(2015)}]{Brax:2015a}
\bibinfo{author}{\bibfnamefont{P.}~\bibnamefont{{Brax}}} \bibnamefont{and}
  \bibinfo{author}{\bibfnamefont{C.}~\bibnamefont{{Burrage}}},
  \bibinfo{journal}{\prd} \textbf{\bibinfo{volume}{91}}, \bibinfo{eid}{043515}
  (\bibinfo{year}{2015}), \eprint{1407.2376}.

\bibitem[{\citenamefont{{Eisenstein} et~al.}(1998)\citenamefont{{Eisenstein},
  {Hu}, and {Tegmark}}}]{Eisenstein1998a}
\bibinfo{author}{\bibfnamefont{D.~J.} \bibnamefont{{Eisenstein}}},
  \bibinfo{author}{\bibfnamefont{W.}~\bibnamefont{{Hu}}}, \bibnamefont{and}
  \bibinfo{author}{\bibfnamefont{M.}~\bibnamefont{{Tegmark}}},
  \bibinfo{journal}{\apjl} \textbf{\bibinfo{volume}{504}},
  \bibinfo{pages}{L57+} (\bibinfo{year}{1998}),
  \eprint{arXiv:astro-ph/9805239}.

\bibitem[{\citenamefont{{Eisenstein} et~al.}(2005)\citenamefont{{Eisenstein},
  {Zehavi}, {Hogg}, {Scoccimarro}, {Blanton}, {Nichol}, {Scranton}, {Seo},
  {Tegmark}, {Zheng} et~al.}}]{Eisenstein2005}
\bibinfo{author}{\bibfnamefont{D.~J.} \bibnamefont{{Eisenstein}}},
  \bibinfo{author}{\bibfnamefont{I.}~\bibnamefont{{Zehavi}}},
  \bibinfo{author}{\bibfnamefont{D.~W.} \bibnamefont{{Hogg}}},
  \bibinfo{author}{\bibfnamefont{R.}~\bibnamefont{{Scoccimarro}}},
  \bibinfo{author}{\bibfnamefont{M.~R.} \bibnamefont{{Blanton}}},
  \bibinfo{author}{\bibfnamefont{R.~C.} \bibnamefont{{Nichol}}},
  \bibinfo{author}{\bibfnamefont{R.}~\bibnamefont{{Scranton}}},
  \bibinfo{author}{\bibfnamefont{H.-J.} \bibnamefont{{Seo}}},
  \bibinfo{author}{\bibfnamefont{M.}~\bibnamefont{{Tegmark}}},
  \bibinfo{author}{\bibfnamefont{Z.}~\bibnamefont{{Zheng}}},
  \bibnamefont{et~al.}, \bibinfo{journal}{\apj} \textbf{\bibinfo{volume}{633}},
  \bibinfo{pages}{560} (\bibinfo{year}{2005}), \eprint{arXiv:astro-ph/0501171}.

\bibitem[{\citenamefont{C. and {Paczynski}}(1979)}]{Alcock1979}
\bibinfo{author}{\bibfnamefont{A.}~\bibnamefont{C.}} \bibnamefont{and}
  \bibinfo{author}{\bibfnamefont{B.}~\bibnamefont{{Paczynski}}},
  \bibinfo{journal}{Nature} \textbf{\bibinfo{volume}{281}},
  \bibinfo{pages}{358} (\bibinfo{year}{1979}).

\bibitem[{\citenamefont{{Valageas}}(2009)}]{Valageas2009}
\bibinfo{author}{\bibfnamefont{P.}~\bibnamefont{{Valageas}}},
  \bibinfo{journal}{\aap} \textbf{\bibinfo{volume}{508}}, \bibinfo{pages}{93}
  (\bibinfo{year}{2009}), \eprint{0905.2277}.

\bibitem[{\citenamefont{{Valageas}}(2014)}]{Valageas2014}
\bibinfo{author}{\bibfnamefont{P.}~\bibnamefont{{Valageas}}},
  \bibinfo{journal}{Phys. Rev. D} \textbf{\bibinfo{volume}{89}},
  \bibinfo{pages}{123522} (\bibinfo{year}{2014}).

\bibitem[{\citenamefont{{Nishimichi} and {Valageas}}(2014)}]{Nishimichi2014}
\bibinfo{author}{\bibfnamefont{T.}~\bibnamefont{{Nishimichi}}}
  \bibnamefont{and}
  \bibinfo{author}{\bibfnamefont{P.}~\bibnamefont{{Valageas}}},
  \bibinfo{journal}{Phys. Rev. D} \textbf{\bibinfo{volume}{90}},
  \bibinfo{pages}{023546} (\bibinfo{year}{2014}).

\bibitem[{\citenamefont{{Kehagias} et~al.}(2014)\citenamefont{{Kehagias},
  {Perrier}, and {Riotto}}}]{Kehagias2014}
\bibinfo{author}{\bibfnamefont{A.}~\bibnamefont{{Kehagias}}},
  \bibinfo{author}{\bibfnamefont{H.}~\bibnamefont{{Perrier}}},
  \bibnamefont{and} \bibinfo{author}{\bibfnamefont{A.}~\bibnamefont{{Riotto}}},
  \bibinfo{journal}{Modern Physics Letters A} \textbf{\bibinfo{volume}{29}},
  \bibinfo{pages}{1450152} (\bibinfo{year}{2014}).

\bibitem[{\citenamefont{{Navarro} et~al.}(1996)\citenamefont{{Navarro},
  {Frenk}, and {White}}}]{Navarro:1996}
\bibinfo{author}{\bibfnamefont{J.~F.} \bibnamefont{{Navarro}}},
  \bibinfo{author}{\bibfnamefont{C.~S.} \bibnamefont{{Frenk}}},
  \bibnamefont{and} \bibinfo{author}{\bibfnamefont{S.~D.~M.}
  \bibnamefont{{White}}}, \bibinfo{journal}{\apj}
  \textbf{\bibinfo{volume}{462}}, \bibinfo{pages}{563} (\bibinfo{year}{1996}),
  \eprint{astro-ph/9508025}.

\bibitem[{\citenamefont{{Valageas} et~al.}(2013)\citenamefont{{Valageas},
  {Nishimichi}, and {Taruya}}}]{Valageas2013}
\bibinfo{author}{\bibfnamefont{P.}~\bibnamefont{{Valageas}}},
  \bibinfo{author}{\bibfnamefont{T.}~\bibnamefont{{Nishimichi}}},
  \bibnamefont{and} \bibinfo{author}{\bibfnamefont{A.}~\bibnamefont{{Taruya}}},
  \bibinfo{journal}{ArXiv e-prints}  (\bibinfo{year}{2013}),
  \eprint{1302.4533}.

\bibitem[{\citenamefont{{Valageas}}(2013)}]{Valageas2013a}
\bibinfo{author}{\bibfnamefont{P.}~\bibnamefont{{Valageas}}},
  \bibinfo{journal}{Phys. Rev. D} \textbf{\bibinfo{volume}{88}},
  \bibinfo{pages}{083524} (\bibinfo{year}{2013}).

\bibitem[{\citenamefont{{Kaiser}}(1986)}]{Kaiser1986}
\bibinfo{author}{\bibfnamefont{N.}~\bibnamefont{{Kaiser}}},
  \bibinfo{journal}{Mont. Not. Roy. Astron. Soc.}
  \textbf{\bibinfo{volume}{222}}, \bibinfo{pages}{323} (\bibinfo{year}{1986}).

\bibitem[{\citenamefont{{Merten} et~al.}(2014)\citenamefont{{Merten},
  {Meneghetti}, {Postman}, {Umetsu}, {Zitrin}, {Medezinski}, {Nonino},
  {Koekemoer}, {Melchior}, {Gruen} et~al.}}]{Merten2014}
\bibinfo{author}{\bibfnamefont{J.}~\bibnamefont{{Merten}}},
  \bibinfo{author}{\bibfnamefont{M.}~\bibnamefont{{Meneghetti}}},
  \bibinfo{author}{\bibfnamefont{M.}~\bibnamefont{{Postman}}},
  \bibinfo{author}{\bibfnamefont{K.}~\bibnamefont{{Umetsu}}},
  \bibinfo{author}{\bibfnamefont{A.}~\bibnamefont{{Zitrin}}},
  \bibinfo{author}{\bibfnamefont{E.}~\bibnamefont{{Medezinski}}},
  \bibinfo{author}{\bibfnamefont{M.}~\bibnamefont{{Nonino}}},
  \bibinfo{author}{\bibfnamefont{A.}~\bibnamefont{{Koekemoer}}},
  \bibinfo{author}{\bibfnamefont{P.}~\bibnamefont{{Melchior}}},
  \bibinfo{author}{\bibfnamefont{D.}~\bibnamefont{{Gruen}}},
  \bibnamefont{et~al.}, \bibinfo{journal}{ArXiv e-prints}
  (\bibinfo{year}{2014}), \eprint{1404.1376}.

\bibitem[{\citenamefont{{Navarro} et~al.}(1997)\citenamefont{{Navarro},
  {Frenk}, and {White}}}]{Navarro1997}
\bibinfo{author}{\bibfnamefont{J.~F.} \bibnamefont{{Navarro}}},
  \bibinfo{author}{\bibfnamefont{C.~S.} \bibnamefont{{Frenk}}},
  \bibnamefont{and} \bibinfo{author}{\bibfnamefont{S.~D.~M.}
  \bibnamefont{{White}}}, \bibinfo{journal}{\apj}
  \textbf{\bibinfo{volume}{490}}, \bibinfo{pages}{493} (\bibinfo{year}{1997}),
  \eprint{arXiv:astro-ph/9611107}.

\bibitem[{\citenamefont{{Mo} et~al.}(2010)\citenamefont{{Mo}, {van den Bosch},
  and {White}}}]{Mo2010}
\bibinfo{author}{\bibfnamefont{H.}~\bibnamefont{{Mo}}},
  \bibinfo{author}{\bibfnamefont{F.~C.} \bibnamefont{{van den Bosch}}},
  \bibnamefont{and} \bibinfo{author}{\bibfnamefont{S.}~\bibnamefont{{White}}},
  \emph{\bibinfo{title}{{Galaxy Formation and Evolution}}}
  (\bibinfo{publisher}{Cambridge University Press}, \bibinfo{year}{2010}).

\bibitem[{\citenamefont{{Binney} and {Tremaine}}(1987)}]{Binney1987}
\bibinfo{author}{\bibfnamefont{J.}~\bibnamefont{{Binney}}} \bibnamefont{and}
  \bibinfo{author}{\bibfnamefont{S.}~\bibnamefont{{Tremaine}}},
  \emph{\bibinfo{title}{{Galactic dynamics}}} (\bibinfo{publisher}{Princeton
  University Press, Princeton, NJ, USA}, \bibinfo{year}{1987}).

\bibitem[{\citenamefont{{Cavaliere} and {Fusco-Femiano}}(1976)}]{Cavaliere1976}
\bibinfo{author}{\bibfnamefont{A.}~\bibnamefont{{Cavaliere}}} \bibnamefont{and}
  \bibinfo{author}{\bibfnamefont{R.}~\bibnamefont{{Fusco-Femiano}}},
  \bibinfo{journal}{\aap} \textbf{\bibinfo{volume}{49}}, \bibinfo{pages}{137}
  (\bibinfo{year}{1976}).

\bibitem[{\citenamefont{{Plionis}}(2008)}]{Plionis2008}
\bibinfo{author}{\bibfnamefont{M.}~\bibnamefont{{Plionis}}},
  \bibinfo{journal}{Lect. Notes Phys.} \textbf{\bibinfo{volume}{740}}
  (\bibinfo{year}{2008}).

\bibitem[{\citenamefont{{Vikhlinin} et~al.}(2009)\citenamefont{{Vikhlinin},
  {Burenin}, {Ebeling}, {Forman}, {Hornstrup}, {Jones}, {Kravtsov}, {Murray},
  {Nagai}, {Quintana} et~al.}}]{Vikhlinin2009b}
\bibinfo{author}{\bibfnamefont{A.}~\bibnamefont{{Vikhlinin}}},
  \bibinfo{author}{\bibfnamefont{R.~A.} \bibnamefont{{Burenin}}},
  \bibinfo{author}{\bibfnamefont{H.}~\bibnamefont{{Ebeling}}},
  \bibinfo{author}{\bibfnamefont{W.~R.} \bibnamefont{{Forman}}},
  \bibinfo{author}{\bibfnamefont{A.}~\bibnamefont{{Hornstrup}}},
  \bibinfo{author}{\bibfnamefont{C.}~\bibnamefont{{Jones}}},
  \bibinfo{author}{\bibfnamefont{A.~V.} \bibnamefont{{Kravtsov}}},
  \bibinfo{author}{\bibfnamefont{S.~S.} \bibnamefont{{Murray}}},
  \bibinfo{author}{\bibfnamefont{D.}~\bibnamefont{{Nagai}}},
  \bibinfo{author}{\bibfnamefont{H.}~\bibnamefont{{Quintana}}},
  \bibnamefont{et~al.}, \bibinfo{journal}{\apj} \textbf{\bibinfo{volume}{692}},
  \bibinfo{pages}{1033} (\bibinfo{year}{2009}), \eprint{0805.2207}.

\bibitem[{\citenamefont{{Zhang} et~al.}(2008)\citenamefont{{Zhang},
  {Finoguenov}, {B{\"o}hringer}, {Kneib}, {Smith}, {Kneissl}, {Okabe}, {Dahle},
  and {Reiprich}}}]{Zhang2008}
\bibinfo{author}{\bibfnamefont{Y.-Y.} \bibnamefont{{Zhang}}},
  \bibinfo{author}{\bibfnamefont{A.}~\bibnamefont{{Finoguenov}}},
  \bibinfo{author}{\bibfnamefont{H.}~\bibnamefont{{B{\"o}hringer}}},
  \bibinfo{author}{\bibfnamefont{J.-P.} \bibnamefont{{Kneib}}},
  \bibinfo{author}{\bibfnamefont{G.~P.} \bibnamefont{{Smith}}},
  \bibinfo{author}{\bibfnamefont{R.}~\bibnamefont{{Kneissl}}},
  \bibinfo{author}{\bibfnamefont{N.}~\bibnamefont{{Okabe}}},
  \bibinfo{author}{\bibfnamefont{H.}~\bibnamefont{{Dahle}}}, \bibnamefont{and}
  \bibinfo{author}{\bibfnamefont{T.~H.} \bibnamefont{{Reiprich}}}, in
  \emph{\bibinfo{booktitle}{The X-ray Universe 2008}} (\bibinfo{year}{2008}),
  p. \bibinfo{pages}{222}.

\bibitem[{\citenamefont{{Mantz} et~al.}(2010)\citenamefont{{Mantz}, {Allen},
  {Ebeling}, {Rapetti}, and {Drlica-Wagner}}}]{Mantz2010}
\bibinfo{author}{\bibfnamefont{A.}~\bibnamefont{{Mantz}}},
  \bibinfo{author}{\bibfnamefont{S.~W.} \bibnamefont{{Allen}}},
  \bibinfo{author}{\bibfnamefont{H.}~\bibnamefont{{Ebeling}}},
  \bibinfo{author}{\bibfnamefont{D.}~\bibnamefont{{Rapetti}}},
  \bibnamefont{and}
  \bibinfo{author}{\bibfnamefont{A.}~\bibnamefont{{Drlica-Wagner}}},
  \bibinfo{journal}{\mnras} \textbf{\bibinfo{volume}{406}},
  \bibinfo{pages}{1773} (\bibinfo{year}{2010}), \eprint{0909.3099}.

\bibitem[{\citenamefont{{Pratt} et~al.}(2009)\citenamefont{{Pratt}, {Croston},
  {Arnaud}, and {B{\"o}hringer}}}]{Pratt2009}
\bibinfo{author}{\bibfnamefont{G.~W.} \bibnamefont{{Pratt}}},
  \bibinfo{author}{\bibfnamefont{J.~H.} \bibnamefont{{Croston}}},
  \bibinfo{author}{\bibfnamefont{M.}~\bibnamefont{{Arnaud}}}, \bibnamefont{and}
  \bibinfo{author}{\bibfnamefont{H.}~\bibnamefont{{B{\"o}hringer}}},
  \bibinfo{journal}{\aap} \textbf{\bibinfo{volume}{498}}, \bibinfo{pages}{361}
  (\bibinfo{year}{2009}), \eprint{0809.3784}.

\bibitem[{\citenamefont{{Ikebe} et~al.}(2002)\citenamefont{{Ikebe}, {Reiprich},
  {B{\"o}hringer}, {Tanaka}, and {Kitayama}}}]{Ikebe2002}
\bibinfo{author}{\bibfnamefont{Y.}~\bibnamefont{{Ikebe}}},
  \bibinfo{author}{\bibfnamefont{T.~H.} \bibnamefont{{Reiprich}}},
  \bibinfo{author}{\bibfnamefont{H.}~\bibnamefont{{B{\"o}hringer}}},
  \bibinfo{author}{\bibfnamefont{Y.}~\bibnamefont{{Tanaka}}}, \bibnamefont{and}
  \bibinfo{author}{\bibfnamefont{T.}~\bibnamefont{{Kitayama}}},
  \bibinfo{journal}{\aap} \textbf{\bibinfo{volume}{383}}, \bibinfo{pages}{773}
  (\bibinfo{year}{2002}), \eprint{astro-ph/0112315}.

\bibitem[{\citenamefont{{Bender} et~al.}(2014)\citenamefont{{Bender},
  {Kennedy}, {Ade}, {Basu}, {Bertoldi}, {Burkutean}, {Clarke}, {Dahlin},
  {Dobbs}, {Ferrusca} et~al.}}]{Bender2014}
\bibinfo{author}{\bibfnamefont{A.~N.} \bibnamefont{{Bender}}},
  \bibinfo{author}{\bibfnamefont{J.}~\bibnamefont{{Kennedy}}},
  \bibinfo{author}{\bibfnamefont{P.~A.~R.} \bibnamefont{{Ade}}},
  \bibinfo{author}{\bibfnamefont{K.}~\bibnamefont{{Basu}}},
  \bibinfo{author}{\bibfnamefont{F.}~\bibnamefont{{Bertoldi}}},
  \bibinfo{author}{\bibfnamefont{S.}~\bibnamefont{{Burkutean}}},
  \bibinfo{author}{\bibfnamefont{J.}~\bibnamefont{{Clarke}}},
  \bibinfo{author}{\bibfnamefont{D.}~\bibnamefont{{Dahlin}}},
  \bibinfo{author}{\bibfnamefont{M.}~\bibnamefont{{Dobbs}}},
  \bibinfo{author}{\bibfnamefont{D.}~\bibnamefont{{Ferrusca}}},
  \bibnamefont{et~al.}, \bibinfo{journal}{ArXiv e-prints}
  (\bibinfo{year}{2014}), \eprint{1404.7103}.

\bibitem[{\citenamefont{{Cole} and {Kaiser}}(1989)}]{Cole1989}
\bibinfo{author}{\bibfnamefont{S.}~\bibnamefont{{Cole}}} \bibnamefont{and}
  \bibinfo{author}{\bibfnamefont{N.}~\bibnamefont{{Kaiser}}},
  \bibinfo{journal}{\mnras} \textbf{\bibinfo{volume}{237}},
  \bibinfo{pages}{1127} (\bibinfo{year}{1989}).

\bibitem[{\citenamefont{{Mo} and {White}}(1996)}]{Mo1996}
\bibinfo{author}{\bibfnamefont{H.~J.} \bibnamefont{{Mo}}} \bibnamefont{and}
  \bibinfo{author}{\bibfnamefont{S.~D.~M.} \bibnamefont{{White}}},
  \bibinfo{journal}{\mnras} \textbf{\bibinfo{volume}{282}},
  \bibinfo{pages}{347} (\bibinfo{year}{1996}), \eprint{arXiv:astro-ph/9512127}.

\bibitem[{\citenamefont{{Sheth} et~al.}(2001)\citenamefont{{Sheth}, {Mo}, and
  {Tormen}}}]{Sheth2001d}
\bibinfo{author}{\bibfnamefont{R.~K.} \bibnamefont{{Sheth}}},
  \bibinfo{author}{\bibfnamefont{H.~J.} \bibnamefont{{Mo}}}, \bibnamefont{and}
  \bibinfo{author}{\bibfnamefont{G.}~\bibnamefont{{Tormen}}},
  \bibinfo{journal}{\mnras} \textbf{\bibinfo{volume}{323}}, \bibinfo{pages}{1}
  (\bibinfo{year}{2001}), \eprint{arXiv:astro-ph/9907024}.

\bibitem[{\citenamefont{{Tinker} et~al.}(2010)\citenamefont{{Tinker},
  {Robertson}, {Kravtsov}, {Klypin}, {Warren}, {Yepes}, and
  {Gottl{\"o}ber}}}]{Tinker2010}
\bibinfo{author}{\bibfnamefont{J.~L.} \bibnamefont{{Tinker}}},
  \bibinfo{author}{\bibfnamefont{B.~E.} \bibnamefont{{Robertson}}},
  \bibinfo{author}{\bibfnamefont{A.~V.} \bibnamefont{{Kravtsov}}},
  \bibinfo{author}{\bibfnamefont{A.}~\bibnamefont{{Klypin}}},
  \bibinfo{author}{\bibfnamefont{M.~S.} \bibnamefont{{Warren}}},
  \bibinfo{author}{\bibfnamefont{G.}~\bibnamefont{{Yepes}}}, \bibnamefont{and}
  \bibinfo{author}{\bibfnamefont{S.}~\bibnamefont{{Gottl{\"o}ber}}},
  \bibinfo{journal}{\apj} \textbf{\bibinfo{volume}{724}}, \bibinfo{pages}{878}
  (\bibinfo{year}{2010}), \eprint{1001.3162}.

\bibitem[{\citenamefont{{Brax} et~al.}(2012{\natexlab{a}})\citenamefont{{Brax},
  {Davis}, {Li}, and {Winther}}}]{Brax2012b}
\bibinfo{author}{\bibfnamefont{P.}~\bibnamefont{{Brax}}},
  \bibinfo{author}{\bibfnamefont{A.-C.} \bibnamefont{{Davis}}},
  \bibinfo{author}{\bibfnamefont{B.}~\bibnamefont{{Li}}}, \bibnamefont{and}
  \bibinfo{author}{\bibfnamefont{H.~A.} \bibnamefont{{Winther}}},
  \bibinfo{journal}{\prd} \textbf{\bibinfo{volume}{86}}, \bibinfo{eid}{044015}
  (\bibinfo{year}{2012}{\natexlab{a}}), \eprint{1203.4812}.

\bibitem[{\citenamefont{Khoury}(2013)}]{Khoury:2013tda}
\bibinfo{author}{\bibfnamefont{J.}~\bibnamefont{Khoury}}
  (\bibinfo{year}{2013}), \eprint{1312.2006}.

\bibitem[{\citenamefont{Brax}(2013)}]{Brax:2013ida}
\bibinfo{author}{\bibfnamefont{P.}~\bibnamefont{Brax}},
  \bibinfo{journal}{Class.Quant.Grav.} \textbf{\bibinfo{volume}{30}},
  \bibinfo{pages}{214005} (\bibinfo{year}{2013}).

\bibitem[{\citenamefont{{Hu} and {Sawicki}}(2007)}]{Hu2007a}
\bibinfo{author}{\bibfnamefont{W.}~\bibnamefont{{Hu}}} \bibnamefont{and}
  \bibinfo{author}{\bibfnamefont{I.}~\bibnamefont{{Sawicki}}},
  \bibinfo{journal}{\prd} \textbf{\bibinfo{volume}{76}}, \bibinfo{eid}{064004}
  (\bibinfo{year}{2007}), \eprint{0705.1158}.

\bibitem[{\citenamefont{Brax et~al.}(2008)\citenamefont{Brax, van~de Bruck,
  Davis, and Shaw}}]{Brax:2008hh}
\bibinfo{author}{\bibfnamefont{P.}~\bibnamefont{Brax}},
  \bibinfo{author}{\bibfnamefont{C.}~\bibnamefont{van~de Bruck}},
  \bibinfo{author}{\bibfnamefont{A.-C.} \bibnamefont{Davis}}, \bibnamefont{and}
  \bibinfo{author}{\bibfnamefont{D.~J.} \bibnamefont{Shaw}},
  \bibinfo{journal}{Phys.Rev.} \textbf{\bibinfo{volume}{D78}},
  \bibinfo{pages}{104021} (\bibinfo{year}{2008}), \eprint{0806.3415}.

\bibitem[{\citenamefont{Sotiriou and Faraoni}(2010)}]{Sotiriou:2008rp}
\bibinfo{author}{\bibfnamefont{T.~P.} \bibnamefont{Sotiriou}} \bibnamefont{and}
  \bibinfo{author}{\bibfnamefont{V.}~\bibnamefont{Faraoni}},
  \bibinfo{journal}{Rev.Mod.Phys.} \textbf{\bibinfo{volume}{82}},
  \bibinfo{pages}{451} (\bibinfo{year}{2010}), \eprint{0805.1726}.

\bibitem[{\citenamefont{{de Felice} and {Tsujikawa}}(2010)}]{Felice2010}
\bibinfo{author}{\bibfnamefont{A.}~\bibnamefont{{de Felice}}} \bibnamefont{and}
  \bibinfo{author}{\bibfnamefont{S.}~\bibnamefont{{Tsujikawa}}},
  \bibinfo{journal}{Living Re. Relativ.} \textbf{\bibinfo{volume}{13}},
  \bibinfo{pages}{3} (\bibinfo{year}{2010}).

\bibitem[{\citenamefont{Gannouji et~al.}(2012)\citenamefont{Gannouji, Sami, and
  Thongkool}}]{Gannouji:2012iy}
\bibinfo{author}{\bibfnamefont{R.}~\bibnamefont{Gannouji}},
  \bibinfo{author}{\bibfnamefont{M.}~\bibnamefont{Sami}}, \bibnamefont{and}
  \bibinfo{author}{\bibfnamefont{I.}~\bibnamefont{Thongkool}}
  (\bibinfo{year}{2012}), \eprint{1206.3395}.

\bibitem[{\citenamefont{{Chiba}}(2003)}]{Chiba2003}
\bibinfo{author}{\bibfnamefont{T.}~\bibnamefont{{Chiba}}},
  \bibinfo{journal}{Physics Letters B} \textbf{\bibinfo{volume}{575}},
  \bibinfo{pages}{1} (\bibinfo{year}{2003}).

\bibitem[{\citenamefont{Gasperini et~al.}(2002)\citenamefont{Gasperini, Piazza,
  and Veneziano}}]{Gasperini:2001pc}
\bibinfo{author}{\bibfnamefont{M.}~\bibnamefont{Gasperini}},
  \bibinfo{author}{\bibfnamefont{F.}~\bibnamefont{Piazza}}, \bibnamefont{and}
  \bibinfo{author}{\bibfnamefont{G.}~\bibnamefont{Veneziano}},
  \bibinfo{journal}{Phys.Rev.} \textbf{\bibinfo{volume}{D65}},
  \bibinfo{pages}{023508} (\bibinfo{year}{2002}), \eprint{gr-qc/0108016}.

\bibitem[{\citenamefont{Brax et~al.}(2010)\citenamefont{Brax, van~de Bruck,
  Davis, and Shaw}}]{Brax:2010gi}
\bibinfo{author}{\bibfnamefont{P.}~\bibnamefont{Brax}},
  \bibinfo{author}{\bibfnamefont{C.}~\bibnamefont{van~de Bruck}},
  \bibinfo{author}{\bibfnamefont{A.-C.} \bibnamefont{Davis}}, \bibnamefont{and}
  \bibinfo{author}{\bibfnamefont{D.}~\bibnamefont{Shaw}},
  \bibinfo{journal}{Phys. Rev.} \textbf{\bibinfo{volume}{D82}},
  \bibinfo{pages}{063519} (\bibinfo{year}{2010}), \eprint{1005.3735}.

\bibitem[{\citenamefont{Brax et~al.}(2011)\citenamefont{Brax, van~de Bruck,
  Davis, Li, and Shaw}}]{Brax:2011ja}
\bibinfo{author}{\bibfnamefont{P.}~\bibnamefont{Brax}},
  \bibinfo{author}{\bibfnamefont{C.}~\bibnamefont{van~de Bruck}},
  \bibinfo{author}{\bibfnamefont{A.-C.} \bibnamefont{Davis}},
  \bibinfo{author}{\bibfnamefont{B.}~\bibnamefont{Li}}, \bibnamefont{and}
  \bibinfo{author}{\bibfnamefont{D.~J.} \bibnamefont{Shaw}},
  \bibinfo{journal}{Phys.Rev.} \textbf{\bibinfo{volume}{D83}},
  \bibinfo{pages}{104026} (\bibinfo{year}{2011}), \eprint{1102.3692}.

\bibitem[{\citenamefont{Hinterbichler and Khoury}(2010)}]{Hinterbichler:2010es}
\bibinfo{author}{\bibfnamefont{K.}~\bibnamefont{Hinterbichler}}
  \bibnamefont{and} \bibinfo{author}{\bibfnamefont{J.}~\bibnamefont{Khoury}},
  \bibinfo{journal}{Phys. Rev. Lett.} \textbf{\bibinfo{volume}{104}},
  \bibinfo{pages}{231301} (\bibinfo{year}{2010}), \eprint{1001.4525}.

\bibitem[{\citenamefont{{Brax} et~al.}(2012{\natexlab{b}})\citenamefont{{Brax},
  {Davis}, {Li}, {Winther}, and {Zhao}}}]{Brax:2012nk}
\bibinfo{author}{\bibfnamefont{P.}~\bibnamefont{{Brax}}},
  \bibinfo{author}{\bibfnamefont{A.-C.} \bibnamefont{{Davis}}},
  \bibinfo{author}{\bibfnamefont{B.}~\bibnamefont{{Li}}},
  \bibinfo{author}{\bibfnamefont{H.~A.} \bibnamefont{{Winther}}},
  \bibnamefont{and} \bibinfo{author}{\bibfnamefont{G.-B.}
  \bibnamefont{{Zhao}}}, \bibinfo{journal}{\jcap}
  \textbf{\bibinfo{volume}{10}}, \bibinfo{eid}{002}
  (\bibinfo{year}{2012}{\natexlab{b}}), \eprint{1206.3568}.

\bibitem[{\citenamefont{Nicolis et~al.}(2009)\citenamefont{Nicolis, Rattazzi,
  and Trincherini}}]{Nicolis:2008in}
\bibinfo{author}{\bibfnamefont{A.}~\bibnamefont{Nicolis}},
  \bibinfo{author}{\bibfnamefont{R.}~\bibnamefont{Rattazzi}}, \bibnamefont{and}
  \bibinfo{author}{\bibfnamefont{E.}~\bibnamefont{Trincherini}},
  \bibinfo{journal}{Phys.Rev.} \textbf{\bibinfo{volume}{D79}},
  \bibinfo{pages}{064036} (\bibinfo{year}{2009}), \eprint{0811.2197}.

\bibitem[{\citenamefont{{Deffayet}
  et~al.}(2009{\natexlab{a}})\citenamefont{{Deffayet}, {Esposito-Far{\`e}se},
  and A.}}]{Deffayet2009}
\bibinfo{author}{\bibfnamefont{C.}~\bibnamefont{{Deffayet}}},
  \bibinfo{author}{\bibfnamefont{G.}~\bibnamefont{{Esposito-Far{\`e}se}}},
  \bibnamefont{and} \bibinfo{author}{\bibfnamefont{V.}~\bibnamefont{A.}},
  \bibinfo{journal}{Phys. Rev. D} \textbf{\bibinfo{volume}{79}},
  \bibinfo{pages}{084003} (\bibinfo{year}{2009}{\natexlab{a}}).

\bibitem[{\citenamefont{{Deffayet}
  et~al.}(2009{\natexlab{b}})\citenamefont{{Deffayet}, S., and
  {Esposito-Far{\`e}se}}}]{Deffayet2009a}
\bibinfo{author}{\bibfnamefont{C.}~\bibnamefont{{Deffayet}}},
  \bibinfo{author}{\bibfnamefont{D.}~\bibnamefont{S.}}, \bibnamefont{and}
  \bibinfo{author}{\bibfnamefont{G.}~\bibnamefont{{Esposito-Far{\`e}se}}},
  \bibinfo{journal}{Phys. Rev. D} \textbf{\bibinfo{volume}{80}},
  \bibinfo{pages}{064015} (\bibinfo{year}{2009}{\natexlab{b}}).

\bibitem[{\citenamefont{{De Felice} and {Tsujikawa}}(2010)}]{De-Felice2010a}
\bibinfo{author}{\bibfnamefont{A.}~\bibnamefont{{De Felice}}} \bibnamefont{and}
  \bibinfo{author}{\bibfnamefont{S.}~\bibnamefont{{Tsujikawa}}},
  \bibinfo{journal}{Phys. Rev. Lett.} \textbf{\bibinfo{volume}{105}},
  \bibinfo{pages}{111301} (\bibinfo{year}{2010}).

\bibitem[{\citenamefont{{De Felice} and {Tsujikawa}}(2012)}]{De-Felice2012}
\bibinfo{author}{\bibfnamefont{A.}~\bibnamefont{{De Felice}}} \bibnamefont{and}
  \bibinfo{author}{\bibfnamefont{S.}~\bibnamefont{{Tsujikawa}}},
  \bibinfo{journal}{\jcap} \textbf{\bibinfo{volume}{3}}, \bibinfo{eid}{025}
  (\bibinfo{year}{2012}), \eprint{1112.1774}.

\bibitem[{\citenamefont{{Barreira}
  et~al.}(2014{\natexlab{b}})\citenamefont{{Barreira}, {Li}, {Hellwing},
  {Lombriser}, {Baugh}, and {Pascoli}}}]{Barreira2014}
\bibinfo{author}{\bibfnamefont{A.}~\bibnamefont{{Barreira}}},
  \bibinfo{author}{\bibfnamefont{B.}~\bibnamefont{{Li}}},
  \bibinfo{author}{\bibfnamefont{W.~A.} \bibnamefont{{Hellwing}}},
  \bibinfo{author}{\bibfnamefont{L.}~\bibnamefont{{Lombriser}}},
  \bibinfo{author}{\bibfnamefont{C.~M.} \bibnamefont{{Baugh}}},
  \bibnamefont{and}
  \bibinfo{author}{\bibfnamefont{S.}~\bibnamefont{{Pascoli}}},
  \bibinfo{journal}{\jcap} \textbf{\bibinfo{volume}{4}}, \bibinfo{eid}{029}
  (\bibinfo{year}{2014}{\natexlab{b}}), \eprint{1401.1497}.

\bibitem[{\citenamefont{{Brax} and {Valageas}}(2013)}]{BraxPV2013}
\bibinfo{author}{\bibfnamefont{P.}~\bibnamefont{{Brax}}} \bibnamefont{and}
  \bibinfo{author}{\bibfnamefont{P.}~\bibnamefont{{Valageas}}},
  \bibinfo{journal}{\prd} \textbf{\bibinfo{volume}{88}}, \bibinfo{eid}{023527}
  (\bibinfo{year}{2013}), \eprint{1305.5647}.

\bibitem[{\citenamefont{{Terukina} and {Yamamoto}}(2012)}]{Terukina2012}
\bibinfo{author}{\bibfnamefont{A.}~\bibnamefont{{Terukina}}} \bibnamefont{and}
  \bibinfo{author}{\bibfnamefont{K.}~\bibnamefont{{Yamamoto}}},
  \bibinfo{journal}{\prd} \textbf{\bibinfo{volume}{86}}, \bibinfo{eid}{103503}
  (\bibinfo{year}{2012}), \eprint{1203.6163}.

\bibitem[{\citenamefont{{Arnold} et~al.}(2014)\citenamefont{{Arnold},
  {Puchwein}, and {Springel}}}]{Arnold2014}
\bibinfo{author}{\bibfnamefont{C.}~\bibnamefont{{Arnold}}},
  \bibinfo{author}{\bibfnamefont{E.}~\bibnamefont{{Puchwein}}},
  \bibnamefont{and}
  \bibinfo{author}{\bibfnamefont{V.}~\bibnamefont{{Springel}}},
  \bibinfo{journal}{\mnras} \textbf{\bibinfo{volume}{440}},
  \bibinfo{pages}{833} (\bibinfo{year}{2014}), \eprint{1311.5560}.

\bibitem[{\citenamefont{{Terukina} et~al.}(2014)\citenamefont{{Terukina},
  {Lombriser}, {Yamamoto}, {Bacon}, {Koyama}, and {Nichol}}}]{Terukina2014}
\bibinfo{author}{\bibfnamefont{A.}~\bibnamefont{{Terukina}}},
  \bibinfo{author}{\bibfnamefont{L.}~\bibnamefont{{Lombriser}}},
  \bibinfo{author}{\bibfnamefont{K.}~\bibnamefont{{Yamamoto}}},
  \bibinfo{author}{\bibfnamefont{D.}~\bibnamefont{{Bacon}}},
  \bibinfo{author}{\bibfnamefont{K.}~\bibnamefont{{Koyama}}}, \bibnamefont{and}
  \bibinfo{author}{\bibfnamefont{R.~C.} \bibnamefont{{Nichol}}},
  \bibinfo{journal}{\jcap} \textbf{\bibinfo{volume}{4}}, \bibinfo{eid}{013}
  (\bibinfo{year}{2014}), \eprint{1312.5083}.

\bibitem[{\citenamefont{{Wilcox} et~al.}(2015)\citenamefont{{Wilcox}, {Bacon},
  {Nichol}, {Rooney}, {Terukina}, {Romer}, {Koyama}, {Zhao}, {Hood}, {Mann}
  et~al.}}]{Wilcox2015}
\bibinfo{author}{\bibfnamefont{H.}~\bibnamefont{{Wilcox}}},
  \bibinfo{author}{\bibfnamefont{D.}~\bibnamefont{{Bacon}}},
  \bibinfo{author}{\bibfnamefont{R.~C.} \bibnamefont{{Nichol}}},
  \bibinfo{author}{\bibfnamefont{P.~J.} \bibnamefont{{Rooney}}},
  \bibinfo{author}{\bibfnamefont{A.}~\bibnamefont{{Terukina}}},
  \bibinfo{author}{\bibfnamefont{A.~K.} \bibnamefont{{Romer}}},
  \bibinfo{author}{\bibfnamefont{K.}~\bibnamefont{{Koyama}}},
  \bibinfo{author}{\bibfnamefont{G.-B.} \bibnamefont{{Zhao}}},
  \bibinfo{author}{\bibfnamefont{R.}~\bibnamefont{{Hood}}},
  \bibinfo{author}{\bibfnamefont{R.~G.} \bibnamefont{{Mann}}},
  \bibnamefont{et~al.}, \bibinfo{journal}{ArXiv e-prints}
  (\bibinfo{year}{2015}), \eprint{1504.03937}.

\bibitem[{\citenamefont{{Lombriser}
  et~al.}(2012{\natexlab{a}})\citenamefont{{Lombriser}, {Schmidt}, {Baldauf},
  {Mandelbaum}, {Seljak}, and {Smith}}}]{Lombriser2012}
\bibinfo{author}{\bibfnamefont{L.}~\bibnamefont{{Lombriser}}},
  \bibinfo{author}{\bibfnamefont{F.}~\bibnamefont{{Schmidt}}},
  \bibinfo{author}{\bibfnamefont{T.}~\bibnamefont{{Baldauf}}},
  \bibinfo{author}{\bibfnamefont{R.}~\bibnamefont{{Mandelbaum}}},
  \bibinfo{author}{\bibfnamefont{U.}~\bibnamefont{{Seljak}}}, \bibnamefont{and}
  \bibinfo{author}{\bibfnamefont{R.~E.} \bibnamefont{{Smith}}},
  \bibinfo{journal}{\prd} \textbf{\bibinfo{volume}{85}}, \bibinfo{eid}{102001}
  (\bibinfo{year}{2012}{\natexlab{a}}), \eprint{1111.2020}.

\bibitem[{\citenamefont{{Clampitt} et~al.}(2012)\citenamefont{{Clampitt},
  {Jain}, and {Khoury}}}]{Clampitt2012}
\bibinfo{author}{\bibfnamefont{J.}~\bibnamefont{{Clampitt}}},
  \bibinfo{author}{\bibfnamefont{B.}~\bibnamefont{{Jain}}}, \bibnamefont{and}
  \bibinfo{author}{\bibfnamefont{J.}~\bibnamefont{{Khoury}}},
  \bibinfo{journal}{\jcap} \textbf{\bibinfo{volume}{1}}, \bibinfo{eid}{030}
  (\bibinfo{year}{2012}), \eprint{1110.2177}.

\bibitem[{\citenamefont{{Schmidt}
  et~al.}(2009{\natexlab{a}})\citenamefont{{Schmidt}, {Lima}, {Oyaizu}, and
  {Hu}}}]{Schmidt2009}
\bibinfo{author}{\bibfnamefont{F.}~\bibnamefont{{Schmidt}}},
  \bibinfo{author}{\bibfnamefont{M.}~\bibnamefont{{Lima}}},
  \bibinfo{author}{\bibfnamefont{H.}~\bibnamefont{{Oyaizu}}}, \bibnamefont{and}
  \bibinfo{author}{\bibfnamefont{W.}~\bibnamefont{{Hu}}},
  \bibinfo{journal}{\prd} \textbf{\bibinfo{volume}{79}}, \bibinfo{eid}{083518}
  (\bibinfo{year}{2009}{\natexlab{a}}), \eprint{0812.0545}.

\bibitem[{\citenamefont{{Moran} et~al.}(2015)\citenamefont{{Moran}, {Teyssier},
  and {Li}}}]{Moran2015}
\bibinfo{author}{\bibfnamefont{C.~C.} \bibnamefont{{Moran}}},
  \bibinfo{author}{\bibfnamefont{R.}~\bibnamefont{{Teyssier}}},
  \bibnamefont{and} \bibinfo{author}{\bibfnamefont{B.}~\bibnamefont{{Li}}},
  \bibinfo{journal}{\mnras} \textbf{\bibinfo{volume}{448}},
  \bibinfo{pages}{307} (\bibinfo{year}{2015}).

\bibitem[{\citenamefont{{Gronke} et~al.}(2015)\citenamefont{{Gronke},
  {Llinares}, {Mota}, and {Winther}}}]{Gronke2015}
\bibinfo{author}{\bibfnamefont{M.}~\bibnamefont{{Gronke}}},
  \bibinfo{author}{\bibfnamefont{C.}~\bibnamefont{{Llinares}}},
  \bibinfo{author}{\bibfnamefont{D.~F.} \bibnamefont{{Mota}}},
  \bibnamefont{and} \bibinfo{author}{\bibfnamefont{H.~A.}
  \bibnamefont{{Winther}}}, \bibinfo{journal}{\mnras}
  \textbf{\bibinfo{volume}{449}}, \bibinfo{pages}{2837} (\bibinfo{year}{2015}),
  \eprint{1412.0066}.

\bibitem[{\citenamefont{{Hammami} et~al.}(2015)\citenamefont{{Hammami},
  {Llinares}, {Mota}, and {Winther}}}]{Hammami2015}
\bibinfo{author}{\bibfnamefont{A.}~\bibnamefont{{Hammami}}},
  \bibinfo{author}{\bibfnamefont{C.}~\bibnamefont{{Llinares}}},
  \bibinfo{author}{\bibfnamefont{D.~F.} \bibnamefont{{Mota}}},
  \bibnamefont{and} \bibinfo{author}{\bibfnamefont{H.~A.}
  \bibnamefont{{Winther}}}, \bibinfo{journal}{\mnras}
  \textbf{\bibinfo{volume}{449}}, \bibinfo{pages}{3635} (\bibinfo{year}{2015}),
  \eprint{1503.02004}.

\bibitem[{\citenamefont{{Lombriser}
  et~al.}(2012{\natexlab{b}})\citenamefont{{Lombriser}, {Koyama}, {Zhao}, and
  {Li}}}]{Lombriser2012a}
\bibinfo{author}{\bibfnamefont{L.}~\bibnamefont{{Lombriser}}},
  \bibinfo{author}{\bibfnamefont{K.}~\bibnamefont{{Koyama}}},
  \bibinfo{author}{\bibfnamefont{G.-B.} \bibnamefont{{Zhao}}},
  \bibnamefont{and} \bibinfo{author}{\bibfnamefont{B.}~\bibnamefont{{Li}}},
  \bibinfo{journal}{\prd} \textbf{\bibinfo{volume}{85}}, \bibinfo{eid}{124054}
  (\bibinfo{year}{2012}{\natexlab{b}}), \eprint{1203.5125}.

\bibitem[{\citenamefont{{Tsujikawa} and {Tatekawa}}(2008)}]{Tsujikawa2008}
\bibinfo{author}{\bibfnamefont{S.}~\bibnamefont{{Tsujikawa}}} \bibnamefont{and}
  \bibinfo{author}{\bibfnamefont{T.}~\bibnamefont{{Tatekawa}}},
  \bibinfo{journal}{Physics Letters B} \textbf{\bibinfo{volume}{665}},
  \bibinfo{pages}{325} (\bibinfo{year}{2008}), \eprint{0804.4343}.

\bibitem[{\citenamefont{{Lam} et~al.}(2013)\citenamefont{{Lam}, {Schmidt},
  {Nishimichi}, and {Takada}}}]{Lam2013}
\bibinfo{author}{\bibfnamefont{T.~Y.} \bibnamefont{{Lam}}},
  \bibinfo{author}{\bibfnamefont{F.}~\bibnamefont{{Schmidt}}},
  \bibinfo{author}{\bibfnamefont{T.}~\bibnamefont{{Nishimichi}}},
  \bibnamefont{and} \bibinfo{author}{\bibfnamefont{M.}~\bibnamefont{{Takada}}},
  \bibinfo{journal}{\prd} \textbf{\bibinfo{volume}{88}}, \bibinfo{eid}{023012}
  (\bibinfo{year}{2013}), \eprint{1305.5548}.

\bibitem[{\citenamefont{{Zu} et~al.}(2014)\citenamefont{{Zu}, {Weinberg},
  {Jennings}, {Li}, and {Wyman}}}]{Zu2014}
\bibinfo{author}{\bibfnamefont{Y.}~\bibnamefont{{Zu}}},
  \bibinfo{author}{\bibfnamefont{D.~H.} \bibnamefont{{Weinberg}}},
  \bibinfo{author}{\bibfnamefont{E.}~\bibnamefont{{Jennings}}},
  \bibinfo{author}{\bibfnamefont{B.}~\bibnamefont{{Li}}}, \bibnamefont{and}
  \bibinfo{author}{\bibfnamefont{M.}~\bibnamefont{{Wyman}}},
  \bibinfo{journal}{\mnras} \textbf{\bibinfo{volume}{445}},
  \bibinfo{pages}{1885} (\bibinfo{year}{2014}), \eprint{1310.6768}.

\bibitem[{\citenamefont{Zhao et~al.}(2011)\citenamefont{Zhao, Li, and
  Koyama}}]{Zhao:2010qy}
\bibinfo{author}{\bibfnamefont{G.-B.} \bibnamefont{Zhao}},
  \bibinfo{author}{\bibfnamefont{B.}~\bibnamefont{Li}}, \bibnamefont{and}
  \bibinfo{author}{\bibfnamefont{K.}~\bibnamefont{Koyama}},
  \bibinfo{journal}{Phys.Rev.} \textbf{\bibinfo{volume}{D83}},
  \bibinfo{pages}{044007} (\bibinfo{year}{2011}), \eprint{1011.1257}.

\bibitem[{\citenamefont{{Ferraro} et~al.}(2011)\citenamefont{{Ferraro},
  {Schmidt}, and {Hu}}}]{Ferraro2011}
\bibinfo{author}{\bibfnamefont{S.}~\bibnamefont{{Ferraro}}},
  \bibinfo{author}{\bibfnamefont{F.}~\bibnamefont{{Schmidt}}},
  \bibnamefont{and} \bibinfo{author}{\bibfnamefont{W.}~\bibnamefont{{Hu}}},
  \bibinfo{journal}{\prd} \textbf{\bibinfo{volume}{83}}, \bibinfo{eid}{063503}
  (\bibinfo{year}{2011}), \eprint{1011.0992}.

\bibitem[{\citenamefont{{Li} et~al.}(2012)\citenamefont{{Li}, {Zhao}, and
  {Koyama}}}]{Li2012}
\bibinfo{author}{\bibfnamefont{B.}~\bibnamefont{{Li}}},
  \bibinfo{author}{\bibfnamefont{G.-B.} \bibnamefont{{Zhao}}},
  \bibnamefont{and} \bibinfo{author}{\bibfnamefont{K.}~\bibnamefont{{Koyama}}},
  \bibinfo{journal}{\mnras} \textbf{\bibinfo{volume}{421}},
  \bibinfo{pages}{3481} (\bibinfo{year}{2012}), \eprint{1111.2602}.

\bibitem[{\citenamefont{{Lombriser} et~al.}(2013)\citenamefont{{Lombriser},
  {Li}, {Koyama}, and {Zhao}}}]{Lombriser2013}
\bibinfo{author}{\bibfnamefont{L.}~\bibnamefont{{Lombriser}}},
  \bibinfo{author}{\bibfnamefont{B.}~\bibnamefont{{Li}}},
  \bibinfo{author}{\bibfnamefont{K.}~\bibnamefont{{Koyama}}}, \bibnamefont{and}
  \bibinfo{author}{\bibfnamefont{G.-B.} \bibnamefont{{Zhao}}},
  \bibinfo{journal}{\prd} \textbf{\bibinfo{volume}{87}}, \bibinfo{eid}{123511}
  (\bibinfo{year}{2013}), \eprint{1304.6395}.

\bibitem[{\citenamefont{{Schmidt}
  et~al.}(2009{\natexlab{b}})\citenamefont{{Schmidt}, {Vikhlinin}, and
  {Hu}}}]{Schmidt2009a}
\bibinfo{author}{\bibfnamefont{F.}~\bibnamefont{{Schmidt}}},
  \bibinfo{author}{\bibfnamefont{A.}~\bibnamefont{{Vikhlinin}}},
  \bibnamefont{and} \bibinfo{author}{\bibfnamefont{W.}~\bibnamefont{{Hu}}},
  \bibinfo{journal}{Phys. Rev. D} \textbf{\bibinfo{volume}{80}},
  \bibinfo{pages}{083505} (\bibinfo{year}{2009}{\natexlab{b}}).

\bibitem[{\citenamefont{{Cataneo} et~al.}(2014)\citenamefont{{Cataneo},
  {Rapetti}, {Schmidt}, {Mantz}, {Allen}, {Applegate}, {Kelly}, {von der
  Linden}, and {Morris}}}]{Cataneo2014}
\bibinfo{author}{\bibfnamefont{M.}~\bibnamefont{{Cataneo}}},
  \bibinfo{author}{\bibfnamefont{D.}~\bibnamefont{{Rapetti}}},
  \bibinfo{author}{\bibfnamefont{F.}~\bibnamefont{{Schmidt}}},
  \bibinfo{author}{\bibfnamefont{A.~B.} \bibnamefont{{Mantz}}},
  \bibinfo{author}{\bibfnamefont{S.~W.} \bibnamefont{{Allen}}},
  \bibinfo{author}{\bibfnamefont{D.~E.} \bibnamefont{{Applegate}}},
  \bibinfo{author}{\bibfnamefont{P.~L.} \bibnamefont{{Kelly}}},
  \bibinfo{author}{\bibfnamefont{A.}~\bibnamefont{{von der Linden}}},
  \bibnamefont{and} \bibinfo{author}{\bibfnamefont{R.~G.}
  \bibnamefont{{Morris}}}, \bibinfo{journal}{ArXiv e-prints}
  (\bibinfo{year}{2014}), \eprint{1412.0133}.

\bibitem[{\citenamefont{{Boubekeur} et~al.}(2014)\citenamefont{{Boubekeur},
  {Giusarma}, {Mena}, and {Ram{\'{\i}}rez}}}]{Boubekeur2014}
\bibinfo{author}{\bibfnamefont{L.}~\bibnamefont{{Boubekeur}}},
  \bibinfo{author}{\bibfnamefont{E.}~\bibnamefont{{Giusarma}}},
  \bibinfo{author}{\bibfnamefont{O.}~\bibnamefont{{Mena}}}, \bibnamefont{and}
  \bibinfo{author}{\bibfnamefont{H.}~\bibnamefont{{Ram{\'{\i}}rez}}},
  \bibinfo{journal}{\prd} \textbf{\bibinfo{volume}{90}}, \bibinfo{eid}{103512}
  (\bibinfo{year}{2014}), \eprint{1407.6837}.

\bibitem[{\citenamefont{{Mak} et~al.}(2012)\citenamefont{{Mak}, {Pierpaoli},
  {Schmidt}, and {Macellari}}}]{Mak2012}
\bibinfo{author}{\bibfnamefont{D.~S.~Y.} \bibnamefont{{Mak}}},
  \bibinfo{author}{\bibfnamefont{E.}~\bibnamefont{{Pierpaoli}}},
  \bibinfo{author}{\bibfnamefont{F.}~\bibnamefont{{Schmidt}}},
  \bibnamefont{and}
  \bibinfo{author}{\bibfnamefont{N.}~\bibnamefont{{Macellari}}},
  \bibinfo{journal}{\prd} \textbf{\bibinfo{volume}{85}}, \bibinfo{eid}{123513}
  (\bibinfo{year}{2012}), \eprint{1111.1004}.

\bibitem[{\citenamefont{{Okada} et~al.}(2013)\citenamefont{{Okada}, {Totani},
  and {Tsujikawa}}}]{Okada2013}
\bibinfo{author}{\bibfnamefont{H.}~\bibnamefont{{Okada}}},
  \bibinfo{author}{\bibfnamefont{T.}~\bibnamefont{{Totani}}}, \bibnamefont{and}
  \bibinfo{author}{\bibfnamefont{S.}~\bibnamefont{{Tsujikawa}}},
  \bibinfo{journal}{\prd} \textbf{\bibinfo{volume}{87}}, \bibinfo{eid}{103002}
  (\bibinfo{year}{2013}), \eprint{1208.4681}.

\bibitem[{\citenamefont{{Bel} et~al.}(2014)\citenamefont{{Bel}, {Brax},
  {Marinoni}, and {Valageas}}}]{Bel2014}
\bibinfo{author}{\bibfnamefont{J.}~\bibnamefont{{Bel}}},
  \bibinfo{author}{\bibfnamefont{P.}~\bibnamefont{{Brax}}},
  \bibinfo{author}{\bibfnamefont{C.}~\bibnamefont{{Marinoni}}},
  \bibnamefont{and}
  \bibinfo{author}{\bibfnamefont{P.}~\bibnamefont{{Valageas}}},
  \bibinfo{journal}{ArXiv e-prints}  (\bibinfo{year}{2014}),
  \eprint{1406.3347}.

\bibitem[{\citenamefont{Jain and VanderPlas}(2011)}]{Jain:2011ji}
\bibinfo{author}{\bibfnamefont{B.}~\bibnamefont{Jain}} \bibnamefont{and}
  \bibinfo{author}{\bibfnamefont{J.}~\bibnamefont{VanderPlas}},
  \bibinfo{journal}{JCAP} \textbf{\bibinfo{volume}{1110}}, \bibinfo{pages}{032}
  (\bibinfo{year}{2011}), \eprint{1106.0065}.

\bibitem[{\citenamefont{{Vikram} et~al.}(2014)\citenamefont{{Vikram},
  {Sakstein}, {Davis}, and {Neil}}}]{Vikram2014}
\bibinfo{author}{\bibfnamefont{V.}~\bibnamefont{{Vikram}}},
  \bibinfo{author}{\bibfnamefont{J.}~\bibnamefont{{Sakstein}}},
  \bibinfo{author}{\bibfnamefont{C.}~\bibnamefont{{Davis}}}, \bibnamefont{and}
  \bibinfo{author}{\bibfnamefont{A.}~\bibnamefont{{Neil}}},
  \bibinfo{journal}{ArXiv e-prints}  (\bibinfo{year}{2014}),
  \eprint{1407.6044}.

\bibitem[{\citenamefont{{Arnold} et~al.}(2015)\citenamefont{{Arnold},
  {Puchwein}, and {Springel}}}]{Arnold2015}
\bibinfo{author}{\bibfnamefont{C.}~\bibnamefont{{Arnold}}},
  \bibinfo{author}{\bibfnamefont{E.}~\bibnamefont{{Puchwein}}},
  \bibnamefont{and}
  \bibinfo{author}{\bibfnamefont{V.}~\bibnamefont{{Springel}}},
  \bibinfo{journal}{\mnras} \textbf{\bibinfo{volume}{448}},
  \bibinfo{pages}{2275} (\bibinfo{year}{2015}), \eprint{1411.2600}.

\bibitem[{\citenamefont{Dvali et~al.}(2011)\citenamefont{Dvali, Giudice, Gomez,
  and Kehagias}}]{Dvali:2010jz}
\bibinfo{author}{\bibfnamefont{G.}~\bibnamefont{Dvali}},
  \bibinfo{author}{\bibfnamefont{G.~F.} \bibnamefont{Giudice}},
  \bibinfo{author}{\bibfnamefont{C.}~\bibnamefont{Gomez}}, \bibnamefont{and}
  \bibinfo{author}{\bibfnamefont{A.}~\bibnamefont{Kehagias}},
  \bibinfo{journal}{JHEP} \textbf{\bibinfo{volume}{08}}, \bibinfo{pages}{108}
  (\bibinfo{year}{2011}), \eprint{1010.1415}.

\bibitem[{\citenamefont{Keltner and Tolley}(2015)}]{Keltner:2015xda}
\bibinfo{author}{\bibfnamefont{L.}~\bibnamefont{Keltner}} \bibnamefont{and}
  \bibinfo{author}{\bibfnamefont{A.~J.} \bibnamefont{Tolley}}
  (\bibinfo{year}{2015}), \eprint{1502.05706}.

\bibitem[{\citenamefont{de~Rham}(2012)}]{deRham:2012az}
\bibinfo{author}{\bibfnamefont{C.}~\bibnamefont{de~Rham}},
  \bibinfo{journal}{Comptes Rendus Physique} \textbf{\bibinfo{volume}{13}},
  \bibinfo{pages}{666} (\bibinfo{year}{2012}), \eprint{1204.5492}.

\end{thebibliography}

\end{document}